\documentclass[fleqn,usenatbib]{mnras}

\usepackage[T1]{fontenc}

\DeclareRobustCommand{\VAN}[3]{#2}
\let\VANthebibliography\thebibliography
\def\thebibliography{\DeclareRobustCommand{\VAN}[3]{##3}\VANthebibliography}

\usepackage{graphicx}	% Including figure files
\usepackage{amsmath}	% Advanced maths commands
\usepackage{amssymb}	% Extra maths symbols

\usepackage{newtxtext,newtxmath}
\usepackage{bm}
\usepackage{verbatim}
\usepackage{mathrsfs}
\usepackage{array}
\newcolumntype{R}{>{$}r<{$}} % math-mode version of "l" column type
\usepackage{ulem}%%%%%%%%%%%%%%%%%%%%%%%%%%%%%%%%%%%%%%%%%%%%%%%%%%

\defcitealias{2020MNRAS.497..498C}{CUBS I}
\defcitealias{2021ApJ...913...18B}{CUBS II}
\defcitealias{2021MNRAS.506..877Z}{CUBS III}

\newcommand{\N}[1]{\ensuremath{N_\mathrm{#1}}}
\newcommand{\kms}{\ensuremath{\mathrm{km\ s}^{-1}}}
\newcommand{\xvh}{\ensuremath{\left\{\left[\mathrm{X}/\mathrm{H}\right]\right\}}}
\newcommand{\xpm}[3]{\ensuremath{#1^{#2}_{#3}}}
\newcommand{\cmx}[1]{\ensuremath{\mathrm{cm}^{#1}}}
\newcommand{\nH}{\ensuremath{n_\mathrm{H}}}
\newcommand{\mstar}{{\mbox{$M_{\rm star}$}}}
\newcommand{\msun}{{\mbox{M$_{\odot}$}}}
\newcommand{\nodata}{\ensuremath{~\cdots~}}% 

\title[The Cosmic Ultraviolet Baryon Survey (CUBS) IV]{The Cosmic Ultraviolet Baryon Survey (CUBS) IV: The Complex Multiphase Circumgalactic Medium as Revealed by Partial Lyman Limit Systems}

\author[Cooper and the CUBS team]{Thomas J.\ Cooper,$^1$\thanks{E-mail: tcooper@carnegiescience.edu},
Gwen C.\ Rudie,$^{1}$,
Hsiao-Wen Chen,$^{2}$
Sean D.\ Johnson,$^{3}$
Fakhri S.\ Zahedy,$^{1}$
\newauthor
Mandy C.\ Chen,$^{2}$
Erin Boettcher,$^{2}$
Gregory L.\ Walth,$^{3}$
Sebastiano Cantalupo,$^{4,5}$
Kathy L.\ Cooksey,$^{6}$
\newauthor
Claude-Andr\'e Faucher-Gigu\`ere,$^{7}$
Jenny E.\ Greene,$^{8}$
Sebastian Lopez,$^{9}$
John S.\ Mulchaey,$^{1}$
\newauthor
Steven V.\ Penton,$^{10}$
Patrick Petitjean,$^{11}$
Mary E.\ Putman,$^{12}$
Marc Rafelski,$^{13,14}$
Michael Rauch,$^{1}$
\newauthor
Joop Schaye,$^{15}$
and Robert A.\ Simcoe$^{16}$
\\
$^{1}$The Observatories of the Carnegie Institution for Science, 813 Santa Barbara Street, Pasadena, CA 91101, USA\\
$^{2}$Department of Astronomy \& Astrophysics, The University of Chicago, Chicago, IL 60637, USA\\
$^{3}$Department of Astronomy, University of Michigan, Ann Arbor, MI 48109, USA\\
$^{4}$Department of Physics, University of Milan Bicocca, Piazza della Scienza 3, 20126 Milano, Italy\\
$^{5}$Department of Physics, ETH Zurich, Wolfgang-Pauli-Strasse 27, 8093, CH-8093 Zurich, Switzerland\\
$^{6}$Department of Physics and Astronomy, University of Hawai`i at Hilo, Hilo, HI 96720, USA\\
$^{7}$Department of Physics \& Astronomy and Center for Interdisciplinary Exploration and Research in Astrophysics (CIERA),\\ Northwestern University, 1800 Sherman Ave, Evanston, IL 60201, USA\\
$^{8}$Department of Astrophysics, Princeton University, Princeton, NJ 08544, USA\\
$^{9}$Departamento de Astronom\'ia, Universidad de Chile, Casilla 36-D, Santiago, Chile\\
$^{10}$Laboratory For Atmospheric and Space Physics, University of Colorado, Boulder, CO 80303, USA\\
$^{11}$Institut d'Astrophysique de Paris, CNRS-SU, UMR 7095, 98bis bd Arago, Paris F-75014, France\\
$^{12}$Department of Astronomy, Columbia University, New York, NY 10027, USA\\
$^{13}$Space Telescope Science Institute, Baltimore, MD 21218, USA\\
$^{14}$Department of Physics \& Astronomy, Johns Hopkins University, Baltimore, MD 21218, USA\\
$^{15}$Leiden Observatory, Leiden University, PO Box 9513, NL-2300 RA Leiden, the Netherlands\\
$^{16}$MIT-Kavli Institute for Astrophysics and Space Research; 77 Massachusetts Ave., Cambridge, MA 02139, USA\\
}

\date{Accepted XXX. Received YYY; in original form ZZZ}

\pubyear{2021}

\begin{document}
\label{firstpage}
\pagerange{\pageref{firstpage}--\pageref{lastpage}}
\maketitle

% Abstract of the paper
\begin{abstract}
    We present a detailed study of two partial Lyman limit systems (pLLSs) of neutral hydrogen column density $\N{H\,I}\approx(1-3)\times10^{16}\,\cmx{-2}$ discovered at $z=0.5$ in the Cosmic Ultraviolet Baryon Survey (CUBS). Available far-ultraviolet spectra from the {\it Hubble Space Telescope} Cosmic Origins Spectrograph and optical echelle spectra from MIKE on the Magellan Telescopes enable a comprehensive ionization analysis of diffuse circumgalactic gas based on resolved kinematics and abundance ratios of atomic species spanning five different ionization stages. These data provide unambiguous evidence of kinematically aligned multi-phase gas that masquerades as a single-phase structure and can only be resolved by simultaneous accounting of the full range of observed ionic species. Both systems are resolved into multiple components with inferred $\alpha$-element abundance varying from $[\alpha/\text{H}]\approx\!{-0.8}$ to near solar and densities spanning over two decades from $\log\nH/\cmx{-3}\approx\!-2.2$ to $<-4.3$. Available deep galaxy survey data from the CUBS program taken with VLT/MUSE, Magellan/LDSS3-C and Magellan/IMACS reveal that the $z=0.47$ system is located 55 kpc from a star-forming galaxy with prominent Balmer absorption of stellar mass $\mstar\approx 2\times 10^{10}\,\msun$, while the $z=0.54$ system resides in an over-dense environment of 11 galaxies within 750 kpc in projected distance, with the most massive being a luminous red galaxy of $\mstar\approx2\times 10^{11}\,\msun$ at 375 kpc. The study of these two pLLSs adds to an emerging picture of the complex, multiphase circumgalactic gas that varies in chemical abundances and density on small spatial scales in diverse galaxy environments.  The inhomogeneous nature of metal enrichment and density revealed in observations must be taken into account in theoretical models of diffuse halo gas.
\end{abstract}


\section{Introduction}

Gas within the extended halos of galaxies, known as the circumgalactic medium (CGM), provides a novel view of the dominant physical processes responsible for galaxy evolution. In addition to serving as the interface regulating gas exchange between galaxies and the intergalactic medium, the CGM provides unique clues about the nature of feedback from star formation, supernovae, and AGN activity within galaxies (see \citealp{2017ARA&A..55..389T}, \citealp{2017ASSL..430.....F}, and \citealp{2017ASSL..434..291C} for recent reviews). The CGM is most readily observed via absorption in the spectra of background quasars, enabling detailed studies of various gas properties and their evolution with redshift. However, interpretation of such absorbers is limited without knowledge of the galactic environment to provide astrophysical context for the absorbers through comparisons of absorber and galaxy properties.

The circumgalactic medium is expected to span several orders of magnitude in density, including a cold ($10^4~$K) high-density phase and a volume filling diffuse hot phase with a temperature set by the gravitational potential of the halo (see, e.g. \citealt{2012MNRAS.419.3319M,2017ApJ...845...80V,2019MNRAS.488.2549S}). Recent observations have confirmed aspects of these models, including finding gas occupying a wide range of temperatures and densities (see, e.g. \citealt{2019MNRAS.484.2257Z, 2021MNRAS.506..877Z}; \citealt{2019ApJ...885...61R}). \citet{2019MNRAS.484.2257Z} found the cold phase of the CGM in halos of $z\sim0.5$ massive ellipticals to vary in both density and metallicity by over a factor of ten. Simulations demonstrate that, as a result of gas dynamics driven by accreting, outflowing, and recycling gas \citep{2012MNRAS.423.2991V, 2014MNRAS.444.1260F,2017MNRAS.470.4698A,2018MNRAS.481..835O,2019MNRAS.488.1248H}, these different phases are seen throughout galaxy halos rather than correlating strongly with distance from galaxy centers \citep{2020ApJ...903...32F}, suggesting that accurate measurements of multiphase properties may be essential to understanding the physics that shapes the CGM and the galaxies it envelopes \citep{2019ApJ...873..129P}.

However, observations of CGM absorption systems are often limited in a fashion that precludes analysis encapsulating the complexities introduced by the multiphase nature and varying abundances of the CGM. For example, \citet{2015ApJ...812...58C} and \citet{2016ApJ...833..270G} used unresolved spectra to study $z\sim3$ Lyman limit systems (LLSs, systems with $17.2\leq\log\,\N{H\,I}/\cmx{-2}\leq19.0)$, measuring bulk column densities of entire absorption complexes rather than components, and estimating gas properties based on these measurements. Most CGM inferences about abundances and densities rely on such assumptions and they may yield bulk metallicity inferences comparable to the average metallicity of the various components that are blended together. However, recent work exploring detailed multi-component and multiphase photoionization modeling underlines the shortcomings in approaches that use such simplifying assumptions \citep{2019MNRAS.484.2257Z,2021MNRAS.506..877Z,2020MNRAS.tmp.3454H,2021MNRAS.501.2112S}. These works demonstrate that gas at a wide range of densities and chemical abundances can be present within the same halo \citep[see also][]{2001A&A...370..729D}. Further, an absorption feature that appears to be a single discrete component may have contributions to the same ion from gas at multiple densities, suggesting either unresolved structure in the absorption profile, or varying densities within a single gaseous structure.

Absorbers with neutral hydrogen column densities of $\N{H\,I}=10^{15}$ - 10$^{17}$ cm$^{-2}$ (partial Lyman limit systems, or pLLSs) are the most likely to represent a typical sightline through the CGM \citep{2011MNRAS.418.1796F,2015MNRAS.449..987F,2017MNRAS.469.2292H}. \citet{2012ApJ...750...67R} find that, at $z\sim2.5$, around 90\% of sightlines within 100 kpc of Lyman break galaxies contain a pLLS, along with weaker components. At lower redshifts, at least half of all pLLSs are within 300 kpc of galaxies \citep[e.g.,][]{2009ApJ...701.1219C,2013ApJ...777...59T,2015MNRAS.449.3263J,2017ApJ...837..169P,2019ApJS..243...24P,2019MNRAS.484..431C}. Therefore, the distribution of LLS and pLLS metallicities provides significant insight into the nature of the CGM, even without corresponding galaxy data \citep{2013ApJ...770..138L,2015ApJ...812...58C,2016ApJ...833..270G,2019ApJ...872...81W,2019ApJ...887....5L}. 

However, our ability to interpret these LLS properties in the lens of galaxy formation and evolution is greatly enhanced by connecting individual absorbers with their galactic environments. Early studies suggest metallicity gradients, decreasing with distance from the ISM to CGM \citep{2005ApJ...620..703C, 2014MNRAS.445..225C, 2016MNRAS.457..903P}, and continuing to decline until reaching abundances typical of the outskirts of halos and into the intergalactic medium \citep[IGM;][]{2006ApJ...637..648S}, although such conclusions require refinement as samples are small ($N_{\rm gal}=6$). Relative abundance gradients have also been discovered, with CGM gas in both quiescent and star-forming galaxies having increased $\alpha$-element abundances, relative to Fe, at larger galactocentric radii \citep{2017MNRAS.466.1071Z}. This evidence of less chemical maturity in the outskirts of halos suggests increased star-formation driven enrichment within the inner halo and/or greater dilution of the outer halo by accretion of gas with more chemically primitive abundance patterns \citep[e.g.,][]{1995ApJ...454...69P,1997ApJ...481..601R,2017MNRAS.468.4170M}. A heuristic picture of gas dynamics with metal-poor accretion along the major axes and enriched outflowing gas along minor axes \citep[e.g.][]{2012MNRAS.423.2991V,2020MNRAS.499.2462P} is challenged by observations demonstrating CGM metallicities appear to not be strongly azimuthally dependent (\citealt{2016MNRAS.457..903P,2019ApJ...883...78P,2019ApJ...886...91K}; but see \citealt{2021MNRAS.502.3733W}). Robust absolute {\it and} relative chemical abundance measurements considered jointly with galactic context are key to understanding the mechanisms that drive such relations.

In addition to identifying a plausible host-galaxy at the redshift of an absorber, further environmental context can be obtained by conducting a survey to measure redshifts of \textit{all} nearby galaxies likely to be at the same redshift. Since the CGM conducts gas between galaxies and the intergalactic medium, it is likely to be sensitive to the galaxy environment. \citet{2020arXiv200906646M} demonstrate the impact of feedback from nearby galaxies on halo gas, with larger galaxies suppressing star-formation in their dwarf neighbors. Several recent works have demonstrated that many absorbers likely arise from complex galaxy structures and galaxy groups, not simply single galaxies evolving in isolation \citep{2019MNRAS.485.1595P,2020MNRAS.492.2347H,2020MNRAS.497..498C, 2021MNRAS.505..738N}. The high incidence of CGM gas within the virial halos of isolated dwarf galaxies \citep{2017ApJ...850L..10J}, coupled with the recent identification of a dwarf galaxy at small projected separation from an absorber initially thought to be associated with a massive galaxy over 300 kpc away \citep{2015ApJ...811..132M,2021MNRAS.500.3987N}, highlights the need for deep galaxy spectroscopy to fully characterize the galactic environment, critical to understanding the true nature of CGM absorbers.

The Cosmic Ultraviolet Baryon Survey (CUBS) is an effort to conduct such a galaxy census in the foreground of 15 $z\sim1$ quasars, primary focusing on CGM studies at redshifts between $z=0.4$--$0.8$ (see \citealp{2020MNRAS.497..498C}, hereafter CUBS I). The CUBS galaxy survey is designed with a tiered approach such that it is deepest and most complete at smaller impact parameters, intending to ensure that relevant lower-mass galaxies are not overlooked in favor of $L_*$ galaxies, while exploring the broader galactic environment traced by brighter galaxies (see \S\ref{sec:data:galaxies} and \citealt{2021ApJ...913...18B}, hereafter CUBS II). In this work we examine all components of two pLLSs at $z=0.47$ and $0.54$, as early CUBS results, along the line-of-sight to QSO J011035.511$-$164827.70 (hereafter J0110$-$1648) at $z_{\rm QSO}=0.7823$.

At $z\sim0.5$, the UV spectra observed for CUBS span a wide range of ionization states, necessary to explore the multiphase properties of the CGM. For example, transitions of comparable strength of all oxygen ions up to O$^{5+}$ are covered, excepting O$^{4+}$, allowing us to examine multiple ionization phases of the CGM without relying on comparisons between different elements, avoiding ambiguities resulting from possible variation in relative chemical abundances. Additionally, the ability to resolve distinct absorbing components reveals the inhomogeneities in ionization state between discrete gaseous structures within a single halo. Leveraging the broad coverage of both ionization states and elements in our data, our analysis focuses on determining which gas properties can be robustly obtained, in light of the complexity and ambiguities inherent in such analyses. We show that combining observed pLLS properties with the galactic environment revealed by the accompanying galaxy survey provides new insight into the connection between galaxies and the diffuse IGM. Notably, the two pLLSs studied here have remarkably different galactic environments, with one within 100 kpc of a galaxy with few neighbors, and the other having no \textbf{$L_*$} galaxy within 300 kpc, but over 20 galaxies within 2 Mpc.

The remainder of the paper is structured as follows. In \S\ref{sec:data} we discuss the data quality and characteristics. In \S\ref{sec:methods} we discuss the methodology used to measure kinematics and column densities, and the approach taken to photoionization modeling. Details specific to each absorption component, including the reliability of measurements and inferred properties, are given in \S\ref{sec:analysis}. For readers less interested in the details of photoionization analysis, Section \S\ref{sec:results} summarizes the derived densities and abundance patterns and presents the galactic environments of the absorbers uncovered by the CUBS galaxy survey. We combine these results with those presented in 
\citet[hereafter CUBS III]{2021MNRAS.506..877Z}
who studied optically thick Lyman limit systems in CUBS, to discuss the nature of high-\N{HI} absorbers in the intermediate redshift CGM. In \S\ref{sec:results} we also discuss ambiguities in the modeling, focusing on outlining which properties can be robustly determined. \S\ref{sec:summary} presents a summary of the paper. Throughout this paper, we assume a $\Lambda$ cosmology with $H_{0} = 67.7$ \kms\  Mpc$^{-1}$, $\Omega_{\rm m} = 0.307$, and $\Omega_{\Lambda} = 0.693$ \citep{2016A&A...594A..13P}. Errors are 1-$\sigma$ and credible intervals contain 68\% of corresponding probability distributions. All distances are expressed in physical units unless stated otherwise. All magnitudes presented in this paper are in the AB system.

\section{Data} \label{sec:data}

In this Section we present the far-UV and optical echelle spectra of QSO J0110$-$1648 and accompanying galaxy survey data, obtained as part of the CUBS program. The CUBS program design and data are discussed in \citetalias{2020MNRAS.497..498C}; here we describe the data characteristics for J0110$-$1648 relevant to the two pLLSs.

\subsection{\texorpdfstring{Spectra of QSO J0110\bm{$-$}1648}{}}
    We acquired far-ultraviolet spectroscopy of J0110$-$1648 with spectral coverage from $\lambda=1100-1800$\AA\ using the Cosmic Origins Spectrograph \citep[COS;][]{2012ApJ...744...60G} aboard the \textit{Hubble Space Telescope} ({\it HST}) during Cycle 25 (GO-CUBS; PID 15163; PI Chen). To achieve continuous spectral coverage, we take spectra using the G130M grating with two central wavelength settings (total $t_\text{exp}=15320$ s) and the G160M grating with four central wavelengths (total $t_\text{exp}=19802$ s), yielding a median signal-to-noise per resolution element of ($S/N$)$_\mathrm{resel}\sim31$. For our analysis, we adopt the COS Lifetime Position 4 line-spread function \citep{2018cos..rept....7F}, which is non-Gaussian with extended wings. Resolution elements have a full width at half-maximum (FWHM) of approximately $v_\mathrm{FWHM}\sim21\;\kms.$

    High-resolution ($v_\mathrm{FWHM}=8~\kms$) optical spectroscopy was obtained with the Magellan Inamori Kyocera Echelle (MIKE, \citealp{2003SPIE.4841.1694B}) spectrograph on UT 2018-11-01, with a 0.7$\arcsec$ slit with $2\times2$ binning. With a total $t_\text{exp}=1800$ s we achieved (S/N)$_\mathrm{resel}\sim25$ at $\lambda=4000$ \AA\ (corresponding to the \ion{Mg}{ii} $\lambda\lambda$ 2796,2803 doublet at $z\sim0.5$), and (S/N)$_\mathrm{resel}\sim10$ at $\lambda=3500$ \AA\ (\ion{Fe}{ii} 2382 \AA). 

    The high-resolution COS FUV and optical echelle spectra together enable studies of resolved component structures within individual absorption systems. In particular, the optical echelle spectrum informs analysis of the FUV spectra (see \S\ref{sec:analysis}). Further details on the observations and reduction process of both spectra can be found in \citetalias{2020MNRAS.497..498C}.

\begin{figure}
\includegraphics[width=\linewidth]{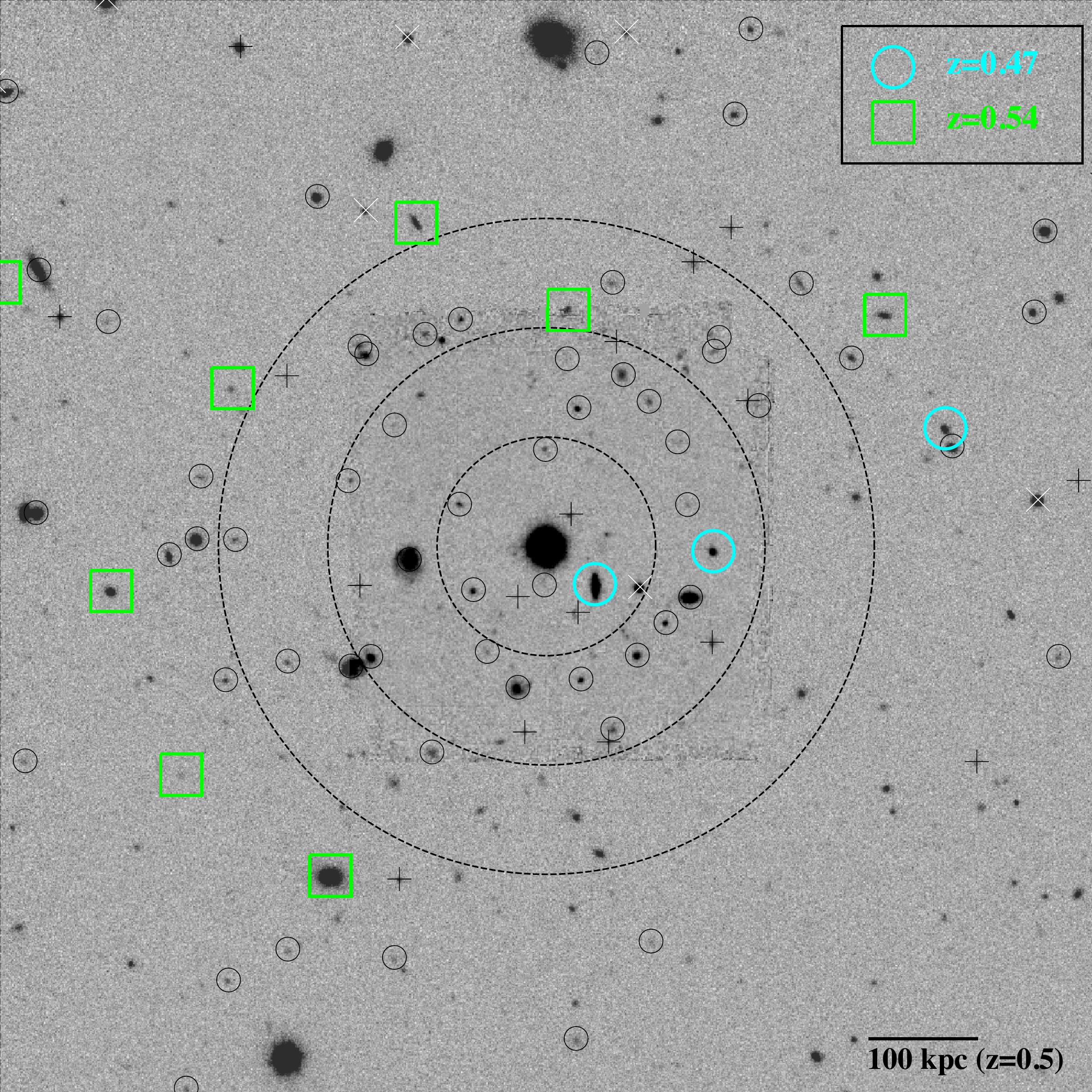}
\caption{Magellan/Fourstar $H$-band image of galaxies in the field of view of QSO J0110$-$1648, with VLT/MUSE pseudo-$r$ overlaid in the center. Galaxies with $\Delta v_g<500~\kms$ relative to the pLLSs at $z=0.47$ and $z=0.54$ are highlighted with cyan circles and green boxes, respectively. Dashed circles, centered on the quasar, indicate impact parameters of  100, 200, and 300 kpc at $z=0.5$. The image is 1 Mpc per side (about 160\arcsec). Galaxies at other known redshifts have black circles, and galaxies with spectral features that disallow either pLLS redshift have crosses. Stars are indicated with white X's. \label{fig:image}}
\end{figure}

\subsection{Galaxy Redshift Survey}\label{sec:data:galaxies}
    A defining characteristic of the CUBS program is quasar selection that is agnostic with respect to nearby galaxies; rather than selecting quasars probing the CGM of particular galaxies, quasars are selected based on observability. Quasars are selected using NUV magnitudes, to select bright QSOs without biasing against LLSs that attenuate FUV magnitudes. The main goal of the galaxy survey is to obtain redshifts for as many galaxies at $0.4<z<z_\mathrm{QSO}$ with impact parameters $d\lesssim R_\mathrm{vir}$ as possible, while minimizing the potential to introduce biases via target selection. With this in mind, the CUBS galaxy survey is designed as a three-tiered approach, with deeper and denser coverage of galaxies at smaller angular separation from the central quasars, enabling a complete view of the galaxy environment of the absorbers. We provide a summary of the galaxy survey here, and discuss observations relevant to J0110$-$1648; future works will provide details of the whole survey.

    The \textit{ultradeep} component of the redshift survey covering the inner 1$\arcmin\times1\arcmin$ region consists of VLT/MUSE \citep{2010SPIE.7735E..08B} wide-field mode, adaptive optics assisted observations (program ID 0104.A-0147, PI:Chen). For J0110$-$1648, spectra were taken over two observing blocks on November 21-22, 2020, each consisting of $3\times900$ s exposure, for a total $t_\mathrm{exp}=5400$~s and a mean FWHM of 0.6$\arcsec$. The data cube reaches a photometric limiting magnitude in pseudo-$r$ of $\approx27$ mag, and enables redshift identification of low mass galaxies that are generally fainter than the thresholds of the wider survey components ($10^8<M_\mathrm{star}/M_\odot<10^9)$, out to about 200 kpc, depending on redshift. PSF subtraction of the QSO light enables galaxy identification down to impact parameters of $\sim8$ kpc, although no galaxy is identified within the QSO PSF in this field. The MUSE cubes are reduced with a combination of the standard pipeline \citep{2014ASPC..485..451W} and \texttt{CUBEXTRACTOR} (described in \citealp{2019MNRAS.483.5188C}).

    The \textit{deep and narrow} component, conducted with Magellan/LDSS-3C in multislit mode, targets $L\gtrsim0.1\,L_*$ galaxies at $z<1$ within $1\arcmin$ in angular radius of the quasar sightline (corresponding to 370 kpc at $z=0.5$). Candidate galaxies are selected based on $g-r$ and $r-H$ colors, to a depth of $r=24$ mag (see also \citetalias{2021ApJ...913...18B}). Guided by the Ultra-VISTA survey \citep{2013ApJ...777...18M}, we estimate this selection identifies over 80\% of galaxies with $\mstar>10^{10}\,\msun$ at $0.4<z<0.8$, and over 50\% at $z=1.0$, while excluding most higher-redshift galaxies. To select galaxies for spectroscopic follow-up, we obtained deep $r-$ and $g-$band imaging using the Inamori Magellan Areal Camera and Spectrograph (IMACS, \citealp{2011PASP..123..288D}), and $H$-band imaging with the FourStar Infrared Camera \citep{2013PASP..125..654P}, both on the Magellan/Baade telescope. LDSS-3C spectra are taken with the VPH-All (400 l/mm) grism using $1\arcsec$ wide slits, resulting in a spectral resolution of $R=650$. The LDSS-3C spectroscopy of galaxies in the J0110$-$1648 field was taken on December 10-11, 2018 and October 27-29, 2019 in 0.6-0.8$\arcsec$ seeing.

    The \textit{wide and shallow} portion of the survey, performed with Magellan/IMACS, targets $L_*$ galaxies with projected distances within several Mpc from the quasar, motivated by the similarity of clustering properties between \ion{H}{i} absorbers and galaxies found in previous cross-correlation studies of galaxies and absorption systems \citep[e.g.,][]{2009ApJ...701.1219C}. IMACS spectra are taken with the short ($f$/2) camera, which has a 15$\arcmin$ radius field of view (5.5 Mpc at $z=0.5$), using the 150 l/mm grism and 1$\arcsec$ wide slits, which yields spectral resolution of $R=550$. The IMACS spectroscopy of galaxies in the J0110$-$1648 field was taken on October 8 and November 7-9, 2018, in 0.5-0.6$\arcsec$ seeing.
    
    Both sets of Magellan galaxy spectra are reduced using the \texttt{CPDMap} routine within the \texttt{CarPy} python distribution, implementing the methods outlined in \citet{2000ApJ...531..159K} and \citet{2003PASP..115..688K}. Redshift measurements are performed by constructing best-fitting (minimum $\chi^2$) model spectra from linear combinations of SDSS-III BOSS eigenspectra \citep{2012AJ....144..144B}, at redshift intervals of $\Delta z=0.0001.$  The resultant redshift is that with the overall lowest $\chi^2$, subject to visual verification and masking of contaminated wavelengths performed independently by at least two separate team members. For a small number of galaxies observed for other purposes where we have higher-resolution MagE spectra in addition to LDSS3 or IMACS spectroscopy, we find that redshifts measured from either lower-dispersion Magellan instrument agree with those measured with MagE  to within $\frac{\Delta z}{1+z} \sim 10^{-4}$. For objects with a single emission line, we infer a small set of potential redshifts based on typical transitions, such that they can be ruled out as associated with absorbers at other redshifts.

    Here we briefly summarize the characteristics of the galaxy survey around QSO J0110$-$1648; further details will be provided in a future publication describing the full galaxy survey. We obtain secure redshifts for 882 galaxies within $11\arcmin$ of the quasar, with 861 at $z<z_\mathrm{QSO}=0.7823$. Focusing on the inner $1\arcmin$ region, there are 45 galaxy candidates\footnote{We refer to targets as galaxy candidates because they may also be stars. We use DES classifications for star-galaxy separation for objects with $r<22.5$.}, 30 of which meet our color-selection criterion. Secure redshifts are identified for 23 of these color-selected galaxies. Ten out of fifteen non-color selected galaxies have secure redshifts, only one of which has $0.4<z<0.8$. In total, we identify 16 (41) foreground galaxies with impact parameters $d<200$ (300) kpc.

    Figure \ref{fig:image} shows the galaxies identified around J0110$-$1648, overlaid on a composite Magellan/Fourstar $H$-band and VLT/MUSE pseudo-$r$-band image. The resolution of the Magellan spectra is insufficient to resolve the [O\,II] $\lambda\lambda\,3727,3729$ doublet, leaving the redshifts of some galaxies with a single emission line and no detected absorption features ambiguous. For this work, single lines are sufficient to determine such galaxies are \textit{not} at either redshift of interest within a projected distance of 1 Mpc. These galaxies are indicated with black crosses. The galaxies highlighted in cyan and green are within 500 \kms\ of the $z=$0.47 and 0.54 pLLSs, respectively. Notably, these two pLLSs have remarkably different galactic environments, as we describe in \S\ref{sec:results}.

\begin{table*}
% \centering
\caption{Column densities measured with the Voigt profile fit to the partial Lyman Limit System at $z=0.4723$.\label{table:z0p47}}
\renewcommand{\arraystretch}{1.3}
\begin{tabular}{l| R| R| R| R| R| R| R| R| R| R}
\hline\hline
& \multicolumn{2}{c|}{Component 1} & \multicolumn{2}{c|}{Component 2} & \multicolumn{2}{c|}{Component 3} & \multicolumn{2}{c|}{Component 4L} & \multicolumn{2}{c}{Component 4H}\\
& \multicolumn{2}{c|}{$\Delta v_\text{c1}={0.0^{+1.4}_{-1.4}}^a$} & \multicolumn{2}{c|}{$\Delta v_\text{c2}=+36.1^{+1.0}_{-1.0}$} & \multicolumn{2}{c|}{$\Delta v_\text{c3}=+56.0^{+1.4}_{-1.5}$} & \multicolumn{2}{c|}{$\Delta v_\mathrm{c4L}=+191.2^{+0.5}_{-0.5}$} & \multicolumn{2}{c}{$\Delta v_\mathrm{4H}=+191.3^{+1.3}_{-1.3}$}\\\hline
\multicolumn{1}{c|}{Ion} & \multicolumn{1}{c|}{$\log N_\text{c1}/\cmx{-2}$} & \multicolumn{1}{c|}{${b_\text{c1}}^b$} & \multicolumn{1}{c|}{$\log N_\text{c2}/\cmx{-2}$} & \multicolumn{1}{c|}{$b_\text{c2}$} & \multicolumn{1}{c|}{$\log N_\text{c3}/\cmx{-2}$} & \multicolumn{1}{c|}{$b_\text{c3}$} & \multicolumn{1}{c|}{$\log N_\text{c4L}/\cmx{-2}$} & \multicolumn{1}{c|}{$b_\text{c4L}$} & \multicolumn{1}{c|}{$\log N_\text{c4H}/\cmx{-2}$} & \multicolumn{1}{c}{$b_\text{c4H}$}\\
\hline
    \ion{H}{I} & 16.03^{+0.02}_{-0.02} & 20.9^{+1.4}_{-1.3}
        & 15.10^{+0.04}_{-0.06} & 22.5^{+2.4}_{-2.7}
        & 14.53^{+0.08}_{-0.09} & 34.5^{+3.0}_{-2.6}
        & 15.14^{+0.03}_{-0.03} & 14.3^{+0.5}_{-0.5}
        & \nodata               & \nodata \\
    \ion{C}{ii}    & 13.26^{+0.06}_{-0.07} & 16.3^{+6.0}_{-6.3}
            & 13.21^{+0.08}_{-0.08} & 11.0^{+7.3}_{-6.0}
            & <13.47          & \text{(\ion{O}{iii})}
            & 13.54^{+0.14}_{-0.08} & 8.1^{+7.2}_{-4.1}
            & \nodata               & \nodata \\
    \ion{C}{III}   & 13.72^{+0.02}_{-0.02} & 30.3^{+1.6}_{-1.6}
            & 13.14^{+0.18}_{-0.18} & 13.9^{+4.3}_{-6.9}
            & 13.69^{+0.05}_{-0.04} & 19.2^{+1.5}_{-1.4}
            & 13.37^{+0.60}_{-0.62} & \text{(\ion{O}{ii})}
            & 13.63^{+0.07}_{-0.08} & \text{(\ion{O}{iv})}\\
    \ion{N}{II}    & <13.03               & \text{(\ion{O}{ii})$^b$}
            & <12.94               & \text{(\ion{O}{ii})}
            & <13.03               & \text{(\ion{O}{iii})}
            & <13.11               & \text{(\ion{O}{ii})}
            & \nodata               & \nodata \\
    \ion{N}{III}   & <13.26                 & \text{(\ion{O}{iii})}
            & <13.32                 & \text{(\ion{O}{iii})}
            & 13.58^{+0.05}_{-0.05}  & 22.9^{+7.5}_{-6.9}
            & 13.71^{+0.09}_{-0.15}  & \text{(\ion{O}{ii})}            
            & <13.84                 & \text{(\ion{N}{iv})} \\            
            % & 13.76^{+0.13}_{-0.07}  & 7.5^{+3.1}_{-2.5}
            % & \nodata                & \nodata \\
    \ion{N}{IV}    & <12.72                 & \text{(\ion{O}{iv})}
            & <13.14                & \text{(\ion{O}{iii})}
            & 13.61^{+0.05}_{-0.05}  & 25.6^{+4.6}_{-4.1}
            & \nodata                & \nodata
            & 13.53^{+0.17}_{-0.10}  & 12.8^{+6.1}_{-4.5} \\
    \ion{O}{I}     & <13.24 & \text{(\ion{O}{ii})}
            & <13.37 & \text{(\ion{O}{ii})}
            & <13.42 & \text{(\ion{O}{iii})}
            & <12.93 & \text{(\ion{O}{ii})}
            & \nodata & \nodata \\
    \ion{O}{II}    & 13.74^{+0.05}_{-0.05} & 14.6^{+5.3}_{-5.6}
            & 12.93^{+0.26}_{-0.35} & 17.2^{+8.7}_{-10.1}
            & <14.11                & \text{(\ion{O}{iii})}
            & 13.50^{+0.06}_{-0.06} & 7.6^{+2.7}_{-1.3}
            & \nodata               & \nodata \\
    \ion{O}{III}   & 14.22^{+0.05}_{-0.07} & 32.7^{+5.7}_{-4.9}
            & 13.97^{+0.21}_{-0.27} & 29.0^{+12.1}_{-15.0}
            & 13.94^{+0.12}_{-0.21} & 12.0^{+5.8}_{-5.5}
            & 13.84^{+0.21}_{-0.34} & \text{(\ion{O}{ii})}
            & 13.94^{+0.12}_{-0.17} & \text{(\ion{O}{iv})} \\
    \ion{O}{IV}    & 14.05^{+0.05}_{-0.05} & 28.6^{+8.4}_{-7.0} 
            & <13.61                & \text{(\ion{O}{iii})}
            & 14.46^{+0.03}_{-0.03} & 23.5^{+2.9}_{-2.5}
            & \nodata               & \nodata
            & 14.38^{+0.03}_{-0.02} & 22.9^{+2.6}_{-2.4} \\
    \ion{O}{VI}    & 13.95^{+0.05}_{-0.06} & 53.9^{+1.5}_{-1.4}
            & <13.96                & \text{(\ion{O}{iii})}
            & 13.96^{+0.06}_{-0.08} & 43.8^{+6.0}_{-6.1}
            & \nodata               & \nodata
            & 14.25^{+0.02}_{-0.02} & 56.4^{+3.9}_{-3.7} \\
    \ion{Si}{II}   & <12.50                & \text{(\ion{Mg}{ii})}
            & <12.46                & \text{(\ion{Mg}{ii})}
            & <12.52                & \text{(\ion{O}{iii})}
            & <12.90                & \text{(\ion{Mg}{ii})}
            & \nodata               & \nodata \\
    \ion{Si}{III}  & 12.83^{+0.07}_{-0.07}  & 24.6^{+7.8}_{-6.5}
            & \nodata                & \nodata
            & \nodata                & \nodata
            & 12.44^{+0.22}_{-0.39}  & \text{(\ion{O}{ii})}
            & <12.63                 & \text{(\ion{O}{iv})} \\
            % & 12.58^{+0.32}_{-0.13}  & 10.2^{+6.4}_{-6.1} \\
    \ion{S}{II}  & 12.56^{+0.18}_{-0.34}  & 10.2^{+6.4}_{-5.8}
            & <12.22                 & \text{(\ion{C}{ii})}
            & <12.32                 & \text{(\ion{C}{iii})}
            & <12.51                 & \text{(\ion{C}{ii})}
            & \nodata                & \nodata\\
    \ion{S}{III}   & <13.52                & \text{(\ion{O}{iii})}
            & <13.41                & \text{(\ion{O}{iii})}
            & <13.27                & \text{(\ion{O}{iii})}
            & \nodata                & \nodata
            & <13.63                & \text{(\ion{O}{iv})} \\
    \ion{S}{IV}    & <13.31                & \text{(\ion{O}{iii})}
            & <13.11                & \text{(\ion{O}{iii})}
            & <13.05                & \text{(\ion{O}{iii})}
            & \nodata                & \nodata
            & <13.46                & \text{(\ion{O}{iv})} \\
    \ion{S}{V}     & <12.16                & \text{(\ion{O}{iii})}
            & <12.41                & \text{(\ion{O}{iv})}
            & <12.50                & \text{(\ion{O}{iv})}
            & \nodata               & \nodata
            & 12.24^{+0.13}_{-0.18}  & \text{(\ion{O}{iv})} \\
    \ion{S}{VI}    & <12.64                & \text{(\ion{O}{iii})}
            & <12.91                & \text{(\ion{O}{iv})}
            & <12.80                & \text{(\ion{O}{iv})}
            & \nodata                & \nodata
            & <12.67                & \text{(\ion{O}{iv})} \\
    \ion{Mg}{I}    & <11.27    & \text{(\ion{Mg}{ii})}
            & <10.90    & \text{(\ion{Mg}{ii})}
            & <11.19    & \text{(\ion{O}{iii})}
            & <11.10    & \text{(\ion{Mg}{ii})}
            & \nodata   & \nodata   \\
    \ion{Mg}{II}   & 11.85^{+0.07}_{-0.09} & 8.7^{+2.6}_{-2.3}
            & 11.66^{+0.09}_{-0.10} & <7.6\;(3.7)^c 
            & <11.68                & \text{(\ion{O}{iii})}
            & 12.02^{+0.05}_{-0.05} & 3.3^{+1.0}_{-0.8} 
            & \nodata               & \nodata \\
    \ion{Fe}{II}   & <12.18    & \text{(\ion{Mg}{ii})}
            & <12.12    & \text{(\ion{Mg}{ii})}
            & <12.29    & \text{(\ion{O}{iii})}
            & <12.26    & \text{(\ion{Mg}{ii})}
            & \nodata   & \nodata   \\
    \hline \hline
    &\multicolumn{2}{c|}{Component 5} & \multicolumn{2}{c|}{Component 6}\\
    & \multicolumn{2}{c|}{$\Delta v_\text{c5}=-40.7^{+6.6}_{-6.9}$} & \multicolumn{2}{c|}{$\Delta v_\text{c6}=+274.0^{+1.6}_{-1.6}$} \\ 
    \cline{1-5}
    \multicolumn{1}{c|}{Ion}& \multicolumn{1}{c|}{$\log N_\text{c5}/\mathrm{cm}^{-2}$} & \multicolumn{1}{c|}{$b_\text{c5}$} & \multicolumn{1}{c|}{$\log N_\text{c6}/\mathrm{cm}^{-2}$} & \multicolumn{1}{c|}{$b_\text{c6}$}\\ 
    \cline{1-5}
    \ion{H}{I}     & 15.34^{+0.08}_{-0.09} & 46.6^{+3.1}_{-3.6}
            & 14.18^{+0.02}_{-0.02} & 35.5^{+2.4}_{-2.2}\\
    \ion{C}{II}    & <12.93                & \text{(\ion{H}{i})}
            & <13.47                & \text{(\ion{H}{i})}\\
    \ion{C}{III}   & <13.59                & \text{(\ion{H}{i})}
            & <12.54                & \text{(\ion{H}{i})}\\
    \ion{N} {I}   & <13.39                 & \text{(\ion{H}{i})}
            & \nodata               & \nodata\\
    \ion{N}{III}   & <13.24                 & \text{(\ion{H}{i})}
            & <13.24                & \text{(\ion{H}{i})}\\
    \ion{N}{IV}    & <12.98                & \text{(\ion{H}{i})}
            & <12.95                & \text{(\ion{H}{i})}\\
    \ion{O}{I}     & <13.35                & \text{(\ion{H}{i})}
            & <13.37                & \text{(\ion{H}{i})}\\
    \ion{O}{II}    & <13.21                & \text{(\ion{H}{i})}
            & <13.04                & \text{(\ion{H}{i})}\\
    \ion{O}{III}   & <13.46                & \text{(\ion{H}{i})}
            & <13.25                & \text{(\ion{H}{i})}\\
    \ion{O}{IV}    & <13.77                & \text{(\ion{H}{i})}
            & <13.18                & \text{(\ion{H}{i})}\\
    \ion{O}{VI}    & <13.28                & \text{(\ion{H}{i})}
            & <13.36                & \text{(\ion{H}{i})}\\
    \ion{Si}{II}   & <12.58                & \text{(\ion{H}{i})}
            & <12.78                & \text{(\ion{H}{i})}\\
    \ion{Si}{III}  & <12.58                & \text{(\ion{H}{i})}
            & <12.34                & \text{(\ion{H}{i})}\\
    \ion{S}{II}    & <12.86                 & \text{(\ion{H}{i})}
            & <12.56                & \text{(\ion{H}{i})}\\
    \ion{S}{III}   & <13.93                & \text{(\ion{H}{i})}
            & <13.85                & \text{(\ion{H}{i})}\\
    \ion{S}{IV}    & <13.73                & \text{(\ion{H}{i})}
            & \nodata               & \nodata\\
    \ion{S}{V}     & <12.25                & \text{(\ion{H}{i})}
            & <12.28                & \text{(\ion{H}{i})}\\
    \ion{S}{VI}    & <12.70                & \text{(\ion{H}{i})}
            & <12.77                & \text{(\ion{H}{i})}\\
    \ion{Mg}{I}    & <11.38                & \text{(\ion{H}{i})}
            & <11.51                & \text{(\ion{H}{i})}\\
    \ion{Mg}{II}   & <12.07                & \text{(\ion{H}{i})}
            & <11.74                & \text{(\ion{H}{i})}\\
    \ion{Fe}{II}   & <12.33                & \text{(\ion{H}{i})}
            & <12.61                & \text{(\ion{H}{i})}\\
\hline
\multicolumn{11}{l}{$^a$ $\Delta v$ and $b$ are in units of \kms.}\\
\multicolumn{11}{l}{$^b$ Parentheses indicate column density was obtained by drawing the Doppler parameter from the posterior distribution of listed ion.}\\
\multicolumn{11}{l}{$^c$ 7.6 \kms\ is a 3-sigma upper limit while the value listed in parenthesis is the best-fit value corresponding to the listed column density.}
\end{tabular}
\end{table*}
\begin{table}
    \caption{Column densities measured with the Voigt profile fit to the partial Lyman Limit System at $z=0.5413$.\label{table:z0p54}}
    \renewcommand{\arraystretch}{1.3}
    \begin{tabular}{l| R| R| R}
    \hline\hline
    & \multicolumn{1}{c|}{Component 1} & \multicolumn{1}{c|}{Component 2} & \multicolumn{1}{c}{Component 3}\\
    & \multicolumn{1}{c|}{$\Delta v_\text{c1}=\xpm{0.0}{+0.2}{-0.2}$} & \multicolumn{1}{c|}{$\Delta v_\text{c2}=19.0$} & \multicolumn{1}{c}{$\Delta v_\text{c3}=\xpm{89.7}{+3.1}{-3.3}$}\\
    & \multicolumn{1}{c|}{$b_{\text{t,c1}}=\xpm{5.6}{+0.5}{-0.5}$} & \multicolumn{1}{c|}{$b_{\text{t,c2}}=\xpm{12.7}{+6.9}{-6.2}$} & \multicolumn{1}{c}{$b_{\text{HI,c3}}=\xpm{33.4}{+4.0}{-4.2}$}\\
    & \multicolumn{1}{c|}{$\log T_\text{c1}=\xpm{3.91}{+0.03}{-0.03}$} & \multicolumn{1}{c|}{$\log T_\text{c2}=\xpm{5.37}{+0.15}{-0.14}$} & \multicolumn{1}{c}{\nodata}\\
    \hline
    \ion{H}{I}     & \xpm{16.51}{+0.02}{-0.02} & \xpm{14.41}{+0.09}{-0.14} & \xpm{14.39}{+0.05}{-0.04}\\ \hline
    \ion{C}{II}    & \xpm{14.21}{+0.19}{-0.16} & <13.96 & <13.35 \\ \hline
    \ion{C}{III}$^a$ & >14.66 (\xpm{15.36}{+0.32}{-0.31}) & \xpm{13.07}{+0.06}{-0.07}& <12.13 \\ \hline
    \ion{N}{II}    & \xpm{13.93}{+0.11}{-0.09} & <13.27 & \nodata  \\ \hline
    \ion{N}{III}   & \xpm{14.40}{+0.13}{-0.11} & \xpm{13.69}{+0.10}{-0.14} & <13.21 \\ \hline
    \ion{N}{IV}    & \xpm{13.96}{+0.24}{-0.22} & \xpm{13.25}{+0.07}{-0.07}& <12.86 \\ \hline
    \ion{O}{II}    & \xpm{14.76}{+0.14}{-0.12} & <13.57& <13.37 \\ \hline
    \ion{O}{III}   & \xpm{14.88}{+0.29}{-0.23} (>14.36) & \xpm{13.88}{+0.08}{-0.10} & <13.06 \\ \hline
    \ion{O}{VI}$^b$ & \xpm{13.22}{+0.39}{-0.84} & \xpm{13.93}{+0.09}{-0.20} & \xpm{13.76}{+0.09}{-0.09} \\ \hline
    \ion{Si}{II}   & <13.89                & <13.08 & <14.37 \\ \hline
    \ion{S}{II}    & \xpm{13.07}{+0.07}{-0.07} & <12.60 & <12.22 \\ \hline
    \ion{S}{III}   & \xpm{13.97}{+0.10}{-0.10} & <13.65 & <12.99 \\ \hline
    \ion{S}{IV}    & \xpm{12.59}{+0.21}{-0.31} & <13.00 & <12.65 \\ \hline
    \ion{S}{V}     & \xpm{12.35}{+0.18}{-0.26} & <12.52 & <12.55 \\ \hline
    \ion{S}{VI}    & <12.57                    & <12.73 & <12.80 \\ \hline
    \ion{Mg}{I}    & \xpm{11.46}{+0.05}{-0.06} & <11.09 & <11.31 \\ \hline
    \ion{Mg}{II}   & \xpm{13.18}{+0.11}{-0.09} & <12.09& <11.76 \\ \hline 
    \ion{Fe}{II}   & \xpm{12.63}{+0.04}{-0.04} & <12.11 & <12.19 \\ \hline
    \end{tabular}
    \footnotesize{$^a$\ion{C}{iii} is likely saturated in c1. We use the lower limit value in the analysis; the maximum likelihood value is given in parenthesis. We use a lower limit for c2 as well, since the fit may be affected by saturation in c1.\vspace{1mm}\\ 
    $^b$\ion{O}{vi} Doppler parameters are fit independently. We find\\
    $\phantom{SPACE}b_{\text{O\,VI,c1}}=\xpm{27.0}{+18.5}{-13.5}$, $b_{\text{O\,VI,c2}}=\xpm{41.3}{+7.5}{-8.5}$, $b_{\text{O\,VI,c3}}=\xpm{41.8}{+11.7}{-8.7}$.\vspace{1mm}\\
    $^c$$\Delta v$, $b_t$, and $T$ are in units of \kms, \kms, and K, respectively.}
\end{table}

\section{Methodology} \label{sec:methods}

To investigate the physical properties of the two pLLSs along the line of sight to J0110$-$1648, we perform a Voigt profile analysis to characterize the multi-component velocity structure and measure column densities. In addition, we carry out detailed photoionization analyses to measure chemical abundances and densities of individual components. In this section we describe the steps we take for these analyses. For the rest of this paper, we refer to singly-ionized species as `low-ionization' and triply-ionized as `high-ionization,' unless otherwise stated.

\subsection{Voigt Profile Analysis of Resolved Component Structure}

\begin{figure*}
\centering
\includegraphics{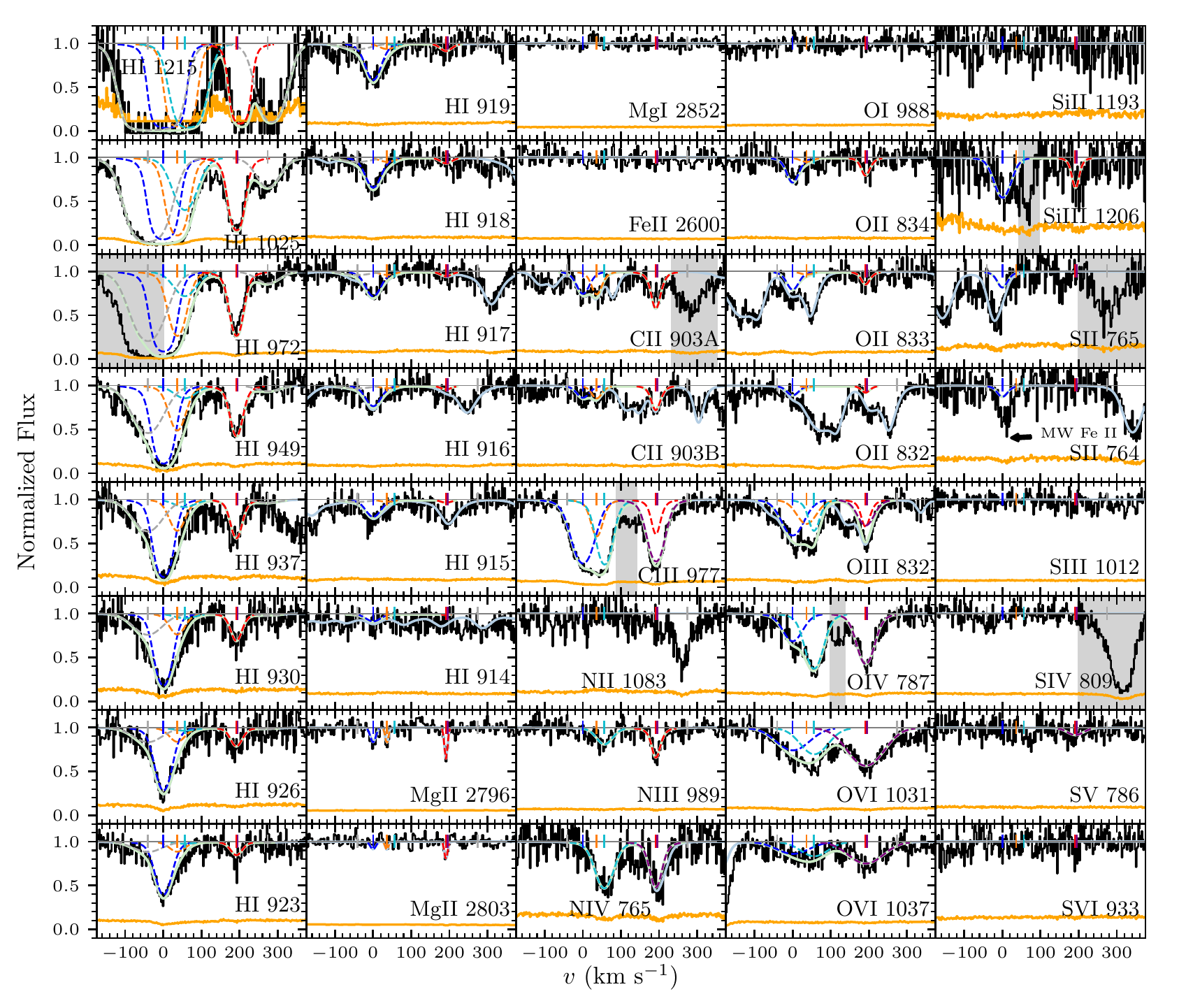}
\caption{Continuum normalized absorption profile of the $z=0.472277$ partial LLS absorber \ion{H}{i} and metals, with velocities centered on $z_\mathrm{c1}$. The \ion{Mg}{i}, \ion{Mg}{ii}, and \ion{Fe}{ii} data come from MIKE; all other species show data from COS. Absorption components c1, c2, and c3 are plotted individually with blue, orange, and cyan dashed lines, respectively, with ticks at unity indicating their centroids.  For component 4, low- and high-ionization species are shown with different colors (red and violet, respectively), and all doubly-ionized species were fit with a two-phase model, having two components at c4, each with Doppler parameters fixed to values from fits to appropriate high- and low-ionization ions. For other components this approach does not yield reliable results, due to their proximity to one another. Components 5 and 6 (grey dashed lines), at $\Delta v=-40$ and 274 \kms\ respectively, have no associated metals. Milky Way \ion{Fe}{ii} absorption is fit from uncontaminated lines, and is accounted for in the fit to \ion{S}{ii} 764 \AA. The joint profile of all components for each ion is shown with a pastel green line, and a pastel blue line shows the net absorption profile (i.e., all ions). Gray shading indicates contamination from absorption unassociated with this pLLS. \label{fig:z0p47_VP}}
\end{figure*}

Absorption components are fit, using custom software, with Voigt profiles parameterized by a velocity centroid $\Delta v_i$, column density $N_i$, and Doppler broadening term $b_i$. An additional parameter necessary to include in our fitting routine is a wavelength shift for the COS spectrum around lines of interest, accounting for slight wavelength calibration issues that are typical with COS \citep[e.g.,][]{2015ApJ...814...40W}. These corrections are typically below $3~\kms$, except for the edges of the spectrum (i.e., Lyman $\alpha$ at $z=0.47$ occurs at $\lambda=1787$\,\AA\ and requires a $10~\kms$ adjustment.)

Velocities are measured relative to the strongest \ion{H}{i} component (placed at $\Delta v=0$), and relative offsets are shared by all ions for a given component. To determine $\Delta v_i$ for discrete absorption components, we first fit the centroids of lines present in the higher-resolution MIKE spectrum. At $z\sim0.5$, the only strong absorption lines at optical wavelengths are low-ionization lines that tend to trace denser gas, such as the \ion{Mg}{ii} doublet. The posteriors of these fits are then used as priors when finding both wavelength adjustments in the COS spectrum and centroiding additional components only detected in \ion{H}{i} and higher ionization lines.

\begin{figure*}
\centering
\includegraphics[width=\linewidth]{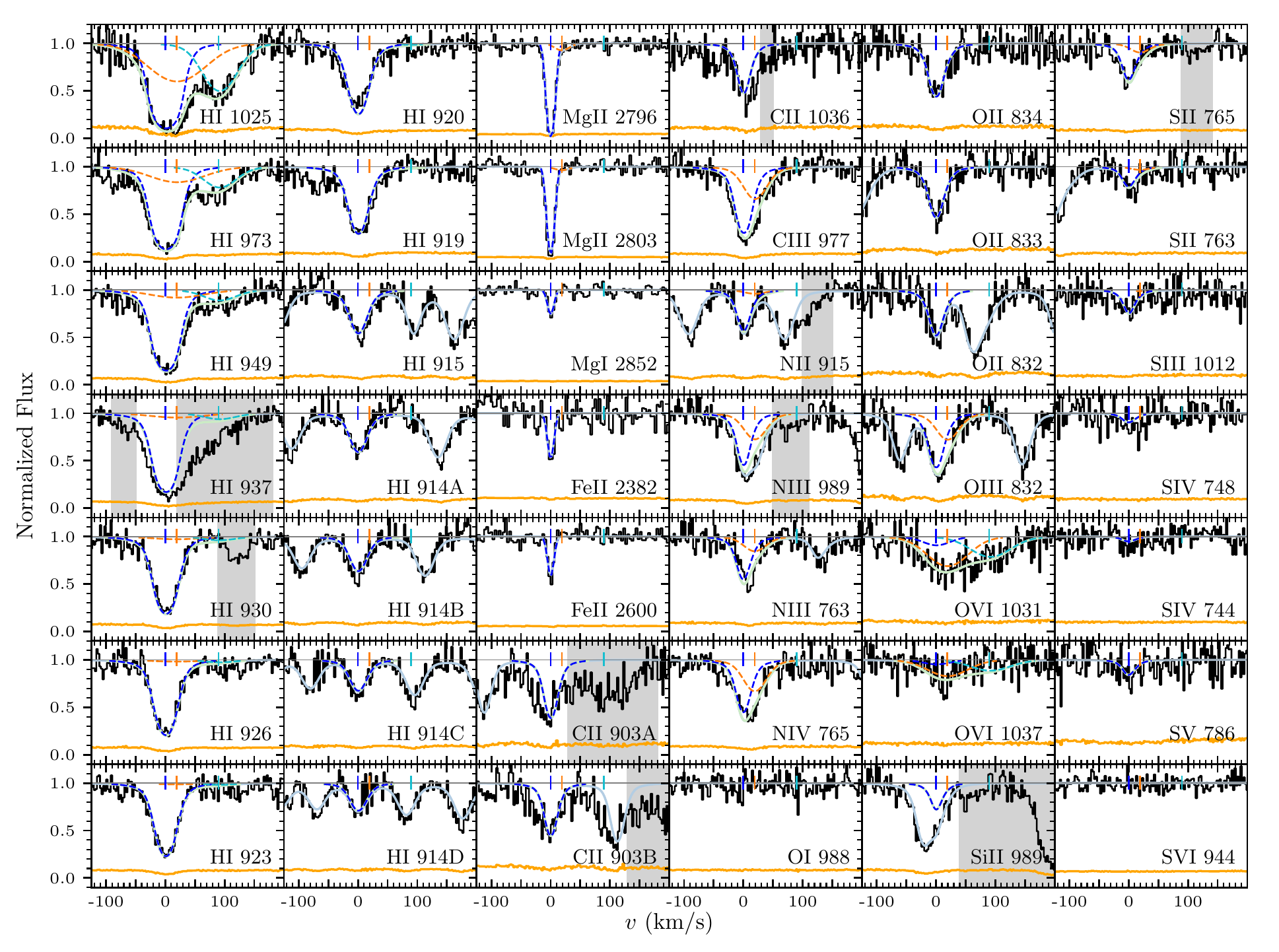}
\caption{Continuum normalized absorption profiles of the $z=0.541288$ pLLS, with the same format described Figure \ref{fig:z0p47_VP}. Absorption components c1, c2, and c3 are plotted individually with blue, orange, and cyan dashed lines, respectively. c1 has strong absorption across from singly- and doubly-ionized species. c2 is more highly ionized, with no absorption from singly-ionized species, and the only metal detected in c3 is \ion{O}{vi}.
\label{fig:z0p54_VP} }
\end{figure*}

The Doppler parameter $b_i$ for a component can be parameterized by a single temperature and turbulent broadening term $b_t$, with the resultant broadening term for each ion being $b^2_{i,\mathrm{ion}}=b^2_{t,i}+2kT_i/m_\mathrm{ion}$. We find that this enables fitting of pLLS at $z=0.54$, but results in inadequate fits for the $z=0.47$ absorber, so we opt to fit the Doppler parameter for each ion independently for the $z=0.47$ pLLS. The lower redshift absorber has a more complex velocity structure, so we attribute this to blended density phases and/or unresolved components (see \S\ref{sec:z0p47_c1}).

The absorber at $z=0.47$ and the Voigt profile components are shown in Figure \ref{fig:z0p47_VP}. We identify at least six distinct \ion{H}{i} components, with metal-bearing components at $\Delta v_{1,2,3,4}=0,+36,+56,+191\,\kms$ and only \ion{H}{i} detected at $\Delta v_{5,6}=-41,+274\,\kms$, respectively. While the \ion{H}{i} component velocity centroids span 315 \kms, detected metals only span 190 \kms\ since neither the bluest nor reddest component has associated metal lines. Components 1 (the dominant component with $\log\N{H\,I}/\cmx{-2}=16.03$), 2, and 4 have detected low-ionization species, with \ion{Mg}{ii} detections\footnote{Component 2 is only well-detected in the stronger 2796\AA\ \ion{Mg}{ii} transition; however, it is corroborated by a more-significant \ion{C}{ii} detection.} allowing for precise redshift measurements. While all three also have \ion{O}{ii} and \ion{O}{iii} detections, only components 1 and 4 have \ion{O}{iv}. Similarly, component 3, in the middle of the absorption complex ($\Delta v_3=56\,\kms$) is clearly more highly ionized, with detections of \ion{O}{iii} and \ion{O}{iv} but not \ion{O}{ii} or \ion{Mg}{ii}.  Column density measurements for components of this pLLS are given in Table \ref{table:z0p47}. $b_\mathrm{Mg\,II}$ is considerably narrower than the COS FWHM (FWHM $=20$ \kms\ corresponding to $b=12$ \kms) and below the range of Doppler parameters that can be resolved by COS ($b\approx9$~\kms\ for the typical S/N ratio in CUBS COS data); for lines that are optically thin, as is the case for all other singly-ionized species, this should not influence column density measurements.

The absorber at $z=0.54$, shown in Figure \ref{fig:z0p54_VP}, is comparatively simpler, with only three \ion{H}{i} components. The strongest \ion{H}{i} absorber ($\log\N{H\,I}/\cmx{-2}=16.51$) is at $z_1=0.53129$\footnote{$z_1=0.531288\pm0.000001$, based on the \ion{Mg}{ii} and \ion{Fe}{ii} profiles.}. A more highly ionized component at $\Delta v_2=19\,\kms$ contains a suite of metal ions (see \S\ref{sec:z0p54_c2} for a discussion on the necessity of this component), while at $\Delta v_3=90\,\kms$ an absorber of comparable \N{H\,I} has no corresponding metal absorption lines aside from \ion{O}{vi}, indicating variation in abundances and/or density between these two clouds. Column density measurements for this pLLS and the associated absorbers are given in Table \ref{table:z0p54}. As detailed in the following discussion, the component structure of the low-density gas is poorly constrained by the data, such that measurements of doubly-ionized lines are somewhat model dependent in c1 and c2.

For all absorbers presented herein, Voigt profile fits are performed using a Markov chain Monte Carlo (MCMC) code, using the {\tt emcee} package \citep{2013PASP..125..306F} as a framework in which we define a likelihood function of $\ln\mathscr{L}\propto-\chi^2$. We adopt the median of the posterior distribution as the reported value for each model parameter, with $68\%$ credible intervals used to approximate 1$\sigma$ errors (see Tables \ref{table:z0p47} and  \ref{table:z0p54}). These uncertainties are marginalized over blending with other components or absorption due to other ions, and in a few instances non-Gaussian posteriors when $b$ is very small or lines are potentially saturated. We note that there is additional model uncertainty due to potential blending with additional unresolved components (or \ion{H}{i}), which is considered in interpreting models of the absorbers. Potential saturation manifests as a `tail' towards large column-densities in posterior distributions, in which case we adopt the 0.05 percentile of the column density posterior distribution as a lower-limit (corresponding to 3$\sigma$). We do not include undetected lines in the initial fit; instead, we run a second iteration with the posteriors from the first iteration set as the priors for detected lines, and for non-detections assume as $b$-priors the posterior of a relevant detected ion. We adopt for these the 99.5th percentile as an upper-limit (again, corresponding to $3\sigma$). 

\subsection{Photoionization Analysis}\label{sec:PI}
We determine gas properties for each absorption component by comparing to models constructed with \texttt{Cloudy} \citep{2013RMxAA..49..137F}. Absorbers are modeled as discrete clouds subject to an external ionizing radiation field, with metal abundance patterns allowed to vary relative to Solar. We use the ultraviolet background radiation (UVB) field prescription from
\citet{2020MNRAS.493.1614F}. This UVB model includes contributions from galaxies and AGN, and was calibrated to a number of recent empirical constraints. We note that the choice of UV background is motivated by its agreement with recent measurements of the $z<1$ neutral hydrogen photoionization rate, $\Gamma_\mathrm{HI}$, and that the abundances derived are robust to variation in the UVB model (see \citetalias{2021MNRAS.506..877Z}; \citealt{2018MNRAS.479.2547C,2019MNRAS.484.2257Z}). 

\begin{figure*}
\centering
\includegraphics{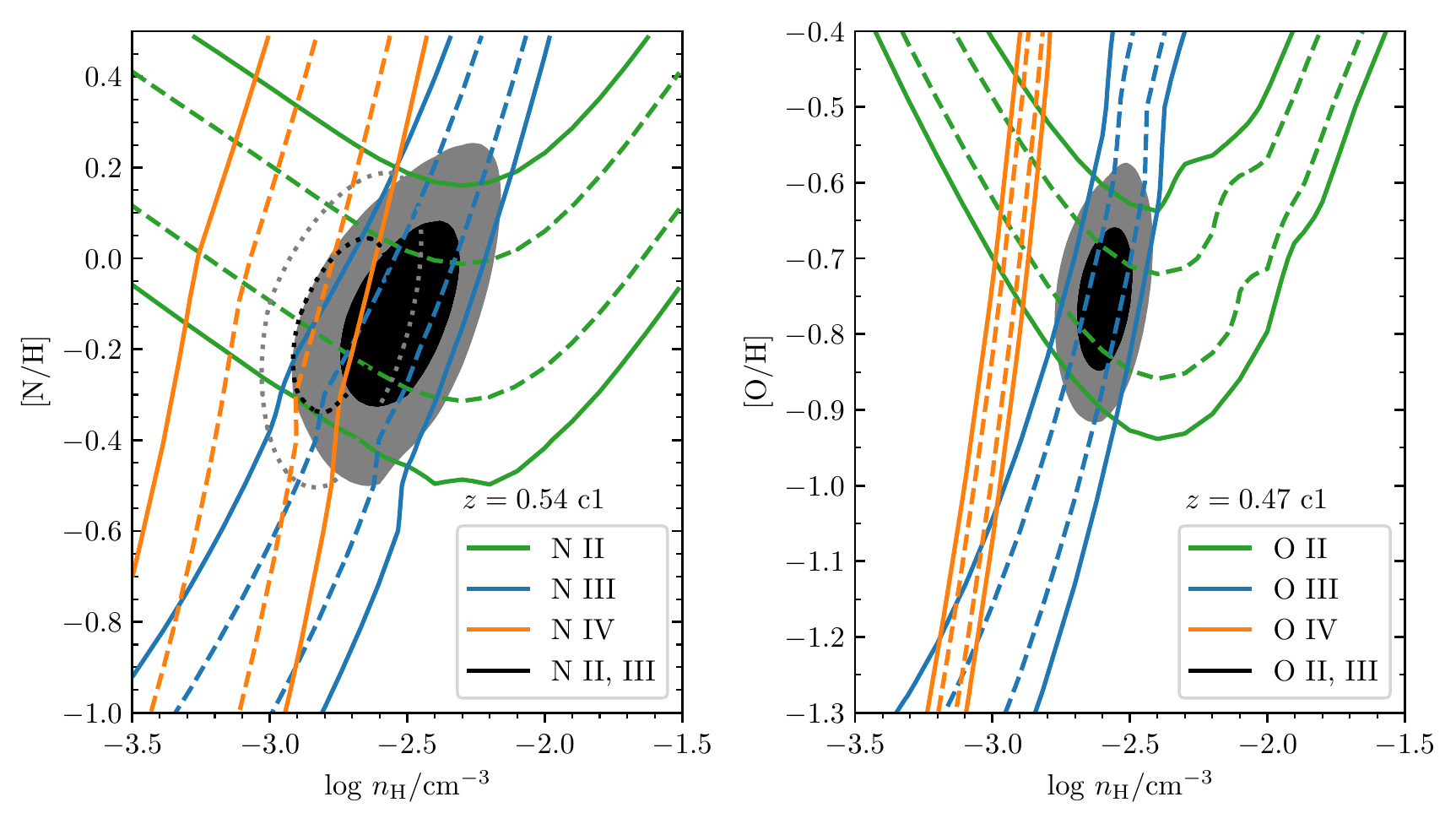}
\caption{{\it Left}: Contours of 68 and 95$\%$ likelihood (solid and dashed lines, respectively) for nitrogen ions detected in the $\log\N{H\,I, c1}/\cmx{-2}=16.5$ pLLS at $z=0.54$ (Component 1), assuming a single phase model. The filled black and gray contours contain 68 and 95\%, respectively, of the joint likelihood of singly- and doubly-ionized N. At the density that best matches the observed ratio of \ion{N}{ii} and \ion{N}{iii}, the [N/H] required to reproduced the observed \N{N\,IV,c1} is nearly ten times solar, well above the abundance that reproduces observed \N{N\,II,c1} and \N{N\,III,c1}. This suggests that the observed absorption of all three ionization states of N, in this single component, do not arise from gas with a single ionization phase. The dotted black and gray contours are maximum likelihood contours if \ion{N}{iv} is included. Note that [N/H] is relatively unaffected by the inclusion of \ion{N}{iv}, while the density changes considerably. 
{\it Right}: Similar contours of oxygen ions detected in the $\log\,\N{H\,I,c1}/\cmx{-2}=16.0$ pLLSs at $z=0.47$ (Component 1). In this absorber multiple phases are unambiguously required to reproduce the observed column density ratios of these ions; there are no densities at which the 95\% likelihood contours of all three ions intersect.
 \label{fig:CI_MP}}
\end{figure*}

Photo-ionization (PI) models are calculated over a grid of densities (\nH) and chemical abundances. The models assume a solar relative abundance pattern; the effect of assuming a reasonable, different abundance pattern is negligible to the ionization fractions of various species, so we introduce varying relative abundances in the comparison step. Since our observations trace elements that span a variety of nucleosynthetic pathways, from core-collapse supernovae (O, Mg), Type Ia supernovae (Fe), and long lived stars (C, N), we do not require metals to have a solar abundance pattern. All abundances are allowed to vary independently, with the exception of the $\alpha$-elements (O, Mg, Si, and S), which we assume to have a fixed, solar ratio. For brevity, we refer to the corresponding list of elemental abundances as \xvh.  We construct likelihood contours in across densities and abundances with the likelihood function:
\begin{equation}
\ln\mathscr{L}\left(n_H,\xvh\right) = \sum\limits_{ions}\left(N_\mathrm{obs}-N_\mathrm{model}\right)^2/\sigma_N^2\label{eq:logL}
\end{equation}
and resulting ionic column densities are input into a Markov-Chain Monte Carlo calculation (implemented using \texttt{emcee}, \citealp{2013PASP..125..306F}), along with \N{H\,I}. Although the \ion{Mg}{i} 2852\AA\ line is well detected we exclude it from photoionization analysis because the dielectric recombination rate is not well known, resulting in possibly inaccurate model predictions \citep[see discussion in][]{2003AJ....125...98C}. This uncertainty is negligible for \ion{Mg}{ii}; since $\N{Mg\,II}/\N{Mg\,I}\gtrsim10-100$ at densities of interest, a modest change to the ionization fraction of \ion{Mg}{i} minimally affects \N{Mg\,II}.

To elucidate the MCMC results, we display contours of constant likelihood in the (\nH,[X/H]) plane for different elemental abundances X, at the central \N{H\,I} value measured; since the absorbers considered in this paper with appreciable uncertainty in \N{H\,I|} are optically thin, changing \N{H\,I} within these uncertainties corresponds largely to an equivalent shift in metallicity, although metal-line cooling begins to become an important factor at metallicities approaching solar. 

Observations of a single element across a range of ionization states should in principle provide higher confidence in the fidelity of PI modeling results, with the density of the gas simply that which reproduces the relative fractions of the different ions, without requiring any assumptions about relative abundances. Figure \ref{fig:CI_MP} shows example contours corresponding to the 68 and 95\% credible intervals in density and abundance within which photoionization models of a single-phase gas can produce the observed column densities. Intersections between model contours for different ionization states indicate the physical conditions where observations can be explained by a single-phase solution.

For the two absorption components shown (unrelated and at different redshifts), the best-fitting densities and abundances inferred from \ion{N}{ii} and \ion{N}{iii} (or \ion{O}{ii} and \ion{O}{iii}) column densities yield \ion{N}{iv} (\ion{O}{iv}) predictions over one order-of-magnitude below the observed value. In the example using N, the $95\%$ credible intervals all overlap, suggesting a single-phase solution may be plausible. Moreover, the derived [N/H] is robust with respect to whether or not \ion{N}{iv} is included. However, the nitrogen measurements alone cannot distinguish between a scenario where there is a single phase, and one in which there are contributions to \ion{N}{iii} from a high-density phase containing \ion{N}{ii} and a low-density phase containing \ion{N}{iv}. In \S\ref{sec:z0p54_c1}, we show that considering all elements favors a multiphase solution.

For the other absorber, where we use measurements of oxygen ions, the only way to reconcile the various column densities is to posit that the singly- and triply-ionized species are not in a single, uniform density gas cloud. The most plausible explanation is rather that we are observing at least two different gaseous absorbers, with similar kinematic profiles that are not resolved in the COS data, with one higher-density component hosting the singly-ionized gas and the other lower-density component hosting the triply-ionized gas\footnote{We note that single phase non-equilibrium models such as rapidly cooling gas \citep[e.g.,][]{2007ApJS..168..213G, 2013MNRAS.434.1043O} and fluctuating radiation fields \citep{2013MNRAS.434.1063O} also do not reproduce the ionization fractions observed. More sophisticated treatments that consider a multiple phase, non-equilibrium gas are beyond the scope of this work.}. This scenario typifies what we refer to as `multiphase' absorbers.

While multiple phases are often necessary to explain observed data, these considerations introduce additional complexity that can make interpretation of results challenging. For example, since the ionization fraction of neutral hydrogen is much smaller in the high-ionization phase, we expect the bulk of the observed \ion{H}{i} to correspond to the low-ionization phase. Resultantly, splitting \ion{H}{i} column density between several phases (when only a single component is evident in the data) generally yields metallicities for the high ionization phases that are highly model/prior dependent and presenting abundances simply based on posteriors can be misleading. Our goal with this analysis is to determine not only gas properties of the absorbers, but to gauge the extent to which various properties can be robustly determined.  Knowing that these absorbers are inherently a blend of multiple components, one can attempt fit the data directly with multiple distinct components. However, we find the blending of these components, as well as blending with kinematically distinct nearby components, often results in large degeneracies that make property inferences highly uncertain. For this reason, we prefer to explore each absorber with a single phase model, introducing additional phases as necessitated by the data, rather than {\it a priori}.

In \S\ref{sec:analysis}, we describe the considerations that go into modeling each component, along with component-specific details on the Voigt profile fitting, explaining what properties are robustly determined and which are ambiguous due to model dependence. The results are presented in \S\ref{sec:results}, alongside general interpretations of the derived densities and abundance patterns, as well as the associated galactic environments.

\section{Details of Photoionization Analysis of Individual Absorption Components}
\label{sec:analysis}

The Voigt profile analysis presented in \S\ \ref{sec:PI} has identified six discrete H\,I components for the pLLS at $z=0.47$ and three for the pLLS at $z=0.54$. All but two components (c5 \& c6 in the $z=0.47$ pLLS; see Table \ref{table:z0p47}) exhibit associated metal lines that cover a broad range of ionic species.  Simultaneous accounting of the full range of observed ion abundances reveals a complex multiphase nature of these pLLSs, showing large variations in density and elemental abundances among different components.  Here we describe the ionization properties of each of these components and the manner in which they are derived.

\subsection{\texorpdfstring{\bm{$z=0.47$} Component 1 \bm{$\left(\Delta v_\mathrm{c1}=0\,\kms\right)$}}{}} \label{sec:z0p47_c1}

    \begin{figure*}
    \centering
    \includegraphics{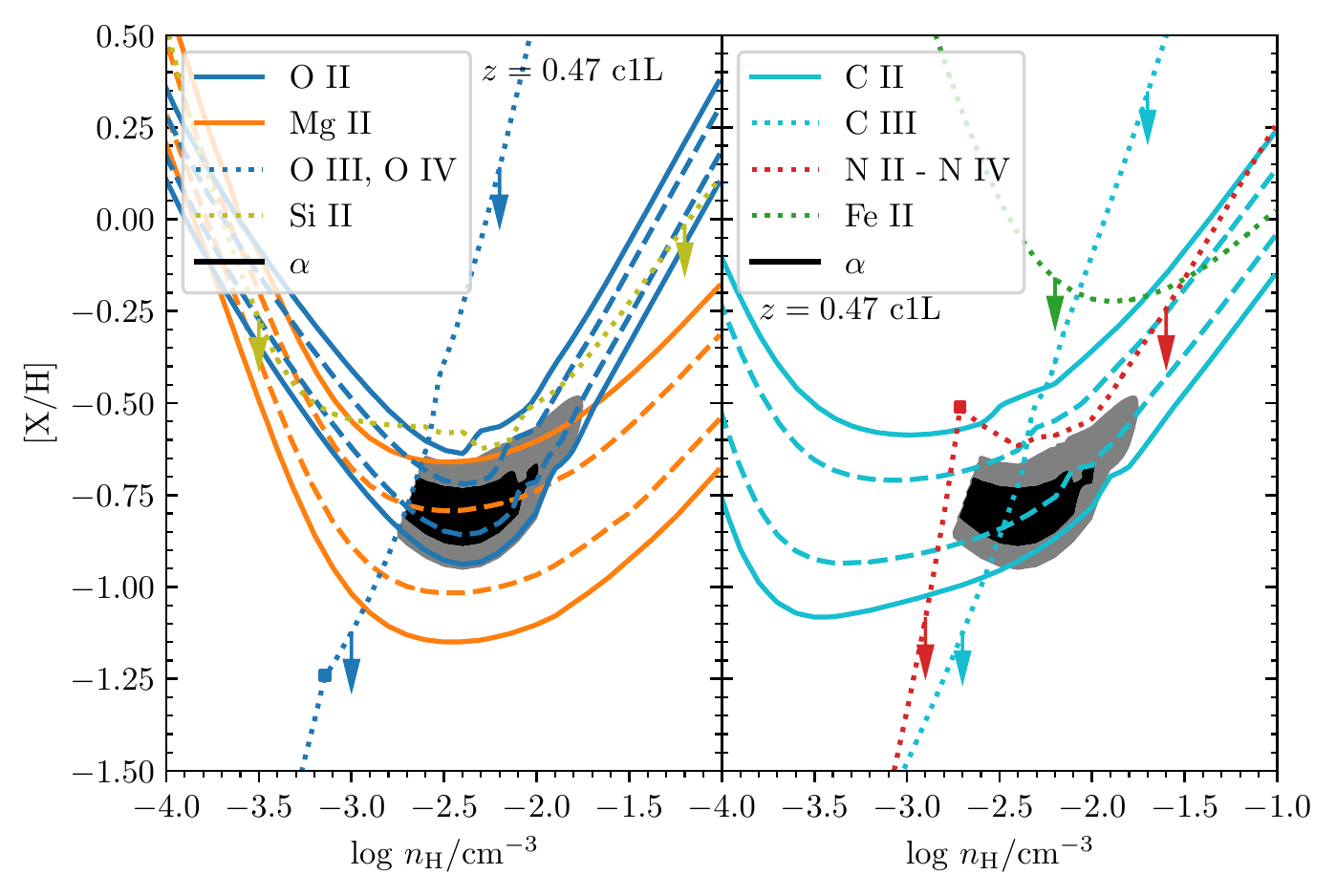}
    \caption{Contours of maximum likelihood for the low-ionization phase of Component 1 at $z=0.47$ (c1L, the dominant H\,I component), using conventions described in Figure \ref{fig:CI_MP}. Dotted lines correspond to 99.5\% column density limits, and boxes indicate where the most constraining ion for each curve changes. We have shown in Figure \ref{fig:CI_MP} that \ion{O}{iv} requires an additional phase. Since this more highly-ionized phase likely contributes some column density to the doubly-ionized species, we treat these measurements as upper-limits for the low-ionization phase. These limits, combined with detections of \ion{O}{ii}, \ion{Mg}{ii}, yield a metallicity of about [$\alpha$/H]$_{\rm c1}\sim-0.75$, with allowed density ranging from $-3\lesssim\log\,n_{\rm H,c1}/\cmx{-3}\lesssim-2$. The low metallicity and solar carbon-to-$\alpha$ elemental abundances, [C/$\alpha$]$_{\rm c1}=0$, are in contrast with the other low-ionization components at $z=0.4723$. \label{fig:z0p47_c1_CI}}
    \end{figure*}
    
    \begin{figure*}
    \centering
    \includegraphics{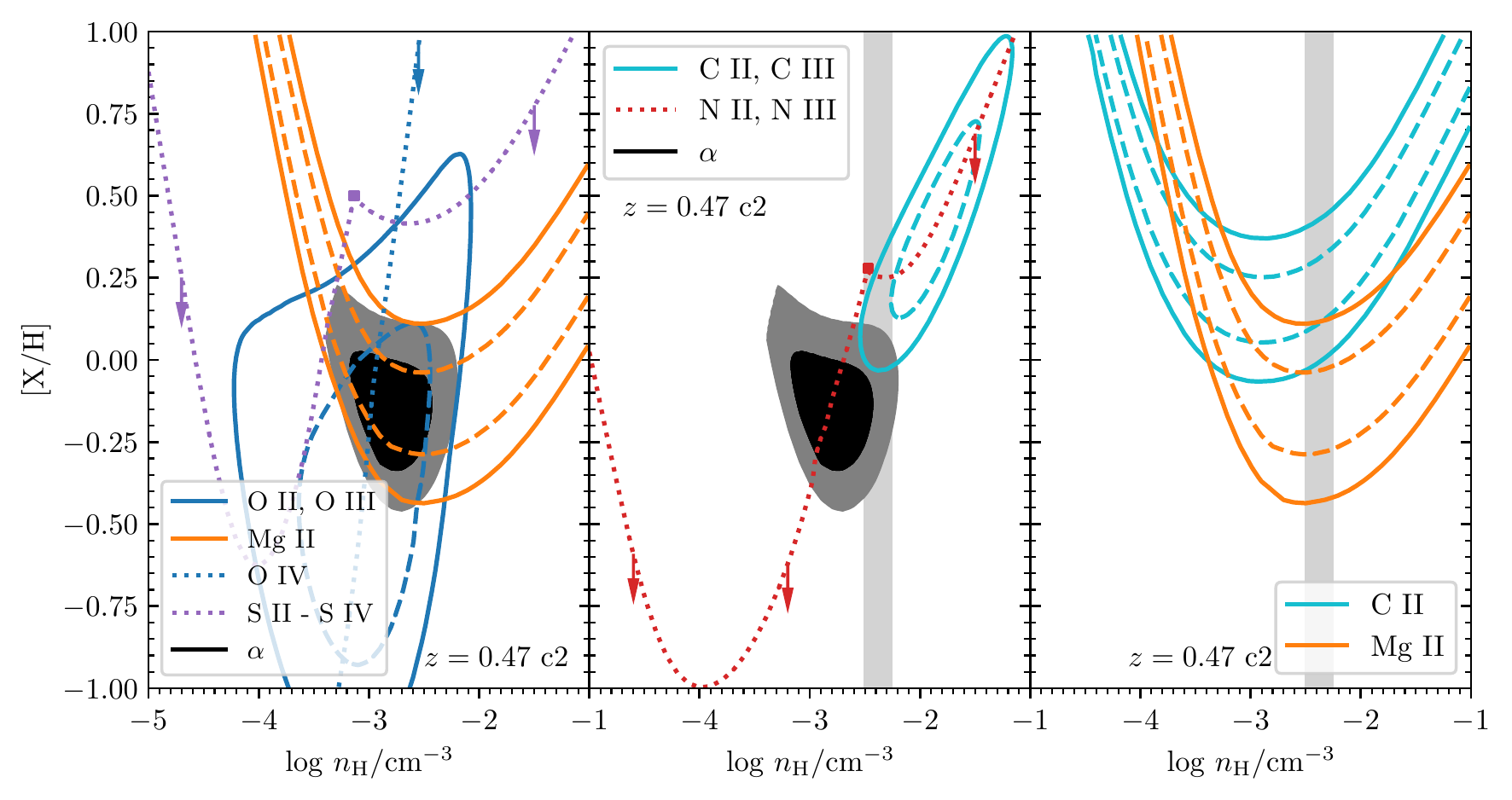}
    \caption{Likelihood contours for component 2 at $z=0.47$, using conventions described in Figure \ref{fig:CI_MP}. The filled contour representing the joint solution for $\alpha$-elements does not include the upper-limit on \ion{O}{iv}, since adjusting the component model may allow for a formal detection rather than a limit; the \ion{O}{iv} \textit{is} included in the abundance calculation. $\alpha$-elements and C ions are not in agreement for $\nH$, although a probable explanation is that \N{C\,III,c2} is uncertain. The vertical gray region is the density obtained by jointly fitting all ions. The rightmost panel demonstrates that \N{C\,II,c2} and \N{Mg\,II,c2} yield [C/$\alpha$]$_{\rm c2}>0$ for all densities not ruled out by the non-detection of \ion{O}{iv} ($\log\,n_{\rm H,c2}/\cmx{-3}\gtrsim-3$). Regardless of the density, \ion{Mg}{ii} makes it clear that c2 enriched relative to c1L.\label{fig:z0p47_c2_CI}}
    \end{figure*}

    Component 1 of the pLLS at $z=0.47$ accounts for $\approx 80$\% of the total $\N{H\,I}$ with $\log\N{H\,I,c1}/\mathrm{cm}^{-2}=16.0$.
    This component also has associated metal absorption lines from a range of ionization states and, as previously shown in Figure \ref{fig:CI_MP}, the column densities of \ion{O}{ii}, \ion{O}{iii}, and \ion{O}{iv} cannot be modeled with a single photoionized gas cloud. Blending between c1, c2 and c3 complicates attempts to directly fit line profiles with two different phases. Instead, we differentiate between the two phases in photoionization modeling.
    
    Before considering a two-phase model, we can gain intuition about the expected result by first considering each phase separately. In the following text, we refer to the low-ionization phase as c1L and the high-ionization phase c1H. We expect the neutral hydrogen and singly-ionized species to largely be in the low-ionization phase. In Figure \ref{fig:z0p47_c1_CI}, we show the corresponding likelihood contours if we treat the column densities of the more highly-ionized species as $3\sigma$ upper-limits. \ion{O}{ii} and \ion{Mg}{ii}, combined with an upper-limit on \ion{O}{iii}, constrain the density of the low-ionization phase to $\log\,n_{\rm H,c1}/\cmx{-3}\sim-2.5$ and the $\alpha$ abundance to [$\alpha/\mathrm{H}]_{\rm c1}\sim-0.75$. Comparing with \ion{C}{ii} in the second panel, it is clear that [C/$\alpha]_{\rm c1}\sim0$. Combining the \N{O\,III,c1} upper-limit with the measurement of \ion{O}{iv} constrains the density of c1H to $\log\,n_\mathrm{H,c1H}/\cmx{-3}<-3.0$, assuming that c1H is optically thin.
    
    We implement a two-phase model by adapting the likelihood function (Eq \ref{eq:logL}) to use the net ionic column densities from both phases:
    \begin{equation}
    N_\mathrm{model}=N_\mathrm{model,c1H}+N_\mathrm{model,c1L}
    \end{equation}
    where c1H is modeled as an optically thin gas. Since $N_\mathrm{H\,I,c1H}$ is unknown, we allow it to vary, subject to priors requiring $n_\text{H,c1H}<n_\text{H,c1L}$,  $N_\mathrm{H\,I,c1H}<N_\mathrm{H\,I,c1L}$, and $-2\leq\rm{[X/H]}\leq1$ for all abundances in $\left\{\rm{[X/H]}\right\}$.
    
    The results for both phases are consistent with expectations from the previous discussion, with $n_\mathrm{H,c1L}=\xpm{-2.28}{+0.14}{-0.07}$, [$\alpha/\mathrm{H}]_{\rm c1L}=\xpm{-0.81}{+0.06}{-0.05}$, and $n_\mathrm{H,c1H}=\xpm{-3.51}{+0.11}{-0.14}$.To evaluate any relation between c1H and c1L, we would like to compare their abundance patterns. Absolute abundances are not readily established for c1H, since $\N{H\,I,c1H}$ is allowed to vary, but {\it relative} abundances can still be constrained. In particular, we find that a conservative 3$\sigma$ lower limit\footnote{This limit is obtained using the 99.5\% upper-limit to [C/$\alpha]_\mathrm{c1L}$ and is not the difference of the values given in Table\,\ref{table:abundances}} on the difference in abundance between the higher and lower ionization components, [C/$\alpha]_\mathrm{c1L}$-[C/$\alpha]_\mathrm{c1H}>0.16$, yields direct evidence of different chemical abundance ratios between these two phases, with c1L relatively carbon enhanced compared with c1H.
    
    \subsection{\texorpdfstring{\bm{$z=0.47$} Component 2 \bm{$\left(\Delta v_\mathrm{c2}=36\,\kms\right)$}}{}} \label{sec:z0p47_c2}

    Component 2 of the absorption complex at $z=0.47$, at $\Delta v_\text{c2}=36\,\kms$, is detected in \ion{H}{i}, \ion{Mg}{ii}, \ion{C}{ii}, \ion{C}{iii} and \ion{O}{iii}, with a marginal \ion{O}{ii} detection. The more highly ionized \ion{N}{iv}, \ion{O}{iv}, and \ion{O}{vi} are all best fit \textit{without} absorption from this component, which seems to have a single, high-density phase. Absorption from this component is necessary to fit the \ion{C}{iii} profiles, but is blended with higher optical depth absorption from the pLLS to the blue and the high-ionization only component 3 to the red, making \N{C\,III,c2} sensitive to model uncertainties.

    In the first two panels of Figure \ref{fig:z0p47_c2_CI} we show likelihood contours for the $\alpha$-elements, and compare with the \nH\ inferred from the C measurements. The \N{C\,II,c2}-to-\N{C\,III,c2} ratio prefers a 10 times larger density than the $\alpha-$elements, but we note that if \N{ C\,III,c2} is underestimated this tension is alleviated, which is likely given the blending with stronger, possibly saturated absorbers. Regardless, the density obtained by jointly considering all ions in the MCMC analysis, $\log\,n_{\rm H,c2}/\cmx{-3}=-2.38\pm0.13$, is consistent within 2$\sigma$ with that derived considering C or $\alpha$ elements alone. The corresponding abundances are close to solar at [$\alpha$/H]$_{\rm c2}=-0.12\pm0.11$ and [C/H]$_{\rm c2}=0.14\pm0.11$.
    
    Since the \ion{C}{ii} 903 doublet is well detected, we can achieve a straightforward estimate of [C/$\alpha$] just by considering the singly-ionized species, regardless of uncertainty in \nH, shown in the rightmost panel. For $\log\,n_{\rm H,c2}/\cmx{-3}\gtrsim-3$, there is a preference for somewhat enhanced C, although solar relative abundances are not disallowed. The lower densities, where \ion{C}{ii} and \ion{Mg}{ii} are consistent with [C/$\alpha$]$_{\rm c2}=0$, are ruled out by the non-detections of more highly ionized species, indicated by the upper limits in the leftmost panel.
    
\subsection{\texorpdfstring{\bm{$z=0.47$} Component 3 \bm{$\left(\Delta v_\mathrm{c3}=56\,\kms\right)$}}{}} \label{sec:z0p47_c3}

    \begin{figure*}
    \centering
    \includegraphics{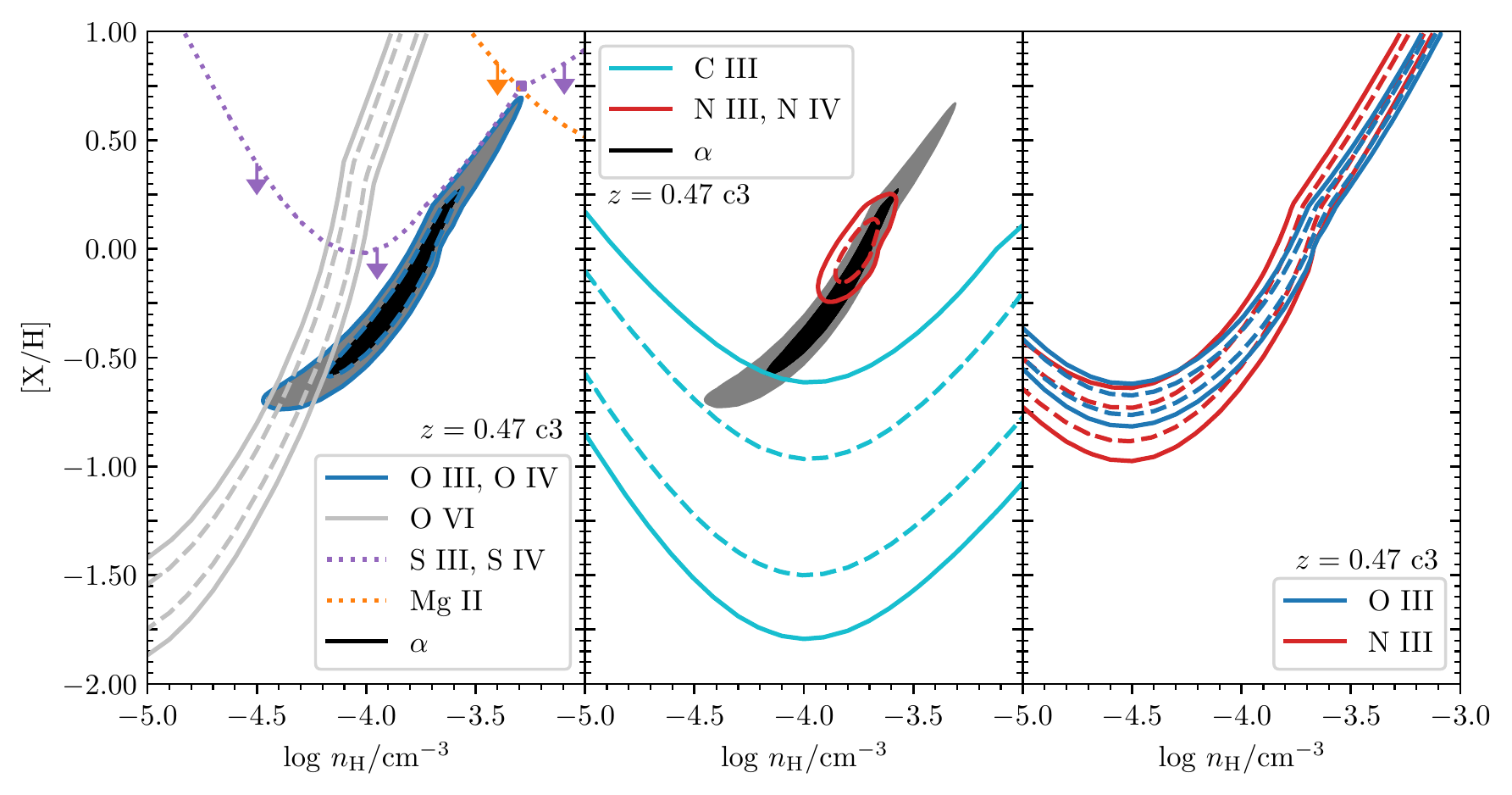}
    \caption{Likelihood contours for component 3 at $z=0.47$, using conventions described in Figure \ref{fig:CI_MP}. \ion{O}{vi} is not included in determining the contour for $\alpha$-elements, since this requires $\nH$ that is inconsistent with \ion{N}{iii} and \ion{N}{iv}. The \ion{O}{vi} absorption likely comes from more highly-ionized gas than the other ions shown. C may be under-abundant relative to other elements, but \N{C\,III} for this component is not robust due to degeneracy with c2. The rightmost panel shows that [N/$\alpha]\sim0$ regardless of \nH.\label{fig:z0p47_c3_CI}}
    \end{figure*} 

    Component 3 of the absorber at $z=0.47$ is only detected in species that are at least doubly-ionized (excepting \ion{H}{i}), opposite to what is seen in c2. While \ion{O}{iii}, \ion{O}{iv} and \ion{O}{vi} are all reproduced by a single-phase photoionized gas at $\log\,n_{\rm H,c3}/\cmx{-3}=-4.28\pm0.05$ and [O/H]$_{\rm c3}=-0.61\pm0.12$, \ion{N}{iii} and \ion{N}{iv} suggest a higher density of $\log\,n_{\rm H, c3}/\cmx{-3}=-3.75\pm0.06$.  The sensitivity of \N{O\,VI,c3} to the density in a photoionized gas, as shown in Figure \ref{fig:z0p47_c3_CI}, suggests a two-phase solution with the more highly-ionized phase bearing the \ion{O}{vi}. Moreover, $b_\mathrm{O\,VI,c3}$=43.8 \kms\ indicates that the \ion{O}{vi} has a temperature of $T>10^6$ K, and is collisionally ionized. The much narrower $b_\mathrm{O IV,c3}=23.5$ \kms\ indicates that the observed \ion{O}{iv} arises predominantly in a lower-ionization phase.

    The right-most panel demonstrates that [$\alpha$/N]$_{\rm c3}\sim0$ irrespective of the density solution, because \ion{N}{iii} and \ion{O}{iii} have similar ionization fractions across the range of plausible densities.  At the same time, this measured \N{CIII} implies an abundance pattern with depleted carbon ([C/$\alpha$]$_{\rm c3}=-1.09\pm0.09$); however, the \ion{C}{iii} profile between 0 - 60 \kms\ may be affected by saturation and blending with contaminating absorption at $\sim$100 \kms, as well as c2.

\subsection{\texorpdfstring{\bm{$z=0.47$} Component 4 \bm{$\left(\Delta v_\mathrm{c4}=191\,\kms\right)$}}{}} \label{sec:z0p47_c4}

   \begin{figure*}
    \centering
    \includegraphics{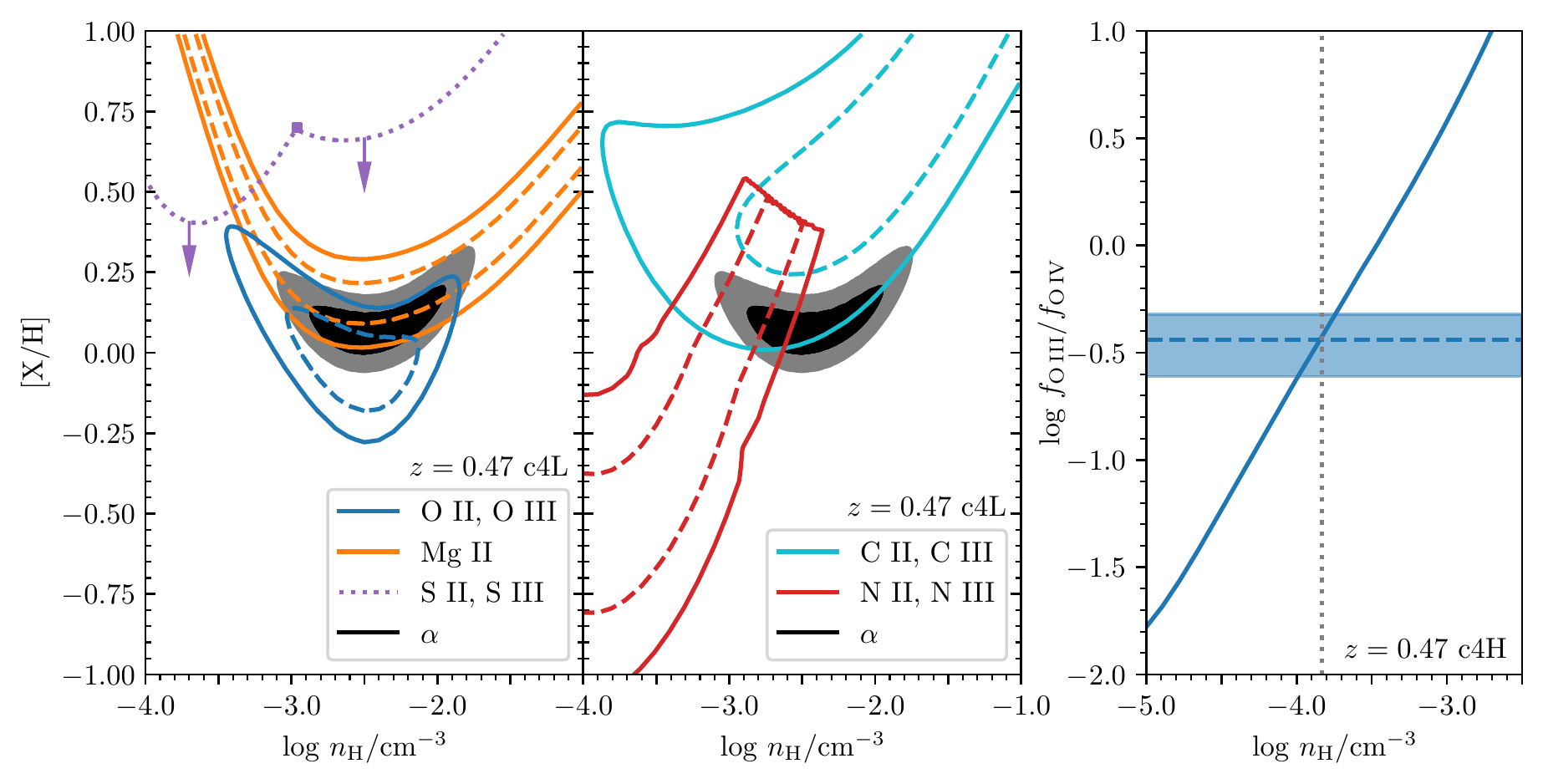}
    \caption{\textit{Left:} Likelihood contours for $z=0.47$ c4L, the low density phase contributing to component 4, using conventions described in Figure \ref{fig:CI_MP}. This phase has solar [$\alpha$/H], with the possibility of a carbon enhancement. Although fitting with two-phases presents model uncertainty, we can see from the \ion{Mg}{ii} contour that [$\alpha$/H] must be at least solar. \ion{Si}{iii} is consistent with the other $\alpha$-elements and is left off the figure for clarity due to its large statistical error. \textit{Right:} Density solution of component c4H at $z=0.47$, obtained from the ratio of \N{O\,III,c4L}-to-\N{c4, O\,IV,c4L}. The shaded region is the ratio of measured column densities, and the solid blue curve is model predictions for an optically thin, photoionized gas at solar metallicity. The vertical dotted line shows the maximum likelihood density, which is about one-tenth that of c4L.\label{fig:z0p47_c4L_CI}}
    \end{figure*}

    The reddest component at $z=0.47$ is well separated kinematically from the other metal absorption lines. We observe an increasing line width with ionization energy in single-component fits to this component (see \S\ref{sec:results} and Figure \ref{fig:b_v_E}). Moreover, the column densities of \ion{O}{ii}, III, and IV from a single component Voigt profile fit, compared with a single-phase photoionization model, result in likelihood contours similar to that of c1 shown in the right panel of Figure \ref{fig:CI_MP}, requiring that the total \N{O\,III,c4} has appreciable contributions from two distinct phases. Since this component is not blended with the other components and because the phases have very different line widths, we are able to directly measure column densities for two different phases, which we denote as Components c4L and c4H, for low and high ionization.

    The line centroids of c4L and c4H are derived from fitting the \ion{Mg}{ii} doublet and \ion{O}{iv} 787\AA, respectively. We find equal velocity offsets for both phases, with $\Delta v_\text{c4L}=+191.2\pm0.5$ and $\Delta v_\text{c4H}=+191.3\pm1.3$. We attribute singly-ionized species to c4L and triply ionized to c4H. 

    Both \ion{C}{iii} and \ion{O}{iii} are too broad to be fit by single profiles with $b_\text{c4L}$. We fit doubly-ionized species jointly to both components using Doppler parameters drawn from the posterior distributions of singly- and triply-ionized species. This yields comparable \N{O\,III} in both c4L and c4H, while the \ion{C}{iii} profile is dominated by c4H. For \ion{Si}{iii} and \ion{N}{iii} the models prefer to attribute the bulk of the absorption to c4L. The comparatively high upper limits on \N{Si\,III,c4H} and \N{N\,III,c4H} reflects that there are lower-likelihood alternative models that place this absorption in c4L.

    In the left panel of Figure \ref{fig:z0p47_c4L_CI} we show that \ion{O}{ii}, \ion{O}{iii}, and \ion{Mg}{ii} yield [$\alpha$/H]$_{\rm c4L}\sim0$. \ion{C}{ii} and \ion{C}{iii} are consistent with this density, but require a slightly enhanced [C/$\alpha$]. Modeling all of these detections together, we find [C/$\alpha$]$_{\rm c4L}=0.33\pm0.03$ and [C/N]$_{\rm c4L}=0.12\pm0.03$, at a density of $\log\,n_{\rm H, c4L}/\cmx{-3}=-2.67\pm0.12$.\footnote{Due to the inherent model uncertainties regarding the inclusion of \ion{C}{iii} and \ion{N}{iii}, as well as the possibility that \ion{C}{iii} 977\AA\ is saturated, for completeness, we also consider a model without these species. This model still finds solar $\alpha$-element abundances and enhanced carbon, with [$\alpha$/H]$_{\rm c4L}=0.08_{-0.05}^{+0.06}$, and [C/$\alpha$]$_{\rm c4L}=0.41\pm0.03$ at $\log\,n_{\rm H,c4L}/\cmx{-3}=-2.46\pm0.22$.}

    For c4H the density and relative abundances are robustly modeled, but we can only obtain a limit to metallicities since \N{H\,I,c4H} cannot be measured. Assuming photoionized, optically thin gas, the \N{O\,III,c4}-to-\N{O\,IV,c4} ratio yields a density of $\log\,n_\mathrm{H,c4H}/\cmx{-3}=-3.85\pm0.14$ (right panel of Figure \ref{fig:z0p47_c4L_CI}) lower than $n_\mathrm{H,c4L}$ by at least a factor of ten. We find relative abundances [C/$\alpha$]$_\mathrm{c4H}=\xpm{-0.28}{+0.18}{-0.17}$ and [N/$\alpha$]$_\mathrm{c4H}=\xpm{0.00}{+0.20}{-0.21}$. Taking the neutral hydrogen column density of c4L as an upper-limit to that of c4H, find a lower limit of $[\alpha/\rm{H}]_{\rm c4H}>-1.19$.
    
\subsection{\texorpdfstring{\bm{$z=0.47$} Components 5 \& 6}{}}
\label{sec:z0p47_c56}

\begin{figure*}
    \centering
    \includegraphics{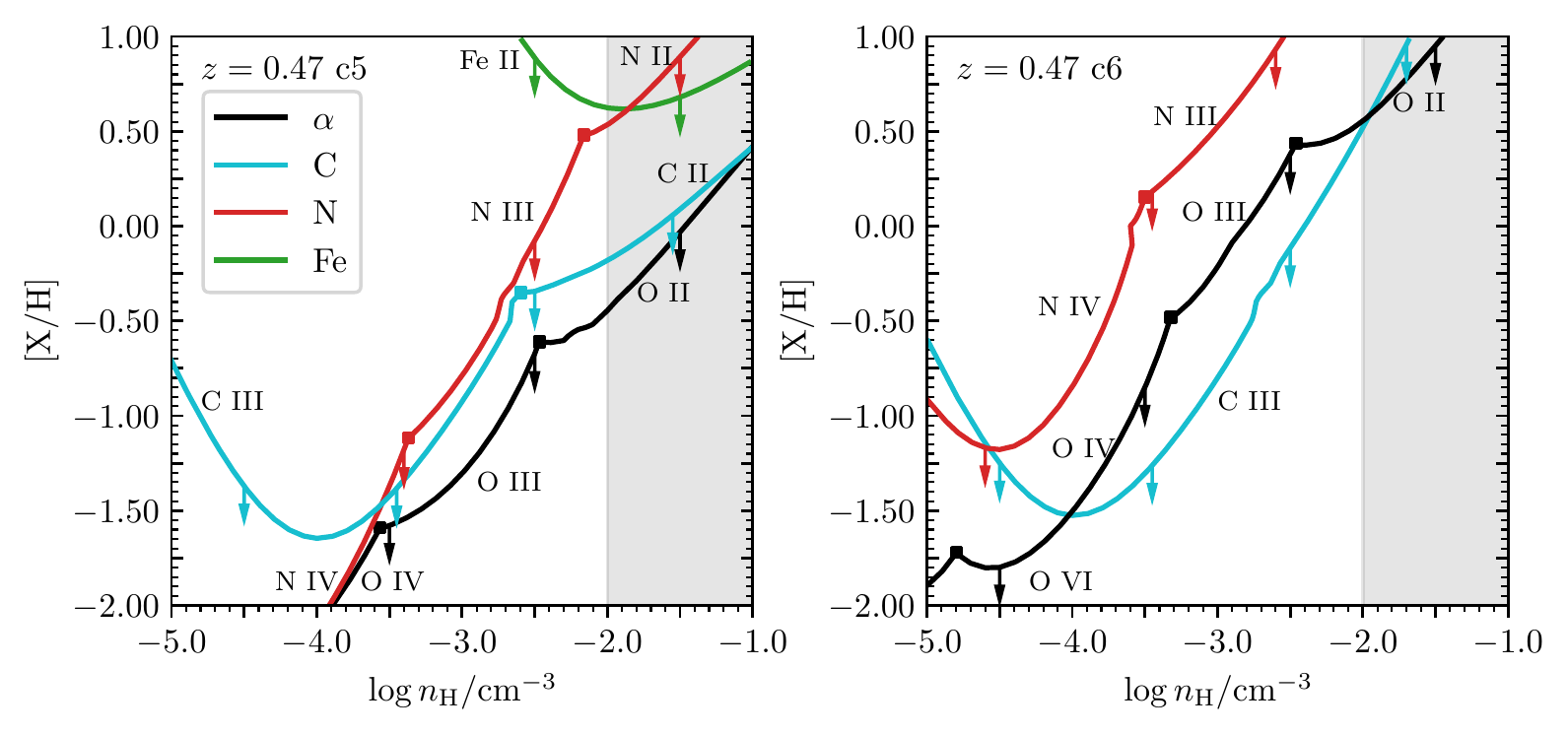}
    \caption{Abundance constraints for c5 and c6 of the $z=0.47$ system (left and right, respectively) based on column density upper-limits. All [X/H] \textit{below} these curves are possible. Squares indicate densities where the ion providing the constraint changes, and segments are labeled by this constraining ion. As we will show in Figure \ref{fig:phases}, $\log n_\mathrm{H} / \mathrm{cm}^{-3} >-2.0$ (indicated with gray shading) is only seen in high \N{HI} components. Thus, upper-limits given in Table \ref{table:abundances} are the largest [X/H] allowed in the range $-5\leq\log n_\mathrm{H}/\,\mathrm{cm}^{-3}\leq-2$.}
    \label{fig:z0p47_c56}
\end{figure*}

Components 5 and 6 are the bluest and reddest \ion{H}{i} absorbers seen in the $z=0.47$ complex, at $\Delta v_\mathrm{c5}=-40\,\kms$ and $\Delta v_\mathrm{6}=274\,\kms$, and neither has any associated metal absorption. In Figure \ref{fig:z0p47_c56} we show contours of the maximum allowed abundances plotted against density, with labels indicating the ion with the most constraining column density limit across different ranges of $n_\mathrm{H}$. 

The abundance limits generally decrease as density decreases, and we opt to obtain conservative metallicity upper-limits, at a maximum value of $n_\mathrm{H}$. Since $\log\nH\,/\cmx{-3}>-2$ is rarely seen outside of optically thick components with $\log N_\mathrm{H\,I}\,/\cmx{-2}>16.0$ (see \S\ref{sec:results:multiphase} and \citetalias{2021MNRAS.506..877Z}) we set this to be the maximum density for c5 and c6.

At $\log N_\mathrm{H\,I}\,/\cmx{-2}=15.34\pm0.09$, c5 is constrained to abundances of [$\alpha/\rm{H}]_\mathrm{c5}<-0.4$, lower than all other components except for c1L. Since c6 has a much lower $\log N_\mathrm{H\,I}\,/\cmx{-2}=14.18\pm0.02$, it yields weaker constraints, simply requiring [$\alpha/\rm{H}]_\mathrm{c6}<0.5$. However, at a density of $\log\nH\,/\cmx{-3}\sim-2.5$, comparable to that seen for other components, [C/H]$_\mathrm{c6}$ is limited to a sub-solar value.

\subsection{\texorpdfstring{\bm{$z=0.54$} Component 1 \bm{$\left(\Delta v_\mathrm{c1}=0\,\kms\right)$}}{}} \label{sec:z0p54_c1}

    \begin{figure*}
    \centering
    \includegraphics{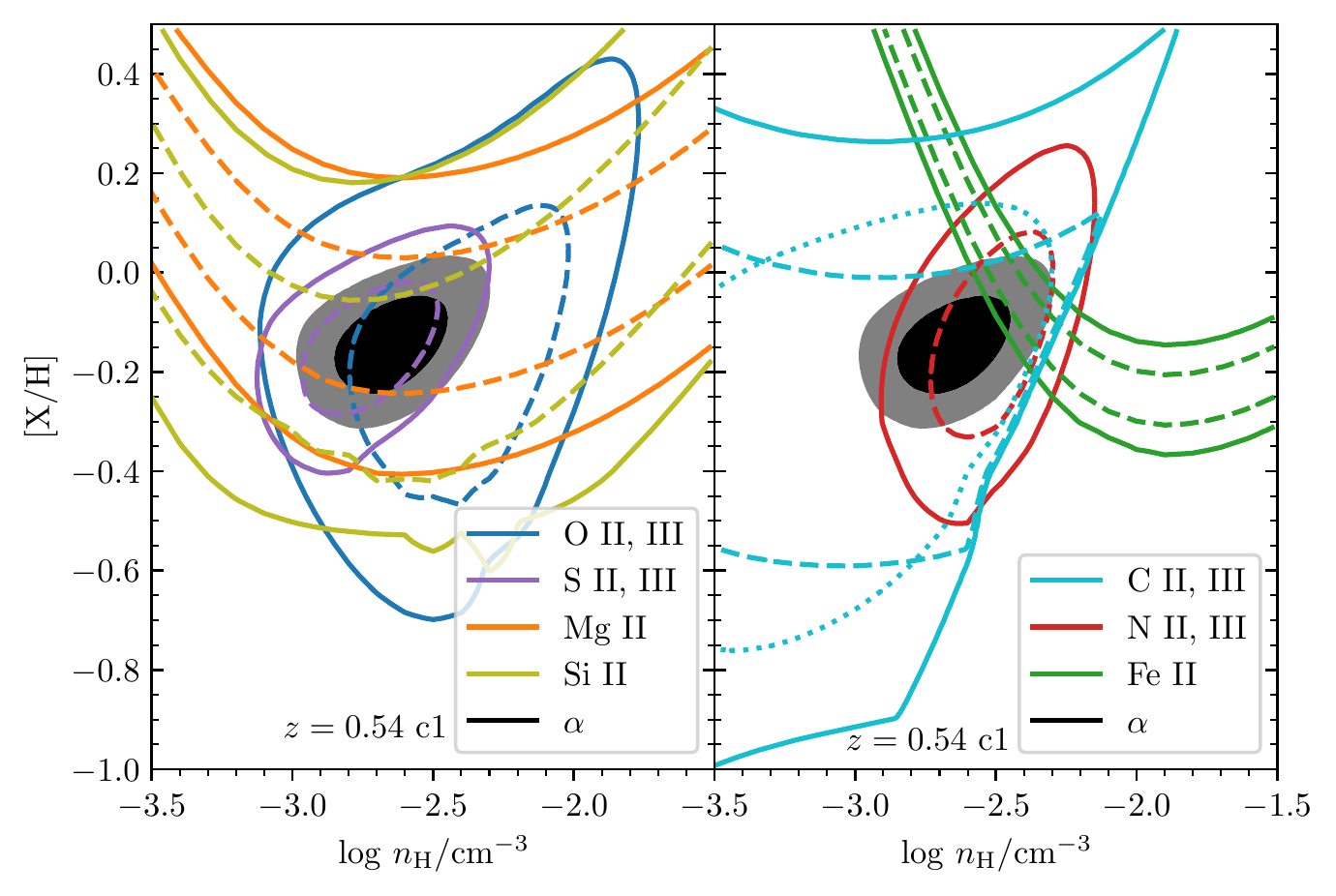}
    \caption{{\it Left}: Contours of 68 and 95\% likelihood, using conventions described in Figure \ref{fig:CI_MP}, for $\alpha$-elements of component 1 at $z=0.54$ with $\log\,\N{ H\,I,c1}/\cmx{-2}=16.5$. The filled contour is a joint fit to all singly- and doubly-ionized $\alpha$-elements. This demonstrates that $\alpha$-element column densities are all in agreement at a density of $\log\,n_\mathrm{H,c1}/\cmx{-3}\approx-2.8$ and an abundance of [$\alpha$/H]$_\mathrm{c1}\approx-0.1$. {\it Right}: Contours of other elements. N is in good agreement with $\alpha$-elements, while Fe may be enhanced by up to $\approx$ 0.2 dex, although [$\alpha$/Fe]$_\mathrm{c1}$=0 is within $95\%$ likelihood intervals. The sharp edge of the C contour is a result of considering \ion{C}{iii} as a lower limit; the dotted interior contour is the 68$\%$ credible interval if we instead treat it as a detection. Although the \ion{N}{iv} is only marginally consistent with this solution (see Figure \ref{fig:CI_MP}), a two-phase model yields the same density and abundances for the low-ionization phase.\label{fig:z0p54_c1_CI}}
    \end{figure*}
    
   \begin{figure*}
    \centering
    \includegraphics{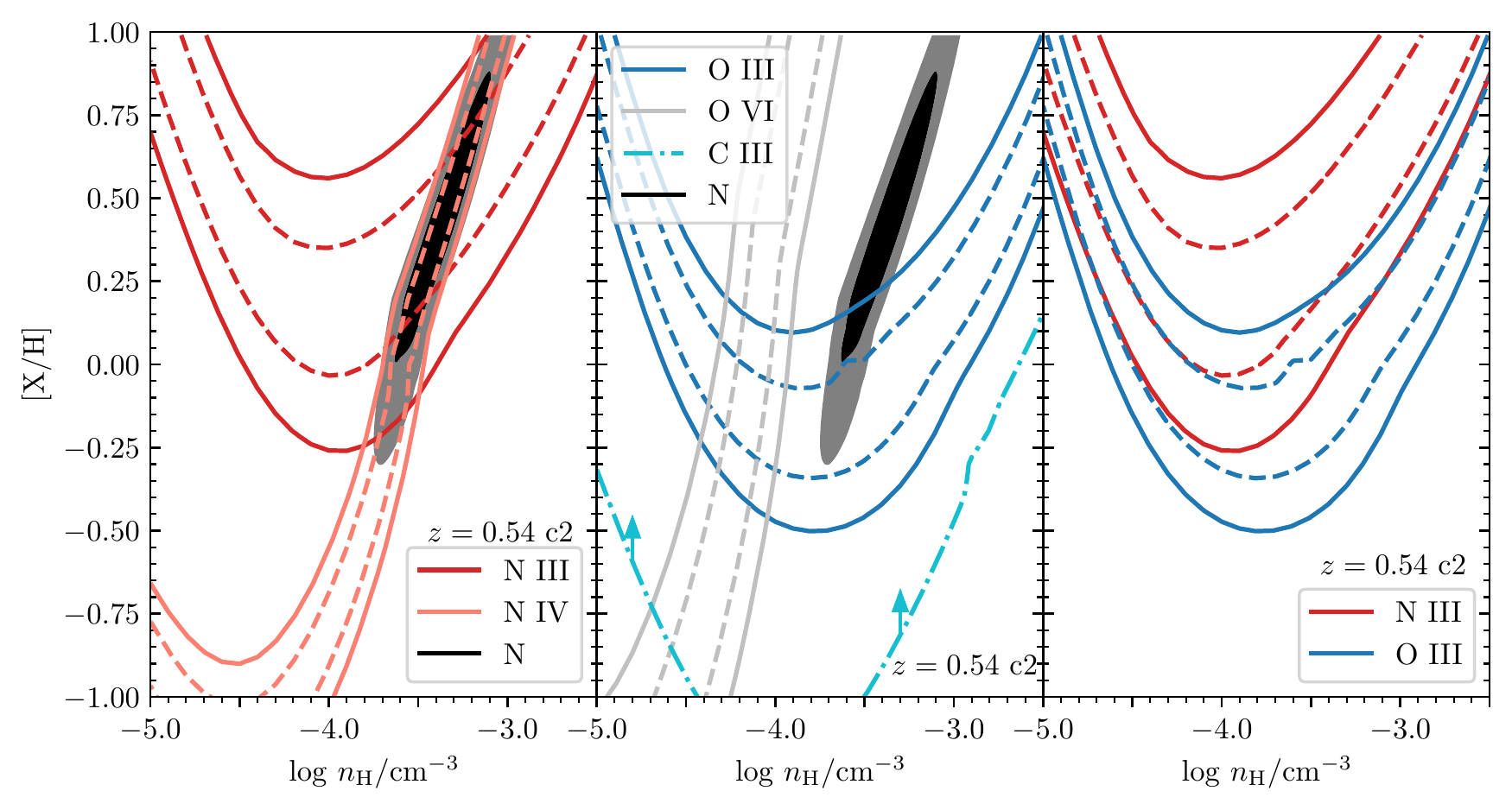}
    \caption{Likelihood contours for $z=0.54$, c2, using conventions described in Figure \ref{fig:CI_MP}. \ion{N}{iii} and \ion{N}{iv}, on the left, yield a slightly supersolar [N/H], at a density of $\log\,n_\mathrm{H,c2}/\cmx{-3}\sim-3.5$. However, \ion{O}{vi} requires the gas to have a density of $\log\,n_\mathrm{H,c2}/\cmx{-3}<-4$ to be consistent with \ion{O}{iii}, shown in the center panel, suggesting that the \ion{O}{vi} arises from a more highly-ionized phase, as is typically expected. The dash-dotted cyan line is a lower limit to [C/H], from \N{C\,III,c2}, which is particularly sensitive to model uncertainties because of potential saturation from c1. It does nonetheless show that, unless \N{C\,III,c2} is underestimated by an order-of-magnitude, [C/$\alpha$]$_{\rm c2}$ is substantially sub-solar, via comparison with the \ion{O}{iii} contour.  \N{N\,III,c2} is also model dependent due to blending and potential saturation however, based on the measured \N{N\,III,c2} and \N{C\,III,c2}, [N/$\alpha$]$_{\rm c2}$ is at least solar, and is favored to be slightly enhanced, as shown by the comparison of \ion{N}{iii} and \ion{O}{iii}.}
 \label{fig:z0p54_c2_CI}
    \end{figure*}

    This component, containing $>98$\% of the total $\N{H\,I}$ of the pLLS at $z=0.54$ with $\log\N{H\,I,c1}/\mathrm{cm}^{-2}=16.5$, has multiple ions detected across a range of ionization stages, ranging from \ion{Mg}{i} to \ion{O}{vi}. \ion{O}{iv} falls at the same wavelength as geocoronal Lyman-$\alpha$ emission from the Milky Way and is excluded from the analysis.

    Figure \ref{fig:z0p54_c1_CI} shows maximum likelihood contours of singly- and doubly-ionized species of all elements detected, assuming they arise from a single density phase. O, S, N, and C all have detected singly- and doubly-ionized species, and are all in agreement with a density of $\log\,n_{\rm H,c1}/\cmx{-3}\sim-2.6$. Moreover, the $\alpha$-elements all have comparable metallicities at this density of $[\alpha/\rm{}H]\sim-0.15$, and C, N, and Fe are consistent with a solar chemical abundance pattern.
    
    While this appears to be a clear solution, \ion{N}{iv}, \ion{S}{iv}, \ion{S}{v}, and \ion{O}{vi} absorption indicates the presence of at least one higher ionization phase which may contribute to the column density measurements of doubly-ionized species and affect the derived density of the low-ionization gas. Regardless, however, of the presence of a high-ionization phase, the abundances derived for the low-ionization phase are robust. Figure \ref{fig:z0p54_c1_CI} shows that based on the singly-ionized species alone, the absorber is highly enriched, with [X/H]$_{\rm c1L}\gtrsim-0.3$.
    
    To account for a higher-ionization component affecting the solution, we implement a two-phase photoionization model, with a high-ionization phase (c1H)  constrained to have a lower density and larger size  than the low-ionization phase (c1L). We also place an upper bound prior on the size of c1H of $200\,\text{kpc}$, a conservative upper bound corresponding roughly to the size of an $L_*$ galaxy halo \citep{2018AstL...44....8K}. Since we do not know \N{H\,I,c1H}, we need to marginalize over a range of plausible \ion{H}{i} column densities, such that even at a given density we cannot estimate $\left\{\rm{[X/H]}\right\}_{\rm c1H}$.
    
    We find that the best solution has $n_\mathrm{H,c1L}=\xpm{-2.56}{+0.10}{-0.08}$, very close to the density of the single-phase model previously considered, and abundances for c1L are unchanged between the single-phase and two-phase model, and $\log n_\mathrm{H,c1H}/\cmx{-3}\lesssim-3$. The allowed $\left\{[\rm{X/H}]\right\}_\mathrm{c1H}$ spans the entire range of our priors, so we cannot compare the chemical enrichment of c1L and c1H. The two phase solution does not reproduce the measured \N{O\,VI}, which results from a still more highly-ionized phase (see, e.g. \citealt{2021MNRAS.501.2112S}) which we do not attempt to model here.

\subsection{\texorpdfstring{\bm{$z=0.54$} Component 2 \bm{$\left(\Delta v_\mathrm{c2}=19\,\kms\right)$}}{}} \label{sec:z0p54_c2}

    Component 2 of the absorption system at $z=0.54$ is slightly redshifted relative to the pLLS, and has less neutral gas, with $\log\N{ H\,I,c2}/\mathrm{cm}^{-2}=14.4$. This component accounts for the excess metal absorption seen only in species that are at least doubly-ionized; fitting absorption profiles of these species with a single component at $\Delta v_0$ (which is well determined by \ion{Mg}{i}, \ion{Mg}{ii}, and \ion{Fe}{ii} in the MIKE spectrum) systematically under-fits profiles redward of $v>15\gtrsim,\kms$. c2 is also necessary to fit the Lyman $\beta$, $\gamma$, and $\delta$ profiles without over-fitting the red wings of higher order Lyman series lines. Since the velocity separation between components c1 and c2 is comparable to a resolution element in the COS spectrum, their profiles are highly blended.

    Due to this blending and the lack of detected low-ions in the MIKE spectrum, the line centroid is not well determined. Since the \ion{N}{iii}, \ion{O}{iii} and \ion{C}{iii} absorption is stronger in  c1, their profiles can be reasonably accommodated by broadening  c1. \ion{N}{iv} has a larger depth for  c2, and cannot be well fit without this additional feature. For this reason we opt to fix the line centroid of c2 at the velocity offset obtained by jointly fitting \ion{N}{iv} and \ion{H}{i}, $\Delta v_\text{c2}=19\,\kms$. We also fit \ion{O}{vi} assuming this component structure.

\begin{figure}
    \centering
    \includegraphics[width=\linewidth]{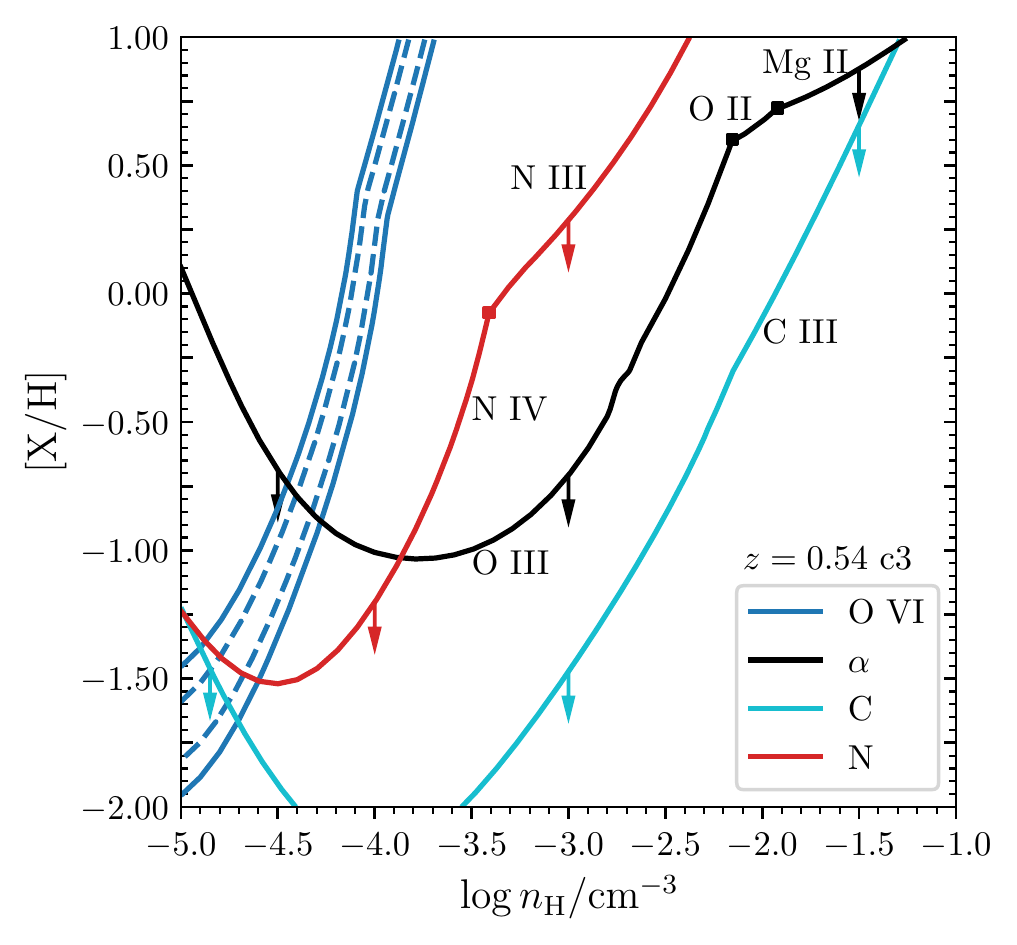}
    \caption{Contours of maximum likelihood for the low-ionization phase of Component 3 at $z=0.54$, using conventions described in Figure  \ref{fig:CI_MP}. Only \ion{O}{vi} and \ion{H}{i} are detected for this component. We indicate upper-limits from other $\alpha$-elements as well as C and N. Including the limits, we can identify limits on $\log n_\mathrm{H,c3}/ \rm{cm}^{-3}\lesssim-4.3$ and $[\alpha/\rm{H}]_\mathrm{c3}\lesssim-0.75$. [C/H]$_\mathrm{c3}$ and [N/H]$_\mathrm{c3}$ are limited to even lower abundances at the constraining density.}
    \label{fig:z0p54_c3_CI}
\end{figure}

    In Figure \ref{fig:z0p54_c2_CI} we show the likelihood contours for different ions detected in this component assuming a single, photoionized phase. The ratio of \ion{N}{iii}-to-\ion{N}{iv} indicates a density of $\log\,n_{\rm H,c2}/\cmx{-3}\sim-3.5$, inconsistent with the \ion{O}{iii}-to-\ion{O}{vi} ratio, so a single phase cannot explain these ions with ionization potentials ranging from 29 eV (\ion{N}{iii}) to 113 eV (\ion{O}{vi}). The most likely scenario is that the \ion{O}{vi} comes from a lower density (and likely collisionally ionized) phase, for which metallicity is highly uncertain without an \ion{H}{i} column density. 

    Our MCMC model considers the the N ions and \ion{O}{iii} as a single phase. We find $\log\,n_{\rm H,c2}/\cmx{-3}=-3.37\pm0.1$, [N/H]$_{\rm c2}=0.44\pm0.24$, and [$\alpha$/H]$_{\rm c2}=-0.03\pm0.2$. Although this suggests a nitrogen enhancement similar to that found for some LLS components in \citetalias{2021MNRAS.506..877Z}, the solution is sensitive to the \ion{N}{iv} column density such that attributing a fraction of it to a lower-density phase would result in roughly solar $\alpha$/N at a density closer to $\log\,n_{\rm H, c2}/\cmx{-3}\sim-4$. The comparatively low \N{C\,III,c2} could indicate depletion of C by a factor of 10, but since the line profile is blended with the saturated \ion{C}{iii} from c1, the measured column density for c2 is likely underestimated, and we treat it as a lower limit for modeling purposes.

    Including the doubly-ionized species in a fit to determine the line centroid yields a centroid of $\Delta v_\mathrm{c2,alt}=6\,\kms$, and larger column densities for the detected species in c2, in order to fit the redward wings of the profiles that are now farther from the line centroid. This suggests that the main result obtained from this component, that the $\alpha$ abundance is nearly solar, is robust regardless of the inherent ambiguity in the exact component structure. Note that the minimum allowed abundances required to explain \ion{N}{iii} and \ion{O}{iii} are [N/H]$_{\rm c2}\approx-0.25$ and [O/H]$_{\rm c2}\approx -0.5$, respectively, showing that c2 cannot have a substantially lower abundance than c1.

\subsection{\texorpdfstring{\bm{$z=0.54$} Component 3 \bm{$\left(\Delta v_\mathrm{c3}=90\,\kms\right)$}}{}} \label{sec:z0p54_c3}

This component has approximately the same $N_\mathrm{H\,I}$ as c2, but the only associated metal absorption is \ion{O}{vi}, implying a markedly lower density and/or metallicity. The measured Doppler parameters are consistent within 68\% credible intervals, $b_\mathrm{H\,I,c3}=\xpm{33.4}{+4.0}{-4.2}\,\kms$ and $b_\mathrm{H\,I,c3}=\xpm{41.8}{+11.7}{-8.7}\,\kms$, and we attribute the \ion{O}{vi} and \ion{H}{i} absorption to the same phase.

The likelihood contours in Figure \ref{fig:z0p54_c3_CI} demonstrate that combining $N_\mathrm{O\,VI}$ and the upper-limit to $N_\mathrm{O\,III}$ requires an abundance that is much lower than the other two components, [$\alpha/\rm{H}]_\mathrm{c3}\lesssim-0.75$ and a density of $\log n_\mathrm{H,c3}/\cmx{-3}\lesssim-4.3$. The upper-limits to $N_\mathrm{N\,IV}$ and $N_\mathrm{C\,III}$ are even more constraining, such that this component must have non-solar abundance ratios, [C/$\alpha]_\mathrm{c3}<0$ and [N/$\alpha]_\mathrm{c3}<0$, unless the density is at $\log n_\mathrm{H,c3}/\cmx{-3}\lesssim-4.6$. We also note that even if the \ion{O}{vi} and \ion{H}{i} are not associated, which allows the \ion{H}{i} to come from gas with higher density, the non-detections of other ions still imply a lower abundance than c1L and c2.

Predicted cloud sizes for c3 range from 30 to 150 kpc, with the largest size corresponding to the lowest density we consider in these models, at $n_\mathrm{H}=10^{-5}\,\cmx{-3}$, which is consistent with volume-filling CGM gas.

\begin{table*}
\centering
\caption{\centering Summary of density and chemical abundance inferences for individual components$^a$\label{table:abundances}}
\renewcommand{\arraystretch}{1.2}
\begin{tabular}{lRRRRRRR}
\hline\hline
 &  \multicolumn{1}{c}{$\log\nH/\cmx{-3}$} & \multicolumn{1}{c}{[$\alpha$/H]} & \multicolumn{1}{c}{[C/H]} & \multicolumn{1}{c}{[N/H]} & \multicolumn{1}{c}{[Fe/H]} & \multicolumn{1}{c}{[C/$\alpha$]} & \multicolumn{1}{c}{[N/$\alpha]$}\\
\hline
$z=0.47$ c1L & \xpm{-2.28}{+0.14}{-0.07} & \xpm{-0.81}{+0.06}{-0.05} & -0.66\pm0.10    & <-0.58 & <-0.06 & 0.14\pm0.10 & <0.28\\
$z=0.47$ c1H & \xpm{-3.51}{+0.11}{-0.14} & >-0.94 & <0.6 & <0.5  & \nodata & <-0.33 & \nodata \\
$z=0.47$ c2 & -2.38\pm0.13 & -0.12\pm0.11 & 0.14\pm0.11 & <0.31 & \nodata & \xpm{0.27}{+0.11}{-0.13} & <0.50\\
$z=0.47$ c3 & -3.77\pm0.05 & -0.07_{-0.13}^{+0.12}& -1.15\pm0.24 & -0.01_{-0.11}^{+0.12}  & \nodata & \xpm{-1.07}{+0.23}{-0.22} & 0.06\pm0.07 \\
$z=0.47$ c4L & -2.67\pm0.11          & 0.08\pm0.05          & 0.41\pm0.14 & 0.30_{-0.15}^{+0.10}    & \nodata & 0.33\pm0.14 & \xpm{0.23}{+0.10}{-0.16}\\
$z=0.47$ c4H & \xpm{-3.85}{+0.15}{-0.13} & >-1.19 & >-1.42 & >-1.40  & \nodata & \xpm{-0.28}{+0.18}{-0.17} & \xpm{0.00}{+0.20}{-0.21} \\
$z=0.47$ c5$^c$ & \nodata & <-0.43 & <-0.14 & <0.51 & \nodata &\nodata &\nodata\\
$z=0.47$ c6$^c$ & \nodata &<0.48 & <0.41 & \nodata & \nodata &\nodata & \nodata\\
\hline\hline
$z=0.54$ c1L & \xpm{-2.56}{+0.10}{-0.08} & -0.12\pm0.06 & \xpm{-0.26}{+0.20}{-0.24} & \xpm{-0.17}{+0.09}{-0.10} & 0.09\pm0.11 & \xpm{-0.15}{+0.20}{-0.23} & \xpm{-0.05}{+0.10}{-0.11}\\
$z=0.54$ c1H & <-2.97 & \nodata & \nodata & \nodata & \nodata & \nodata & \nodata\\
$z=0.54$ c2 & -3.37_{-0.14}^{+0.13} & -0.03_{-0.21}^{+0.19} & >-1.2 & 0.44\pm0.24  & \nodata & >-1.04 & \xpm{0.50}{+0.19}{-0.21}\\
$z=0.54$ c3 & <-4.33 & <-0.77 & <-1.22 & <-1.22 & \nodata& \nodata& \nodata \\\hline
\multicolumn{8}{l}{$^a$ Errors represent 68\% credible intervals.}\\
\multicolumn{8}{l}{$^c$ For components without detected metals, abundance limits are obtained assuming $-5\leq n_\mathrm{H}\leq-2$.}\\
\end{tabular}
\end{table*}
\section{Discussion} \label{sec:results}

    Combining {\it HST} COS FUV spectra and Magellan MIKE optical echelle spectra has enabled a comprehensive ionization analysis of diffuse circumgalactic gas based on resolved kinematics and abundance ratios of atomic species that span five different ionization stages.  Our analysis of two $z\approx 0.5$ pLLSs shows that kinematically aligned multi-phase gas can masquerade as a single-phase structure and can only be resolved by simultaneous accounting of the full range of observed ionic species.  A summary of the inferred gas density and elemental abundances is presented in Table \ref{table:abundances}.  
    
    Both pLLSs in this study exhibit some degree of variations in relative elemental abundances and densities ranging from $\log\nH/\cmx{-3}<-4.3$ to $\log\nH/\cmx{-3} \sim -2.3$ among different components. The absorption components in the $z=0.47$ pLLS exhibit a spread in [$\alpha$/H] of $>$0.8 dex as well as variation in [C/H] and [N/H], requiring that they have disparate physical origins. Conversely, the two strongest components modeled in the $z=0.54$ pLLS have consistent [$\alpha$/H], suggesting they may share a common origin, with only component 3 (which is detected only in \ion{H}{i} and \ion{O}{vi}) appearing to have a lower degree of chemical enrichment.
    
    In this section, we first discuss the advantages and limitations of different approaches adopted for the ionization analysis, followed by comparisons of kinematic profiles of different ionic transitions providing empirical evidence for the presence of multiphase gas in these pLLSs.  Finally, we discuss the chemical and kinematic properties of the gas in relation to their galaxy environments.

    \subsection{\texorpdfstring{Evaluation of photoionization analysis based on integrated \bm{$\N{H\,I}$} and \bm{$\N{ion}$}}{}}

    Due to limitations in wavelength coverage and/or data quality, chemical abundance analyses of the diffuse CGM/IGM often only examine low-ionization gas associated with the bulk of the \ion{H}{i}, and net column densities measured across an entire absorption complex are commonly used in lieu of considering components separately. Although this approach avoids model dependent measurements resulting from unclear component structure in medium-resolution spectra, it is unable to obtain key information about the relationship between components. Here we briefly consider how the interpretation of the two pLLSs presented in this work would differ following such an analysis.
    
    The inferred elemental abundances for the two strongest components of the comparatively simpler pLLS at $z=0.54$ would be minimally affected, since column densities for all species except \ion{O}{vi} are all significantly larger in c1 than c2. However, the main difference between the two components, their very different densities ($n_\mathrm{H, c1}\sim6\,n_\mathrm{H, c2}$), would be overlooked, as well as the possible nitrogen enhancement in c2. Such an analysis would also fail to recognize the difference in enrichment in c3, which shows significantly lower [$\alpha$/H] than c1 and c2.
    
    The higher complexity pLLS at $z=0.47$ has an ambiguity even in a simpler approach: should the clearly distinct absorption at $+200$ \kms\, (c4) be included in summed column densities? If we combine the column densities for components 1, 2, and 3, only considering species with ionization energy $E_i<40$ eV, we find a mean gas density of $\log\,\nH/\cmx{-3}\sim-2.7$ and a mean $\alpha$-elemental abundance of [$\alpha$/H]$\sim-0.6$ with similar [C/H] and [N/H]. The derived properties of the combined pLLS components are then comparable to our results for c1L, the dominant \N{H\,I} component, but we have no remaining indication that it is accompanied by absorption from solar metallicity gas. Component based analysis in this case helps recover a much richer view of the CGM capturing the larger range of physical processes which sculpt galaxy evolution. 
    
    Lastly we comment that critical information is encoded not just by the overall metallicity, but also in the elemental abundance ratios between components, which inform the likely origin of the gas as discussed in \S\ref{sec:results:abundances} (see also \citetalias{2021MNRAS.506..877Z} and \citealt{2019MNRAS.484.2257Z}).

\subsection{The Benefits and Limitations of Resolved Multi-Component and Multi-Phase Ionization Modeling}

\label{sec:results:robust}

  Complex multi-component, multi-phase modeling such as that presented herein has been shown to successfully reproduce observed column densities and/or absorption profiles. But even with this more sophisticated approach, the limited number of distinct elements and available ionization states can still require simplifying assumptions such as a solar relative abundance pattern \citep{2021MNRAS.501.2112S}, or identical elemental abundances for gas in different phases \citep{2020MNRAS.tmp.3454H}. Furthermore, there are often multiple viable models that can satisfactorily explain the data \citep{2020MNRAS.tmp.3454H}. The approach presented in this paper makes neither of these assumptions. This section highlights which results from the photoionization analysis are most robust, and which results are more model or prior dependent. For further details, see \S\ref{sec:analysis}.
    
  As demonstrated by Figures \ref{fig:z0p47_c1_CI}, \ref{fig:z0p47_c2_CI}, \ref{fig:z0p47_c4L_CI}, and  \ref{fig:z0p54_c1_CI}  in \S\ \ref{sec:analysis}, the ionization fractions of singly-ionized species and \ion{H}{i} tend to scale with density such that for $-3\lesssim\log\nH/\cmx{-3}\lesssim-1$, inferred abundances are relatively static. \citet{2016ApJ...831...95W} showed that ionization corrections needed to obtain [Mg/H] from \N{Mg\,II}/\N{H\,I} vary by up to 0.5 dex across the range of densities typically seen for absorbers with $16.5<\log\N{H\,I}/\cmx{-2}<18.0$ \citep[see also][]{2019ApJ...887....5L,2019ApJ...872...81W}.
  
  Our analysis shows that with \N{Mg\,II} precisely measured from the high-resolution optical spectra and all available singly-ionized species, including O$^+$, Si$^+$, and S$^+$, robust constraints for [$\alpha$/H] can be obtained for components c1L, c2, and c4L at $z=0.47$, and c1L at $z=0.54$ to within 0.1 dex uncertainties. For these same absorbers, however, the column densities of more highly-ionized species are required to correctly infer the density, based on the ratio of, e.g., \N{O\,III}/\N{O\,II}, rendering densities more dependent on the manner in which the multi-phase gas is modeled. Relative elemental abundances are also often robustly measured despite uncertainty in \nH. For example, Figure \ref{fig:z0p47_c3_CI} shows that, for optically thin low density gas, the ratio of \ion{O}{iii}-to-\ion{N}{iii} can be used to determine whether N is enriched or depleted relative to $\alpha$-elements. Figure \ref{fig:z0p47_c2_CI} likewise shows that for gas with $-4\lesssim\log\nH/\cmx{-3}\lesssim-2$, \ion{C}{ii} and \ion{Mg}{ii} can be used to gauge [C/$\alpha$].
  
  While the overall degree of enrichment for high-ionization components which are kinematically aligned with lower-ionization components (c1H and c4H in the $z=0.47$ system and c1H in the $z=0.54$ system) is uncertain due to their unconstrained \N{H\,I}, their relative abundances can offer important clues about their origin and relation to each other. Notably, for c1H in the $z=0.47$, even without an \N{H\,I} measurement, we can constrain [C/$\alpha$] using the density solution and measured \N{O\,IV} and \N{C\,III}. Critically, the [C/$\alpha$] derived for the high and low-ionization phases of this component are discrepant, a point we will return to in Section  \ref{sec:results:abundances}.
  
  Conversely, we consider [C/H] of high-ionization components c3 at $z=0.47$ and c2 at $z=0.54$, to be uncertain due to model dependency, and present their seemingly substantially sub-solar carbon abundances, with both around [C/$\alpha]\sim-1.1$,  as lower limits. This is largely due to the fact the \ion{C}{iii} is a single observed transition, which is easily saturated in these components at moderately high density and metallicity. This uncertainty is compounded by the COS line spread function which introduces additional error because narrow saturated profiles can often appear unsaturated with profiles that do not reach near zero transmitted flux.
  
  Another uncertainty inherent to analysis of medium-resolution data relates to the uncertain component structure for blended absorbers, as outlined in \S\ref{sec:methods}. The component structure for absorbers with \ion{Mg}{ii} detected in the high-resolution MIKE spectrum have far more robust Voigt profile models, with precise line centroids and, in some cases, Doppler parameters tied to the MIKE data. 
    
  However, not all absorbers have strong enough associated \ion{Mg}{ii} absorption to be detected.  c3 at $z=0.47$ and c2 at $z=0.54$ only have more highly-ionized species, with profiles that are heavily blended with other nearby components in the COS spectrum. Comparatively small shifts in the velocity centroid of these components introduce large changes in the fits to doubly-ionized species in both these components and those nearby. If c3 at $z=0.47$ is fixed at several \kms\ blueward of its nominal velocity, the best-fit profile to the overall absorption complex exchanges some optical depth with c2, resulting in changes to \N{C\,III} for both components larger than uncertainties at the nominal velocity.  On the other hand, a redward shift leads to best-fit profiles in which \ion{C}{iii} is saturated.  For c2 at $z=0.54$, a blueward shift results in a similar exchange with c1. Coverage of the full Lyman series somewhat alleviates this, but centroids of the comparatively weaker \ion{H}{i} in more highly ionized components are not always well determined; c3 at $z=0.47$ only has an appreciable contribution to the \ion{H}{i} profile up to about Ly$\eta$, and is heavily blended with both c1 and c2.
    
  \begin{figure*}
    \includegraphics{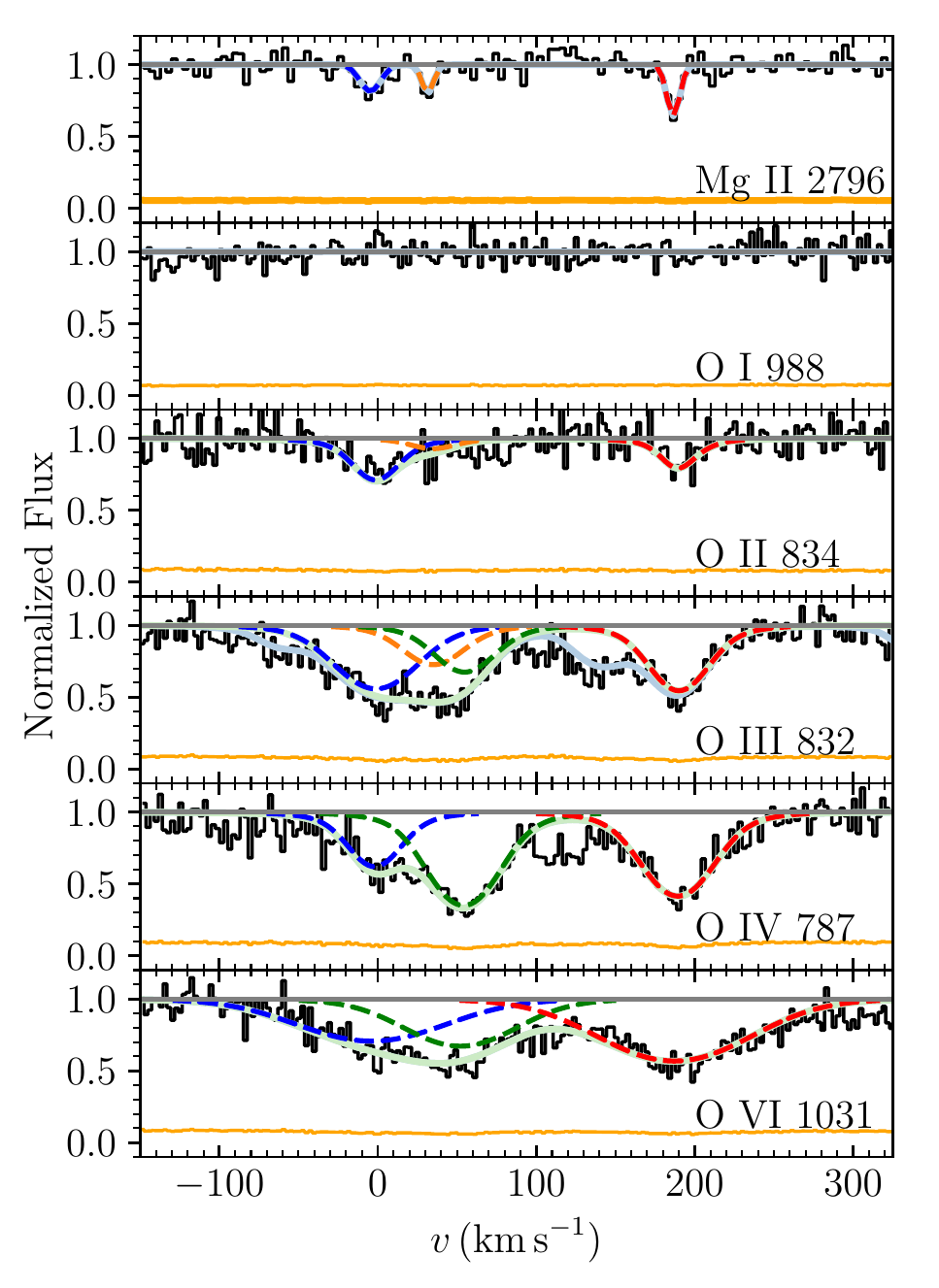}
    \includegraphics{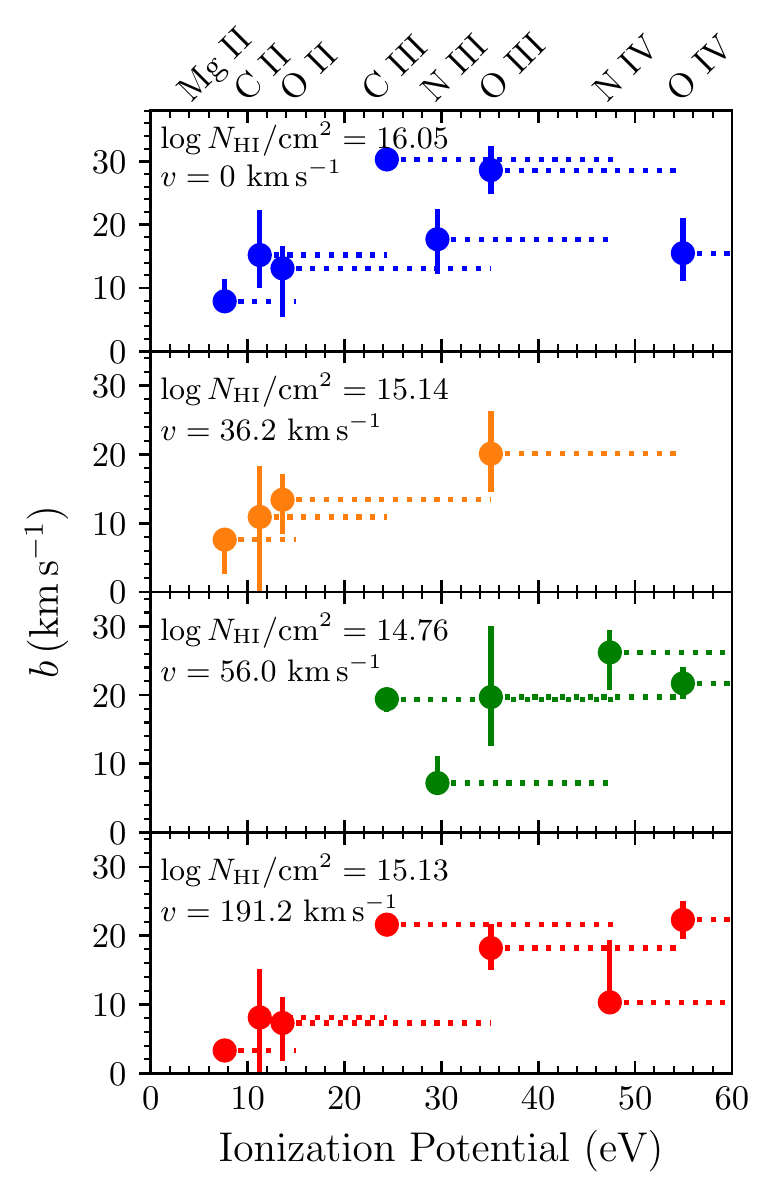}
    \caption{{\it Left}: Absorption profiles of oxygen ions in the partial Lyman limit system at $z=0.47$, overlaid with a Voigt profile fit with the minimum number of components that can fit the data. Focusing on component 4, fit in red, one can see by eye that the width of the oxygen ions rises with increasing ionization potential. 
    {\it Right}: Doppler parameters as a function of ionization potential. Points are placed at the energy required to form this ion from a lower state, with dotted lines extending to the energy required to ionize the indicated ion. There is an overall trend of increasing line widths with ionization energy. Note only the \ion{Mg}{ii} absorption profiles (blue, orange, and red) are from the MIKE spectrum with resolution of 8 \kms. The labels at the top indicate the ions of the data points; some are slightly offset for clarity.}
    \label{fig:b_v_E}
    \end{figure*}
 
    Data with spectral coverage of the \ion{C}{iv}$\lambda\lambda$ 1548,1550 doublet have the potential to enable confident density and [C/$\alpha$] measurements in high-ionization gas. The doublet ratio can be used to ascertain saturation, and line profiles are less likely than \ion{C}{iii} to be contaminated with absorption from nearby lower-ionization components.
    
    There is also uncertainty inherent in the choice of the UVB.  \citet{2015MNRAS.446...18C} developed a technique, since adopted and adapted in some other analyses, to fit modifications to the UVB jointly with absorber models. However, in a multi-phase model with variable abundance ratios, introducing these additional parameters increases the risk of over-fitting the relatively small number of measured column densities. As noted in \citet{2019MNRAS.484.2257Z}, the UVB of \citet{2012ApJ...746..125H}, which is considerably harder than that used here, systematically predicts higher metallicities with the largest difference of 0.7 dex seen for components with $\log\N{H\,I}/\cmx{-2}\lesssim15$ (see also \citetalias{2021MNRAS.506..877Z} \S3.2).

    \subsection{The multiphase nature of the CGM}
    \label{sec:results:multiphase}

    The two pLLSs presented herein provide compelling evidence that the CGM is multiphase with relatively little analysis. The left panel of Figure \ref{fig:b_v_E} shows the absorption profiles of oxygen ions in the $z=0.47$ absorption complex. Most clearly seen in c4, which is not blended with other identified components, these data show clear visual evidence of increasing line width as a function of their ionization potential, even for single components with a common velocity centroid. To quantify this with minimal assumptions, we fit the data with Doppler parameters which are allowed to vary independently for each of the different ionization states. The right panel shows the result, with components 1, 2, and 4 exhibiting clearly increasing widths with ionization state, retaining approximately the same centroids (see also \citealt{2020MNRAS.tmp.3454H} and \citetalias{2021MNRAS.506..877Z}). This trend is also seen quantitatively by comparing measured line widths of ions spanning a range of ionization potentials $E_\mathrm{i}$, with a tendency for species that have $E_\mathrm{i}>20$ eV having $b>20\,\kms$ and those with lower ionization potentials being closer to $b\sim10\,\kms$.

    If this were simply a comparison between line widths typical for different ionization states, a ready explanation would be that the ions are tracing distinct regions of the CGM, with more highly ionized species residing in regions of higher temperature or gas turbulence. However, the shared kinematic structure suggests a more complicated picture, where absorption by gas at different densities occurs within what seems to be a single, discrete absorption feature. Absorption by a single gas phase, without any aligned higher or lower density gas does also exist, such as the low ionization c2 at $z=0.47$. 
    
    Photoionization modeling of components with detected ions spanning a range of ionization states further clarifies the multiphase nature. As discussed in \S\ref{sec:analysis} and shown in Figure \ref{fig:CI_MP}, modeling absorbers detected in such a wide range of ionization states using a single-density photoionization model often results in ionization fractions inconsistent with observations. Components only detected in ions with $E_\mathrm{i}<40$ eV (i.e., up to and including triply-ionized states of commonly observed species) can often be well modeled by a single phase gas, but such solutions are not necessarily robust without incorporating constraints for more highly-ionized species. 
  
    In c1 at $z=0.47$ (the dominant H\,I component), we see that the two phase solution necessitated by the inclusion of \ion{O}{iv} divides the \ion{O}{iii} column density between the two phases. Compared to a single phase model without \ion{O}{iv}, the c1L \ion{O}{iii} ionization fraction is lower, and the resulting density is $\log,n_{\rm H,c1}/\cmx{-3}=-2.28$, twice the density of $\log\,n_{\rm H,c1,alt}/\cmx{-3}\sim-2.6$ obtained by only considering \ion{O}{ii} and \ion{O}{iii} (Figure \ref{fig:CI_MP}). This also means that the size inferred by the single-phase model is too large by a factor of two.
    
    However, c2 of the same absorption system demonstrates this to not be a universal feature, as it does not exhibit absorption from triply-ionized species, despite c1 and c2 having comparable densities. The lower density c3 shows a markedly different ionization pattern, with no absorption seen from ions with $E_\mathrm{i}<20$ eV (i.e., singly-ionized). Nonetheless, the absorption signature of c3 seems to be comprised of multiple phases, since \N{O\,III,c3}, \N{O\,IV,c3}, and \N{O\,VI,c3} are inconsistent with a single phase, and the \ion{O}{vi} profile is wider. Noting that c1 also has \ion{O}{vi}, it seems that c1 is a three-phase absorption component, while c3 may be very similar, except with the highest density phase missing or physically smaller. This result highlights that the multiphase nature of the CGM, often discussed in the context of varying density between components, is also clear through separate analysis of individual components.
    
\begin{figure}
    \includegraphics[width=\linewidth]{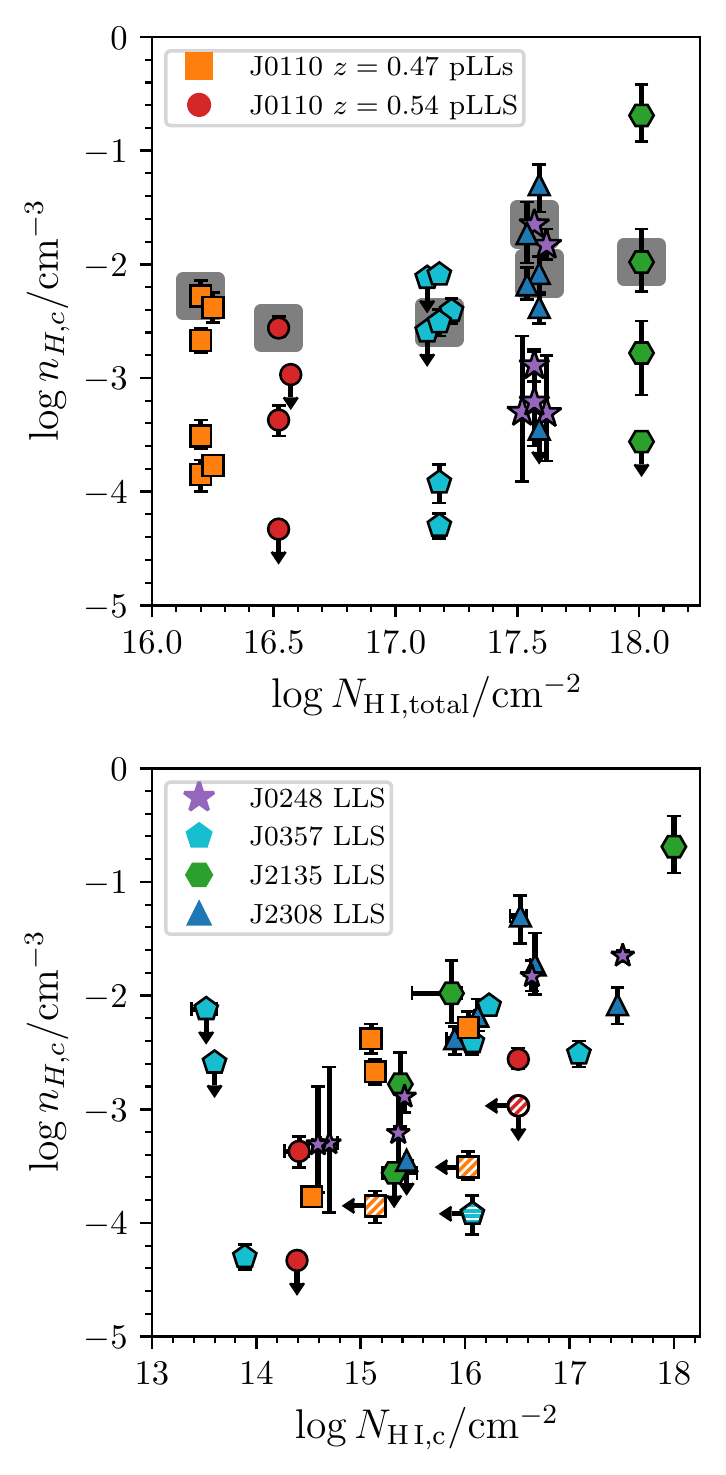}
    \caption{Individual component densities and column densities are compared for LLSs from \citetalias{2021MNRAS.506..877Z} and pLLSs from this work. The top panel shows the range of component densities detected along single sightlines and is plotted against the total \N{H\,I} of the LLS/pLLS. The dominant \N{H\,I} component of each absorber is highlighted by a grey box. The bottom panel shows the strong degree of correlation between the \N{H\,I} column densities of individual components and their derived densities. Hatched markers indicate the high-ionization phase of components with multiphase models, and are placed at \N{H\,I} of the low-ionization phase. The analysis of these CUBS sightlines shows the wide range of CGM gas densities found even within single galaxy halos. \label{fig:phases}}
    \end{figure}

    \citetalias{2021MNRAS.506..877Z} conduct a component-by-component analysis, similar to that presented in this work, of four newly identified Lyman Limit systems in the CUBS data set, with $\log\N{H\,I}/\cmx{-2}\gtrsim 17.2$. Combining the \citetalias{2021MNRAS.506..877Z} analysis with the results presented herein provides a compelling view of multi-phase gas in the intermediate redshift CGM. \citetalias{2021MNRAS.506..877Z} also identify one component, at $\log\N{H\,I,c}/\,\cmx{-2}=16.02$, which requires a two-phase solution, with \N{O\,III} and \N{N\,III} comparable between the two phases, similar to the multi-phase absorber identified in this paper. Overall, as is shown in Figure \ref{fig:phases}, the CUBS data show that single sightlines pass through gas with a wide range of densities from $-4 \lesssim \log(\nH/\cmx{-3})\lesssim 0$ \citep[see also][]{2019MNRAS.484.2257Z}, with the highest densities typically corresponding to those with the largest \N{H\,I,c}. This could reflect the known inverse correlation between impact parameter and total \N{H\,I} that has previously been established \citep{1998ApJ...498...77C,2001ApJ...559..654C,2012ApJ...750...67R,2014ApJ...792....8W,2017ApJ...837..169P}, but also a wide range of physical processes as discussed in Sections \ref{sec:results:abundances} and \ref{sec:results:galaxies}. Regardless, however, of their physical origin, these measurements provide robust guidance to simulations of the range of densities which must be resolved in order to accurately capture the phase structure and physical properties of the CGM.
    
    In both \citetalias{2021MNRAS.506..877Z} and this work, \ion{O}{vi} is often kinematically offset from all other ions, and even two-phase models of components where \ion{O}{vi} is aligned cannot reproduce \N{O\,VI}. Given the difference in line-widths observed in \ion{O}{vi}, we favor a scenario in which \ion{O}{vi} is produced by still more highly ionized and hotter or more turbulent gas. However, for \ion{O}{vi} absorbers with similar line widths like the lower-ionization phases (such as that discussed in \citetalias{2021MNRAS.506..877Z} \S5.1), local hard ionizing sources may play an important role. Alternatively, additional high-energy photons in excess of the assumed UVB may be present within the true background, as the shape of the UVB is less observationally constrained at high energies \citep[see e.g.][]{2016ApJ...833...54W}.

    \begin{figure*}
    \centering
    \includegraphics[width=\linewidth]{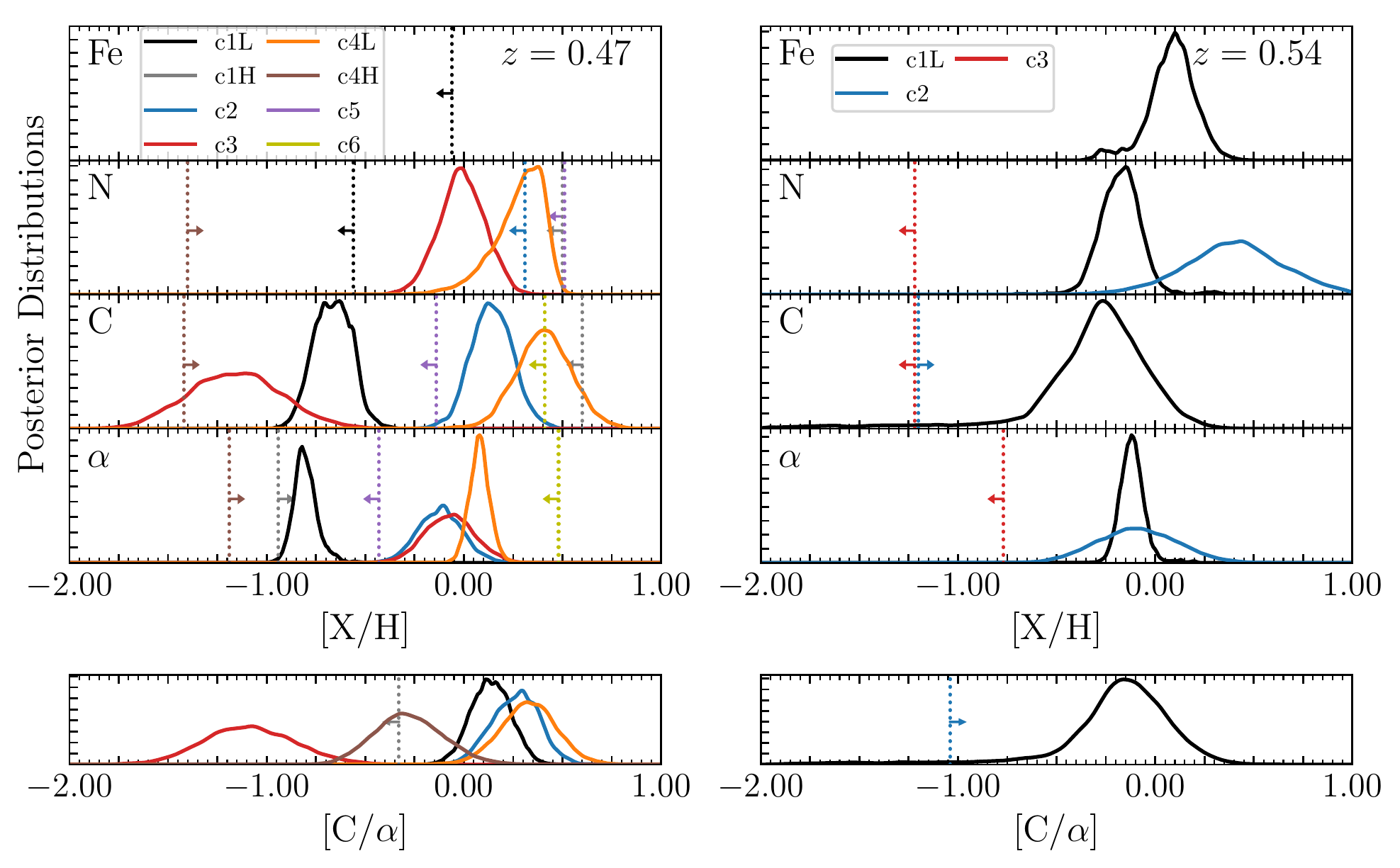}    
    \caption{Kernel density estimations of chemical abundance posterior distributions from MCMC photoionization analysis. Unconstrained posteriors yielding limits are shown by vertical dotted lines, with arrows indicating the direction of the limit. {\it Left:} c2, c3, and c4L in the $z=0.47$  show consistent abundances in [$\alpha$/H] and [N/H] with c2 and c4L having overlapping 95\% credible intervals of [$\alpha$/H],[C/H], and [N/H]. While the inferred [C/H]$_\mathrm{c3}$ is significantly lower, the carbon abundance of c3 is particularly model dependent due to blending and possible saturation of \ion{C}{iii} (see Section \ref{sec:z0p47_c3}) and we treat it hence forth as a lower limit. In contrast to the high-metallicity of c2, c3, and c4L, c1L, the dominant H\,I component, has significantly lower abundances in all measured species, consistent only with the upper-limit to c5. The bottom panel compares the measured [C/$\alpha$] ratios. Notably, c1L and c1H have discrepant values, implying they do not share a common origin.  
    However, c4H and c1H, two more highly ionized absorbers, appear to have consistent abundance patterns. {\it Right:} The two higher-\N{HI} components in the $z=0.54$ complex show consistent [$\alpha$/H] and [C/H]. [N/H]$_\text{c2}$ is enhanced compared to c1, the dominant H\,I component; however, we note that the measured column densities of both \ion{C}{iii} and \ion{N}{iii} for c2 are somewhat model dependent due to blending with c1 and saturation. c3 shows much lower abundance in $\alpha$, N, and C than c1L or c2.}
    \label{fig:posteriors}
    \end{figure*}

    \begin{figure*}
    \includegraphics[width=\linewidth]{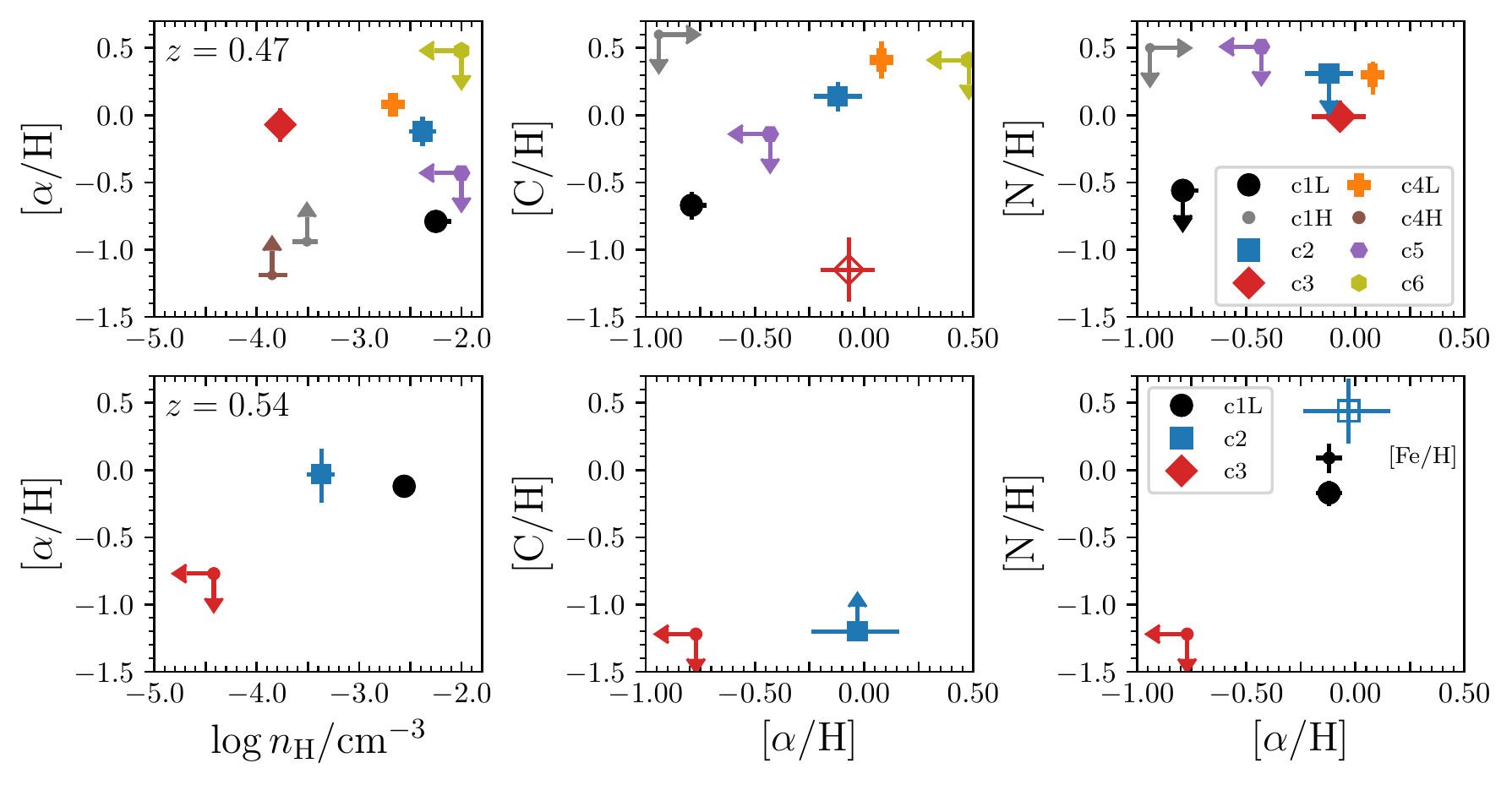}
    \caption{Densities and chemical abundances resulting from photoionization modeling. To the left is $\alpha$-element abundances versus density for each component absorption component, with C and N abundances shown against $\alpha$ to the right. The absorber at $z=0.47$ is on the top row, and $z=0.54$ is on the bottom. [C/H] for $z=0.47$ c3 and [N/H] for $z=0.54$ c2 are plotted with open symbol to indicate that they particularly model-dependent. [Fe/H] for the $z=0.54$ c1 is included in the bottom right panel as a smaller black circle. The strongest \ion{H}{i} component at $z=0.47$ (c1L) is markedly less enriched than the other components with well-measured abundances, whereas the strongest component at $z=0.54$ (c1L) has roughly solar [$\alpha$/H], comparable to c2,. In the latter pLLS, only the high ionization component c3 has significantly sub-solar $\alpha$/H. \label{fig:abundances}}
    \end{figure*}

    \subsection{CGM Abundances and Densities}
    \label{sec:results:abundances}

    The range of derived elemental abundances, summarized in Figure \ref{fig:posteriors} and Table \ref{table:abundances}, makes clear the complexity of the intermediate redshift CGM. Inferred elemental abundance patterns and overall degree of enrichment vary between different components within the same halo, and even vary within the different phases of single kinematically aligned components.

    Figure \ref{fig:posteriors} shows that [$\alpha$/H] is well-measured with narrow posteriors compared to other elements. This is due to the numerous ions contributing to the photoionization analysis. In Section \ref{sec:analysis} we demonstrated that measurements of the various $\alpha$-element in individual absorbers are in good agreement, with overlapping 68\% likelihood contours.

    Figure \ref{fig:abundances} compares the abundances and densities of each component in the two absorption systems. Considering [$\alpha$/H] shown in the left panels, the two absorption complexes tell very different stories. The dominant \ion{H}{i} component (c1L) at $z=0.47$ has $\sim10\times$ lower [$\alpha$/H] than its accompanying nearly solar satellite components, whereas the dominant \ion{H}{i} component (c1L) at $z=0.54$ has approximately solar [$\alpha$/H], as does its lower density satellite component c2. The $z=0.47$ pLLS also shows a spread of relative abundances, with c1L and its two high-density (c2 and c4L) satellites having slightly super-solar [C/$\alpha$] (bottom panel), whereas the low-density c3 may have highly depleted C, but this result is fairly uncertain due to model dependencies, as discussed in \S\ref{sec:z0p47_c3}. Interestingly, the high-ionization phases of component 1 and 4, c1H and c4H, also have low [C/$\alpha$], with [C/$\alpha$]$_\mathrm{c1H}<-0.33$ and [C/$\alpha$]$_\mathrm{c4H}=\xpm{-0.28}{+0.18}{-0.17}$ inconsistent with [C/$\alpha$]$_{\rm c1L}=0.14\pm0.10$ and [C/$\alpha$]$_{\rm c4L}=0.33\pm0.14$.
    
    Across a wide range of elements, the 95\% credible intervals of [$\alpha$/H], [C/H], and [N/H] for c2 and c4L of the $z=0.47$ all overlap, suggesting they likely have a common origin distinct from that of other components. As we discuss in Section \ref{sec:z0p47_c3}, the carbon abundance of component c3 is particularly model dependent due to blending and possible saturation of \ion{C}{iii}.  If we exclude the anomalous [C/$\alpha]_\mathrm{c3}$, c3 would have consistent abundances with those of c2 and c4L suggesting it may be more highly-ionized gas with the same origin.  c1H and c4H appear to have lower [C/$\alpha$] and are both highly-ionized, suggesting that these absorption components may arise from gas with a shared history.  c1L with its low [$\alpha$/H] and roughly solar [C/$\alpha$] appears unique among the components in the system, with only the relatively poorly-constrained c5 and c6 having consistent limits. Collectively, this suggests at least three distinct origins of gas are present within the single halo at $z=0.47$.
    
    \begin{table*}
\caption{Galactic environment within 1 Mpc and $\Delta v_g<500\kms$ of the either pLLS.\label{table:galaxies}}
\renewcommand{\arraystretch}{1.2}
\begin{tabular}{llrRrrrrRRc}
\hline\hline
\multicolumn{1}{c}{ID} & \multicolumn{1}{c}{$z_g$} & \multicolumn{1}{c}{$d$} & \multicolumn{1}{c}{$\Delta v_g$} & \multicolumn{1}{c}{$m_r$} & \multicolumn{1}{c}{$\log\frac{M_\text{star}}{M_\odot}$}  & \multicolumn{1}{c}{$R_\mathrm{vir}$} & \multicolumn{1}{c}{$\theta$} & \multicolumn{1}{c}{$\Delta\alpha$} & \multicolumn{1}{c}{$\Delta\delta$} & \multicolumn{1}{c}{Type$^{a}$}\\
 &  & \multicolumn{1}{c}{(kpc)} & \multicolumn{1}{c}{(\kms)} & \multicolumn{1}{c}{(mag)} & & \multicolumn{1}{c}{(kpc)} & \multicolumn{1}{c}{($\arcsec$)} & \multicolumn{1}{c}{($\arcsec$)} & \multicolumn{1}{c}{($\arcsec$)} & \\

 \multicolumn{1}{c}{(1)} & \multicolumn{1}{c}{(2)} & \multicolumn{1}{c}{(3)} & \multicolumn{1}{c}{(4)} & \multicolumn{1}{c}{(5)} & \multicolumn{1}{c}{(6)}& \multicolumn{1}{c}{(7)} & \multicolumn{1}{c}{(8)} & \multicolumn{1}{c}{(9)} & \multicolumn{1}{c}{(10)} & \multicolumn{1}{c}{(11)}\\

 \hline
\multicolumn{11}{c}{Galaxies near $z_\mathrm{abs}$=0.4723}\\
\hline
J011035.02$-$164833.2 & 0.4725 & 55  & +45     & 21.7 & \xpm{10.34}{+0.04}{-0.07} & 210 & 9.0   & 7.4  & 5.5   & E \& A\\
J011033.82$-$164828.4 & 0.4731 & 148 & +168    & 23.6 & \xpm{9.02}{+0.14}{-0.18}  & 126 & 24.3  & 25.4 & 0.7   & E\\
J011031.47$-$164810.5 & 0.4736 & 370 & +280    & 22.3 & \xpm{9.72}{+0.10}{-0.12}  & 158 & 60.5  & 60.6 & -17.2 & E\\
J011029.59$-$164923.0 & 0.4713 & 617 & -189 & 23.2 & \xpm{8.57}{+0.13}{-0.12}  & 109 & 101.4 & 88.8 & 55.3  & E\\
\hline
\multicolumn{11}{c}{Galaxies near $z_\mathrm{abs}$=0.5413}\\
\hline
J011035.29$-$164753.3 & 0.5413 & 226 & +2     &  23.4 & \xpm{8.90}{+0.16}{-0.15}  & 121 & 34.5 & 3.4 & -34.4 & E \\
J011036.83$-$164740.6 & 0.5401 & 332 & -231 & 23.0 & \xpm{9.93}{+0.10}{-0.13}  & 172 & 50.8 & -19.8 & -47.1 & E\\
J011038.69$-$164804.7 & 0.5411 & 335 & -27  & 23.3 & \xpm{9.21}{+0.12}{-0.16}  & 134 & 51.1 & -47.7 & -23.0 & E\\
J011037.70$-$164915.6 & 0.5406 & 375 & -124 & 20.5 & \xpm{11.31}{+0.01}{-0.01} & 582 & 57.3 & -32.8 & 47.9 & A\\
J011032.08$-$164754.0 & 0.5429 & 392 & +314    & 22.4 & \xpm{9.75}{+0.11}{-0.15}  & 160 & 59.7 & 51.4 & -33.7 & E\\
J011039.21$-$164900.9 & 0.5425 & 411 & +245    & 23.9 & \xpm{8.73}{+0.19}{-0.20}  & 115 & 62.6 & -55.5 & 33.2 & E\\
J011039.92$-$164834.2 & 0.5393 & 416 & -387 & 22.7 & \xpm{10.29}{+0.06}{-0.12} & 204 & 63.6 & -66.1 & 6.5 &  A \\
J011041.05$-$164749.3 & 0.5396 & 578 & -328 & 23.3 & \xpm{9.84}{+0.28}{-0.34}  & 247 & 88.3 & -83.1 & -38.4 & E  \\
J011041.42$-$164752.6 & 0.5420 & 602 & +138    & 22.0 & \xpm{9.61}{+0.14}{-0.12}  & 152 & 91.8 & -88.6 & -35.1 & E \\
J011042.22$-$164757.8 & 0.5398 & 659 & -280 & 20.7 & \xpm{10.58}{+0.07}{-0.04} & 166 & 100.8 & -100.6 & -29.9 & E \\
J011027.80$-$164758.6 & 0.5413 & 750 & +2      & 21.6 & \xpm{9.72}{+0.14}{-0.11}  & 158 & 114.5 & 115.7 & -29.1 & E\\
\hline
\multicolumn{11}{l}{Galaxies are classified as emission-dominated (E), absorption-dominated (A), or both absorption and emission (A \& E).}
\end{tabular}
\end{table*}
    
    The two strongest absorbers in the $z=0.54$ pLLS (c1L and c2) appear to plausibly share a chemical history when considering the [$\alpha$/H] and [C/H] abundances, whereas the measured [N/H] across the two components are discrepant at the $2\sigma$ level. The N enhancement of c2 results nearly directly from the measured values of  \N{N\,III,c2} and \N{O\,III,c2}, and does not depend on the density, as demonstrated in Figure \ref{fig:z0p54_c2_CI}. However, both \N{N\,III,c2} and \N{O\,III,c2} are somewhat model-dependent due to blending with c1, making column densities sensitive to the precise component structure, which is uncertain. c3 of the $z=0.54$ system is constrained to have much lower [$\alpha$/H] and \nH, suggesting this complex also contains gas with diverse origins.
    
    In summary, the chemistry of the two $z=0.54$ strong components are very similar, suggesting they may have a common origin. In contrast, the low abundance of the dominant \ion{H}{i} component (c1) at $z=0.47$ indicates it likely does \textit{not} share a common origin with its high-metallicity satellite components or even with its kinematically-aligned high-ionization phase. \citetalias{2021MNRAS.506..877Z} found that abundances of individual absorption components associated with LLSs also vary widely. Taken together, the CUBS data show compelling evidence that the CGM is complex, with multiple physical processes influencing the gas distribution, abundance pattern, and ionization within single halos.
 
   \subsection{Galaxy Environment}
    \label{sec:results:galaxies}
    
\begin{figure}
\includegraphics[width=\linewidth]{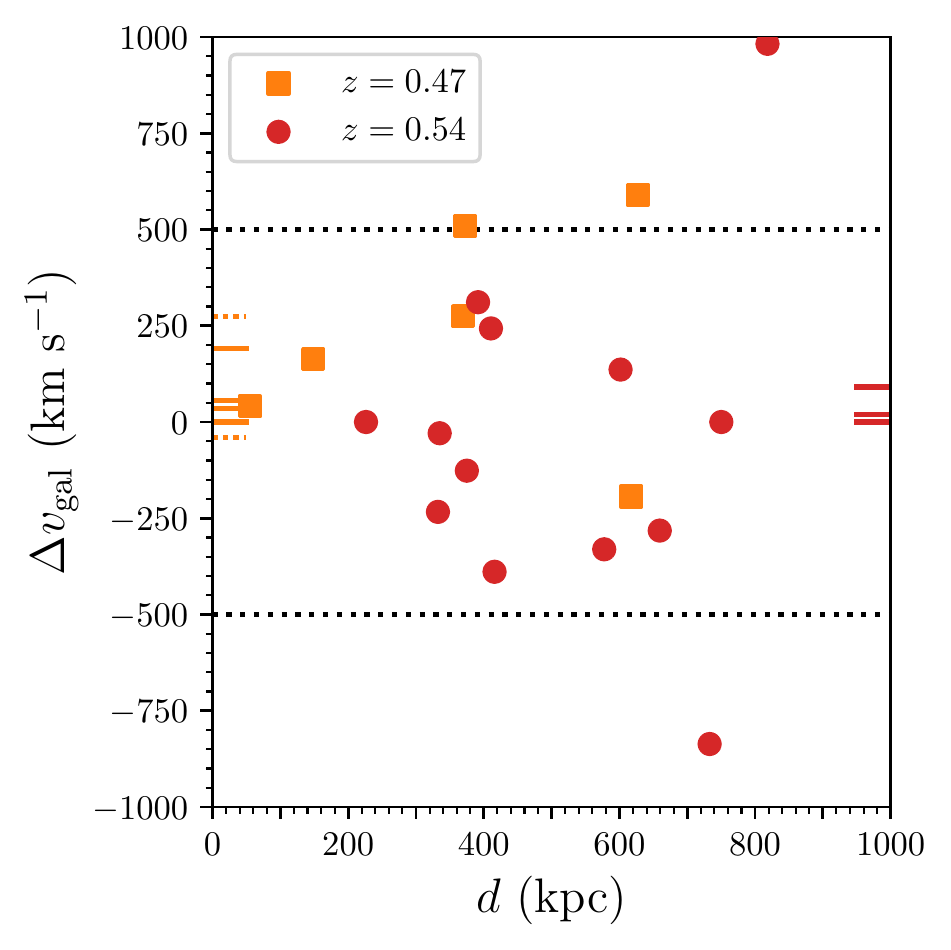}
\caption{Impact parameter and redshift relative to strongest \ion{H}{i} component of the two absorption systems. Short lines at the left and right vertical axes indicate the relative velocities of components in the $z=0.47$ and $z=0.54$ pLLSs, respectively. The dotted orange lines are \ion{H}{i} components that do not have associated metals. Dotted black lines at $\pm500$ \kms\ bracket the galaxies considered in this analysis. Completeness decreases at $\approx$ 375 kpc, corresponding to a 1\arcmin\ separation from the QSO line of sight.}
\label{fig:galaxy_seps}
\end{figure}

    Critical context for the nature and origin of these CGM absorbers is provided by the detailed view of their nearby galaxies as well as the full galactic environment of each absorber uncovered by the tiered CUBS galaxy survey. In Figure \ref{fig:galaxy_seps} we show impact parameters and velocity offsets of identified galaxies within $d<1000\,\rm{kpc}$ and $\left|\Delta v_g\right|<1000\,\kms$ based on secure redshift measurements. The velocity is relative to the centroid of the dominant \ion{H}{i} component (c1) for each system. At $z=0.54$ there are 11 galaxies within 500 \kms, in addition to two with $750<\left|\Delta v_g\right|/\kms<1000$, both at $d>700\,\rm{kpc}$. There is also a 300 kpc gap in impact parameter between the last shown galaxy within 500 \kms\ and the next closest one at 1050 kpc (not shown).\footnote{We note that the galaxy survey is less complete at separations larger than 1\arcmin, about 375 kpc at $z=0.57$.} Hence, we consider only the galaxies within 500 \kms\ in the rest of this work, as they may be a grouping of galaxies. We apply the same selection criterion to galaxies at $z=0.47$. There are four galaxies within 617 kpc with $\Delta v_g$ ranging from $-189$ to 280 \kms, with  additional two at $d>350$ kpc and $\Delta v_g\gtrsim500\,\kms$.  There are no other galaxies within 1 Mpc and $\left|\Delta v_g\right|<\,1000$\,\kms, with the next closest having $d=1200$ kpc.
    
 \begin{figure}
\includegraphics[width=\linewidth]{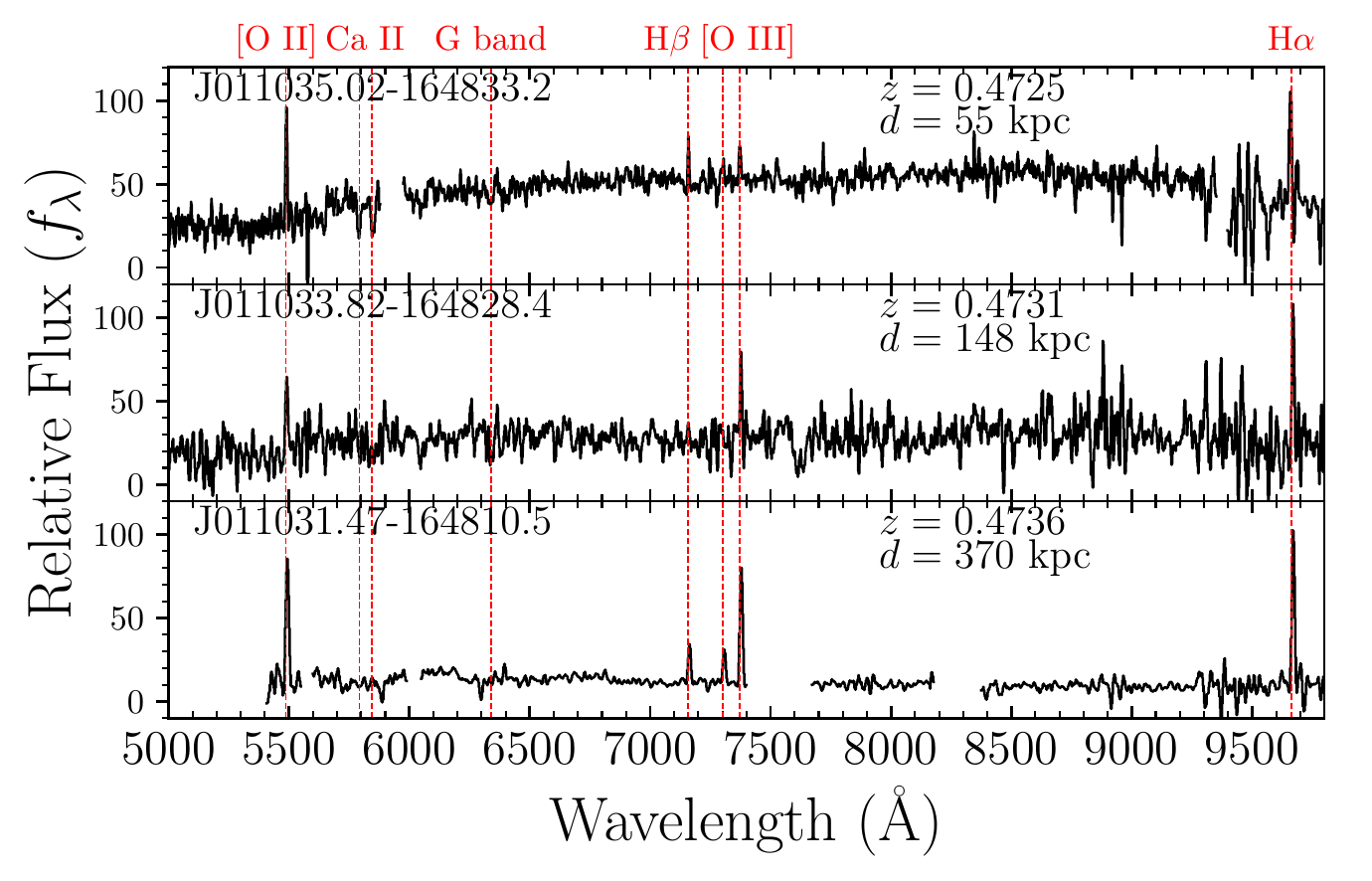}
\caption{Spectra of galaxies with $\left|\Delta v_g\right|<500$ \kms\ relative to the pLLS at $z=0.4723$ and impact parameters of $d<500$ kpc, obtained with the Magellan telescopes. The top panel is a MUSE spectrum, with Magellan data filling in wavelength coverage gaps at 5700--5800\AA\ and above 9400\AA. All three galaxies have emission lines indicative of star formation. The nearest galaxy at $d=55$ kpc has emission lines showing it is forming stars; however, it also exhibits strong Balmer absorption lines. \label{fig:z0p47_spectra}}
\end{figure}

\begin{figure}
\includegraphics[width=\linewidth]{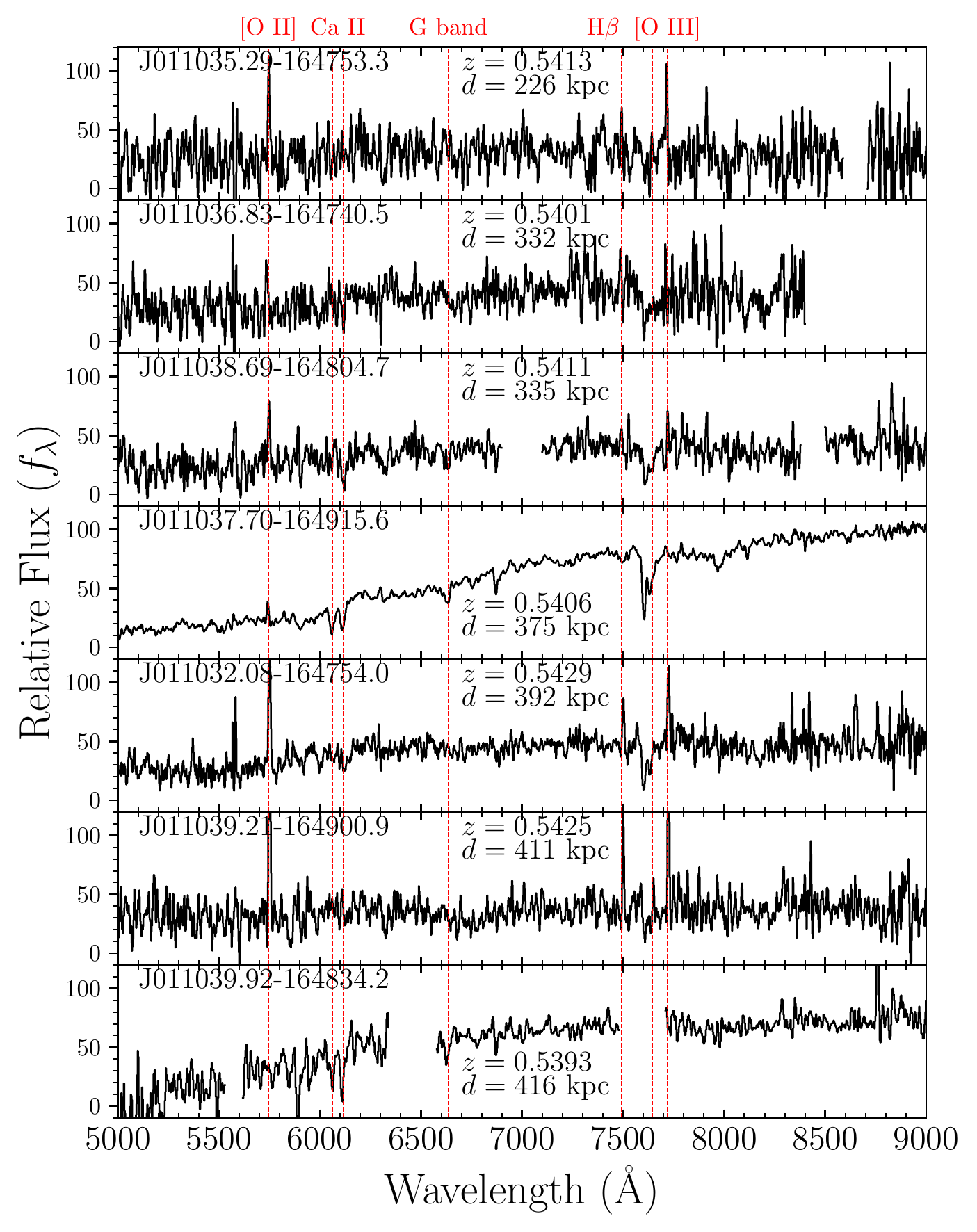}
\caption{Same as Figure \ref{fig:z0p47_spectra}, for the pLLS at $z=0.5413$. The nearest galaxy, at the top, has a projected separation from the quasar of 226 kpc, larger than the assumed extent of its CGM. The nearby galaxy environment is still rich, with a mixture of star-forming and passive galaxies, including an LRG at a distance of 375 kpc.\label{fig:z0p54_spectra}}
\end{figure}

    \begin{figure*}
    \includegraphics[width=\linewidth]{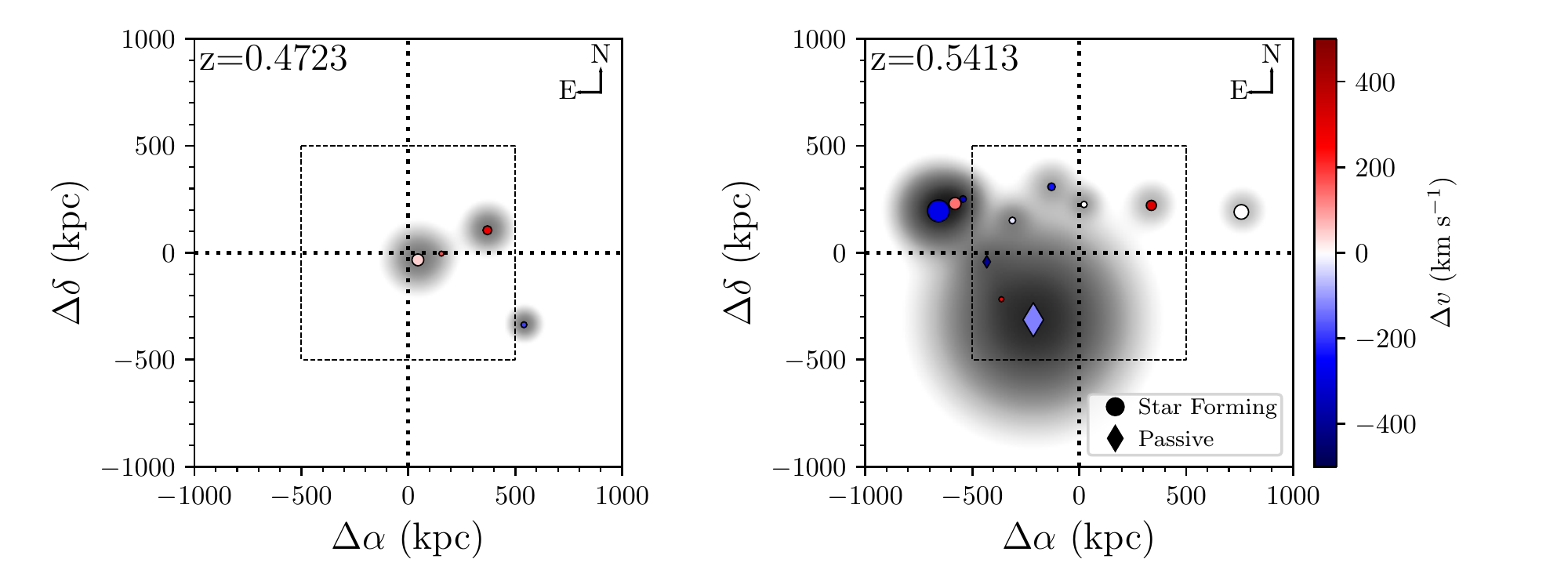}
    \caption{Plots of the galactic environment at the redshift of each pLLS. Marker colors correspond to velocity offset relative to the pLLS. To visualize the extent of each galaxies CGM, we overlay a 2D Gaussian with FWHM$=R_\mathrm{vir}$, and height scaling with luminosity. Marker size is scaled relative to $\log L/L_*$ The dotted box indicates the region shown in Figure \ref{fig:image}. At $z=0.47$ the nearby star-forming galaxy is likely associated with the absorption, while at $z=0.54$ only the LRG has $d<R_\mathrm{vir}$.} \label{fig:haloplots}
    \end{figure*}
    
The properties of CUBS galaxies within 500 \kms\ and 1 Mpc of the sightline are listed in Table \ref{table:galaxies} including the galaxy redshift (2), impact parameter to the QSO line of sight (3), velocity with respect to the pLLS (4), $r$-band magnitude (5), stellar mass (6), inferred dark matter halo virial radius (7), angular separation from the QSO (8), RA separation from the QSO (9) and declination offset from the QSO (10). Each galaxy is classified as either emission-dominated (E), absorption-dominated (A), or a combination (E \& A), based on spectra obtained with the Magellan telescopes, as listed in column 11. Spectra of galaxies at $z=0.47$ and $z=0.54$ within 500 kpc are shown in Figures \ref{fig:z0p47_spectra} and \ref{fig:z0p54_spectra}. 

To estimate the stellar masses of the galaxies, we perform stellar population synthesis fits to the available $grizH$ photometry using \texttt{Bagpipes} \citep{2018MNRAS.480.4379C}. In summary, Bagpipes uses MultiNext \citep{2013AIPC.1553..106F} and PyMultiNest \citep{2014A&A...564A.125B} to find the best-fit \citet{2003MNRAS.344.1000B} stellar population synthesis model and corresponding credibility intervals on the model parameters. We adopt exponentially declining $\tau$ models for the star formation histories with $\tau$ between 10 Myr and 15 Gyr, a \citet{2000ApJ...533..682C} dust law with $A_V$ between 0 and 2 mags, and stellar metallicity between 0 and 3.5 times solar. We report the resulting stellar mass estimates in Table \ref{table:galaxies} (column 6) assuming an initial mass function from \citet{2001MNRAS.322..231K}. Virial mass and radius estimates are then obtained following the relation given in \citet{2018AstL...44....8K}.
    
The closest identified galaxy to the $z=0.47$ pLLS is an $M_\mathrm{star}=10^{10.3} \msun$ star-forming galaxy with strong Balmer absorption located at an impact parameter of only 55 kpc, and a velocity offset of $\Delta v_g=45$ \kms\ (top panel of Figure \ref{fig:z0p47_spectra}). The H$\alpha$ emission is too strong for this galaxy to be a classical post-starburst \citep{2018ApJ...862....2F}, but with strong H$\delta$ absorption ($>5$ \AA) and D$_n$(4000)$>1$ there is clear evidence of a significant intermediate age stellar population, and current sub-dominant star-formation. \citet{2019MNRAS.489.5709C} showed that galaxies with these H$\alpha$ equivalent widths and H$\delta$ values are well-modeled by exponentially declining star-formation histories with a 300 Myr e-folding time. A second lower-mass ($M_\mathrm{star}=10^{9.0}M_\odot$) galaxy which is star-forming is located at $d=148$ kpc, and two additional $M_\mathrm{star}<10^{10.0}M_\odot$ star-forming galaxies are located still farther from the line of sight at $d=370$ and 617 kpc. The left panel of Figure \ref{fig:haloplots} provides a qualitative visual representation of the relative scale of the dark matter halos we expect these four galaxy to occupy using a modified version of the approach taken in \citet{2013MNRAS.434.1765J}. Each galaxy is represented by a Gaussian whose FWHM is equal to the $R_\mathrm{vir}$, and amplitudes that scale linearly with luminosity. These profiles are added together, such that the image appears darker where galaxies' halos overlap.

Given the proximity of the $z=0.47$ $M_\mathrm{star}=10^{10.3}M_\odot$ galaxy to the line of sight and its higher stellar mass, we expect some or all of the components in the pLLS complex are directly related to this galaxy. Given that it is currently star-forming and that its stellar absorption spectrum suggests it was likely more vigorously forming stars within the past Gyr, previous episodes of galactic winds \citep[e.g.,][]{2019MNRAS.488.1248H} may be responsible for the three absorption components with roughly solar metallicity (c2, c3,  c4L, Figure \ref{fig:abundances}).  Two of these, c2 and c4L, also show and a high degree of chemical maturity as measured by [C/$\alpha$], further cementing this likelihood.

The dominant \ion{H}{i} component at $z=0.47$ (c1L), with only 15\% solar $\alpha$ abundance and elemental abundance pattern in C and N consistent with solar ratios, is more likely to be either accreting gas or halo gas that is a mix of accretion and previous outflow/stripped ISM. As [C/$\alpha$] of the kinetically aligned higher-ionization phase of c1 is inconsistent with that of the low-ionization phase, we favor c1H as originating from the interaction of ambient, likely hotter, halo gas with the infalling low-metallicity material which gives rise to c1L.
   
  The low degree of enrichment of c1L is particularly interesting in light of its galaxy counterpart which appears to have a declining star-formation history. In order for the nearby galaxy to evolve into a quiescent system, gas such as this absorber must be prevented from forming stars. \citet{2018MNRAS.479.2547C} report that high-\N{H\,I} absorbers are commonly found at close separation from nearby massive and quiescent luminous red galaxies (LRGs), with \citet{2019MNRAS.484.2257Z} finding that the gas is often, but not uniformly, highly enriched and shows variation within a single halo, similar to this pLLS. Taken together, this suggest that the presence and properties of cool gas in the CGM is not driven entirely by the current star formation activity of nearby galaxies, and that the intermediate-redshift CGM is a complex system with many physical processes contributing.
  
     \begin{figure}
     \centering
     \includegraphics[width=\linewidth]{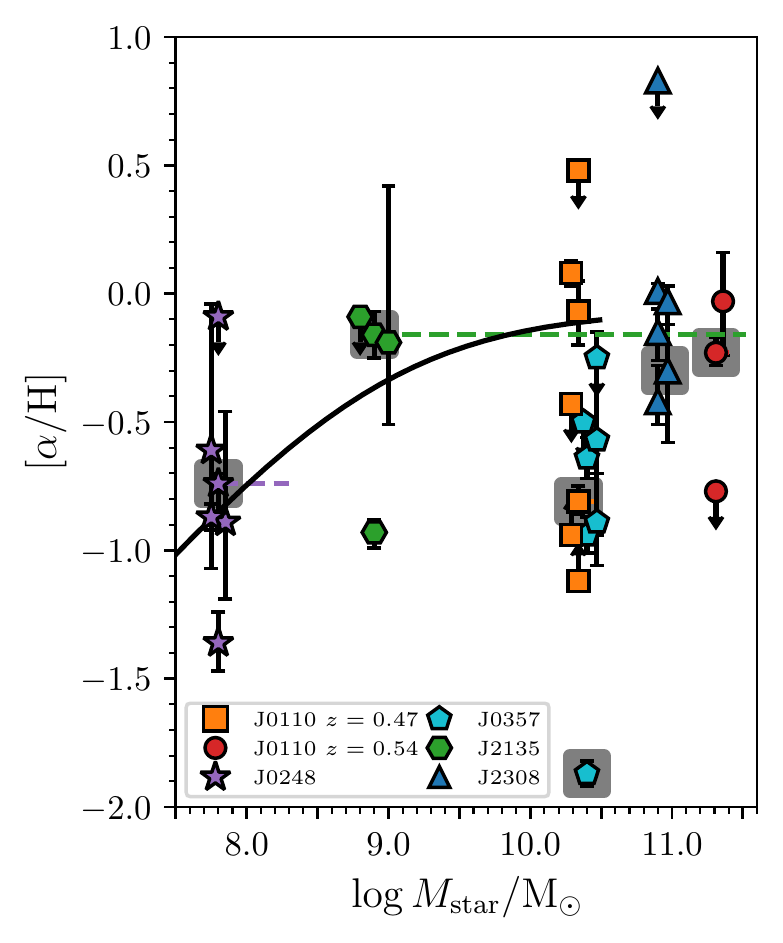}
     \caption{$\alpha$-element abundances for each component of six strong N$_\mathrm{H\,I}$ absorbers in the CUBS sample, plotted against stellar mass of their likely host galaxy (see \citetalias{2020MNRAS.497..498C} for LLS galaxy details). The LLS along the J2135 sightline is in a rich galaxy environment, having 10 galaxies within 250 kpc with $\log M_\mathrm{star}/M_\odot$ ranging from 8.9 to 11.2; we place this absorber at the mass of the closest galaxy. Dashed lines extend to the total stellar mass of all identified galaxies with 250 kpc and 500 \kms; this is only appreciably larger than the host galaxy mass for two of the absorbers. The largest N$_\mathrm{H\,I}$ component of each absorber is highlighted with a gray square; there is no trend with $M_\mathrm{star}$. The relationship between ISM gas-phase metallicity (mostly sensitive to [O/H]) versus stellar mass (commonly referred to as the mass-metallicity relationship) from \citet{2013ApJ...765..140A} is plotted in the black curve.}
     \label{fig:MZR}
 \end{figure}
 
  The galaxy environment at $z=0.54$, where we see the simpler absorber with only three metal components, is considerably more crowded, with 11 galaxies within an impact parameter of 750 kpc. The nearest galaxy to the sightline (226 kpc) is relatively low mass with $M_\mathrm{star}=10^{8.9}M_\odot$; however a massive $M_\mathrm{star}=10^{11.3}M_\odot$ LRG lies at 375 kpc separation.  The right panel of Figure \ref{fig:haloplots} makes clear that the LRG, even at this larger separation, is the only galaxy at $z=0.54$ with $d<R_\mathrm{vir}$. The solar $\alpha$ abundance and high degree of chemical maturity implied by the near solar [C/$\alpha$], [N/$\alpha$], and [Fe/$\alpha$] of c1 and c2 of the pLLS lends further support to this plausible association (but see \citealt{2017MNRAS.466.1071Z} regarding  an [Fe/$\alpha$] gradient around massive ellipticals). Interestingly,  c3, which is much higher ionization (only detected in \ion{H}{i} and \ion{O}{vi}) could represent a more diffuse volume filling phase in the outskirts of this LRG. If so, then the limit of $[\alpha/\rm{H}]<-0.77$ could imply the typical degree of enrichment for gas in the outer halo in these systems is low.
  
 These two systems also highlight the complexity of the relationship between \ion{Mg}{II} absorption and galaxies at intermediate redshift, particularly in a group environment. For the pLLS at $z=0.47$, the associated \ion{Mg}{ii} absorption is weaker than one would generally expect at $d=55$ kpc from a typical $L_*$ galaxy. Previous studies \citep[e.g.,][]{2010ApJ...714.1521C,2011ApJ...743...10B,2021MNRAS.502.4743H,2013ApJ...776..114N} would predict a Mg\,II absorber of rest-frame absorption equivalent width $W_r(2796)\gtrsim 0.1$ \AA. This is in contrast to a relatively weak Mg\,II absorption feature associated with the pLLS at $z=0.47$ with $W_r(2796)=0.075\pm 0.014$ \AA. At the same time, the $z=0.54$ pLLS associated with a massive LRG at $d = 375$ kpc exhibits a Mg\,II feature of $W_r(2796)=0.192\pm 0.005$ \AA, stronger than the Mg\,II absorption associated with the $z=0.47$ pLLS at closer projected distance to a galaxy.
  
 We also note that \citet{2016MNRAS.455.1713H} find that the covering fraction of \ion{Mg}{ii} with $W_r(2796)>0.3\,$\AA\ is a few percent at $d=$300-400 kpc around LRGs, however no systematic study has been conducted of these weaker \ion{Mg}{II} absorbers. Given the large separation of the pLLS and LRG, the results from \citet{2016MNRAS.455.1713H}, and the high degree of enrichment, we expect the richness of the galaxy environment surrounding this system may play an important role in the presence of this gas. Indeed \citet{2021MNRAS.502.4743H} have also shown that the observed mean $W_r(2796)$ vs.\ $d$ relation exhibits a larger scatter in more overdense galaxy environment, suggesting a more complex metal enrichment profile in galaxy groups. The observed galaxy environments for both the pLLS at $z=0.47$ and the absorber at $z=0.54$ reinforce this dynamic picture.

  Figure \ref{fig:MZR} presents a summary of the connection between the chemical abundances of diffuse circumgalactic gas traced by strong \N{HI} systems in CUBS with their galactic environment. The [$\alpha$/H] abundances of absorbers are plotted against the measured stellar masses of their presumed host galaxies. The total stellar mass of all nearby neighbors is indicated by the dashed horizontal lines. We include the absorbers studied in this work and in \citetalias{2021MNRAS.506..877Z}. This is compared to the relationship between ISM [O/H] abundances and stellar mass (the mass-metallicity relation) from \citet{2013ApJ...765..140A}, represented by the black curve. 
  
While the maximum abundance of CGM absorbers appears to be weakly correlated with stellar mass contained in their environments, overall, we see that individual halos exhibit a much wider range in abundance than the range of maximum values that gives rise to this visual correlation. Additionally, most of these systems (4/6) include at least one absorption component with a higher level of enrichment than would be typical of the ISM of their nearby galaxies and all contain absorption components with [$\alpha$/H] more than 0.5 dex below the predicted ISM metallicity. Taken together with Figure \ref{fig:phases} which shows the enormous dynamic range of gas densities captured in these systems, we see clear evidence that the CGM is complex, with multiple physical processes contributing to a single halo.
 
This range of abundances and abundance patterns, combined with the galactic environments of these absorbers, can help to disentangle the plausible origins of multi-phase gas within the intermediate CGM. One physical scenario that is predicted to result in multi-phase gas is the interaction of cool or cold gas with a surrounding hotter halo. \citet{2003ApJS..146..165S} describe such a scenario in relation to interfaces between the high-velocity clouds of the Milky Way and the hot halo. They suggest kinematic misalignment between different phases is due to sheering/shredding of dense gas through ram-pressure stripping. Such configurations would be expected for any cool dense gas moving through the hot halo, whether they are in-falling or gas resulting from a tidal interaction. Another scenario predicted to form multi-phase gas is radiative cooling in galactic winds. \citet{2018ApJ...862...56S} describe the formation of outflowing cool gas in simulations which predict co-moving cold ($T\sim10^4$ K) and hot gas along with gas at intermediate temperatures.
 
    The absorbers studied in this work likely represent both of these scenarios, with the solar and nearly solar [$\alpha$/H] absorbers in both the $z=0.47$ and $z=0.54$ systems resulting from some combination of outflows and galaxy interactions. The multi-phase gas that is kinematically aligned with the 15\% solar component at $z=0.47$ (c1H and c1L) likely results from the interaction of diffuse halo gas with lower-metalllicity infalling gas. The fact that both of the studied gaseous halos remain chemically inhomogeneous strongly implies that either mixing is inefficient within the CGM  or that the physical processes giving rise to this absorption are still ongoing at intermediate redshift.

    \section{Summary}
    \label{sec:summary}
    
    In this paper, we presented a detailed analysis of two partial Lyman limit systems (pLLSs) with $\N{H\,I}\approx (1-3)\times 10^{16}$ cm$^{-2}$ discovered at $z=0.47$ and $z=0.54$ in the Cosmic Ultraviolet Baryon Survey (CUBS). Using Bayesian Voigt profile fitting, we measure the absorption detected in {\it HST}/COS FUV spectra and optical echelle spectra from Magellan/MIKE. These data provide unambiguous evidence of kinematically aligned multi-phase gas that masquerades as a single-phase structure and can only be resolved by simultaneous accounting of the full range of observed ionic species.  These include five ionization stages of oxygen as well as states of many other elements (Mg, N, C, Si, S, Fe).  The full coverage provides a rigorous test of photoionization models.  We combine these absorption data with a detailed view of the small and large-scale galactic environment enabled by the tiered CUBS galaxy survey.  The galaxy survey, employing VLT/MUSE, Magellan/LDSS3-C, and Magellan/IMACS spectroscopy, is sensitive at $z=0.5$ to $L<0.01L_*$ galaxies within $\sim$ 250 kpc of the sightline, $L<0.1L_*$ galaxies within 370 kpc, and $L_*$ galaxies out to $\sim 4$ Mpc. Our findings are summarized as follows:
    
    \begin{enumerate}
    
        \item The pLLS at $z=0.47$ reveals gas with complex kinematics and phase structure 55 kpc from a star-forming galaxy of $\mstar\approx 2\times 10^{10}\,\msun$, which exhibits strong Balmer absorption and is best characterized by a declining star-formation history  (Section \ref{sec:results:galaxies}, Figures \ref{fig:image}, \ref{fig:galaxy_seps}, \ref{fig:haloplots}).  The pLLS is resolved into four separate metal-bearing components in addition to two detected only in \ion{H}{i} (Figure \ref{fig:z0p47_VP}).   While the dominant \ion{H}{i} component has $[\alpha/\rm{H}]_\text{c1L}=-0.81_{-0.05}^{+0.06}$ and comparably low [C/H] and [N/H], three metal-bearing satellite components exhibit a high degree of chemical enrichment ([$\alpha$/H] $> -0.12$) and, where well-measured, solar or super-solar C and N (Sections \ref{sec:z0p47_c1} -- \ref{sec:z0p47_c56}, \ref{sec:results:abundances} and Figure \ref{fig:abundances}). 
        
        Intriguingly, the high-ionization species kinematically aligned with the dominant \ion{H}{i} component does not have the same abundance pattern (Sections \ref{sec:z0p47_c1} and \ref{sec:results:abundances} and Figures \ref{fig:posteriors} and \ref{fig:abundances}), implying that kinematically aligned high- and low-ionization phases do not always share a common origin.
        
        We favor a scenario where the high-metallicity satellite absorbers originate in outflows from the galaxy at 55 kpc, while the dominant \ion{H}{i} component with 15\% solar metallicity results from infalling dense streams interacting with the diffuse halo gas. The presence of a significant reservoir of cool gas challenges our expectations of the gaseous environments of galaxies with declining star-formation histories. 

        \item The pLLS at $z=0.54$ reveals a simpler gaseous environment in a galaxy group anchored by a distant LRG of $\mstar\approx 2\times 10^{11}\,\msun$ at 375 kpc and 10 additional lower-mass galaxies within 750 kpc and 500 \kms\ of the absorber (Section \ref{sec:results:galaxies}, Figures \ref{fig:image}, \ref{fig:galaxy_seps}, \ref{fig:haloplots}).  The pLLS is resolved into two principle components separated by only 20 \kms\ (Figure \ref{fig:z0p54_VP}). While the components differ in density by nearly an order of magnitude, both have a high degree of chemical enrichment with $[\alpha/\rm{H}]>-0.15$. The dominant \ion{H}{i} component has nearly solar abundance of C and N, and super-solar Fe. A third component at $\Delta v_{\rm c3}=90$ \kms, detected only in \ion{H}{i} and \ion{O}{vi}, has a much lower metallicity with [$\alpha$/H]$<-0.77$ (Sections \ref{sec:z0p54_c1}, \ref{sec:z0p54_c2}, and \ref{sec:results:abundances} and Figure \ref{fig:abundances}). We interpret the two solar-metallicity components and high-ionization lower metallicity gas as originating in either the outer CGM of the LRG, or the poorly-mixed intra-group medium.

        \item The CGM of individual galaxies and galaxy groups is inhomogeneous in both density and abundance pattern (Section \ref{sec:results:abundances}, Figures \ref{fig:phases}, \ref{fig:posteriors}, \ref{fig:abundances}), and does not adhere to the ISM mass-metallicity relation (Figure \ref{fig:MZR}). Even kinematically aligned components which exhibit multiple phases of gas do not always share a common abundance pattern. The multiphase nature is clearly demonstrated by the increasing line width with ionization potential of the observed ions (Figure \ref{fig:b_v_E}, Section \ref{sec:results:multiphase}).  This result provides both an opportunity and a significant burden in that the complexity of the CGM is measurable, but only through careful detailed analysis that models each sub-component of the absorption separately and considers contributions from multiple gaseous phases to the intermediate ionization stages. 
        
        The high degree of chemical inhomogeneity within the CGM requires either inefficient mixing within the CGM or that the physical processes responsible for sculpting the CGM are still ongoing at $z=0.5$.
    
        \item Abundance determinations for absorbers detected in singly-ionized phases such as \ion{O}{ii} and \ion{Mg}{ii} are generally robust because the ionization fractions of hydrogen and these ions respond similarly to variations in density (Section \ref{sec:analysis}, Figures \ref{fig:z0p47_c1_CI}, \ref{fig:z0p47_c2_CI}, \ref{fig:z0p47_c4L_CI}, \ref{fig:z0p54_c1_CI}). Density determinations for these same absorbers may be subject to larger uncertainties, especially for those absorbers whose doubly ionized phase is found to be contributed to by both lower and higher ionization gas. Further uncertainties in density and metallicity arise from uncertainty in the UVB (Section \ref{sec:results:robust}, \citetalias{2021MNRAS.506..877Z}).
        
        \item The multiphase absorption phenomenon is emergent in diverse galaxy environments and varies on small spatial scales, providing a critical test of CGM simulations.  Realistic numerical models of the CGM must reproduce the inhomogeneous nature of metal enrichment and density shown in Figures \ref{fig:phases} and \ref{fig:abundances}.
        
  \end{enumerate}
  
  The data presented herein are a small subset of the full CUBS data set. Future papers will explore the properties and kinematics of gas surrounding galaxies with diverse environments, star formation histories and stellar masses. Collectively, these analyses will characterize the CGM of galaxies during the precipitous decline in the cosmic star formation rate density at $z<1$.

%%%%%%%%%%%%%%%%%%%%%%%%%%%%%%%%%%%%%%%%%%%%%%%%%%
\section*{Acknowledgements}

We thank the anonymous referee for a constructive report which improved the content and clarity of the manuscript. TJC and GCR acknowledge support from HST-GO- 15163.015A. HWC, EB, and MCC acknowledge partial support from HST-GO-15163.001A and NSF AST-1715692 grants. FSZ is grateful for the support of a Carnegie Fellowship from the Observatories of the Carnegie Institution for Science. SC gratefully acknowledges support from Swiss National Science Foundation grant PP00P2\_190092 and from the European Research Council (ERC) under the European Union’s Horizon 2020 research and innovation programme grant agreement No 864361. KLC acknowledges partial support from NSF AST-1615296. CAFG was supported by NSF through grants AST-1715216 and CAREER award AST-1652522; by NASA through grant 17-ATP17-0067; by STScI through grant HST-AR-16124.001-A; and by the Research Corporation for Science Advancement through a Cottrell Scholar Award and a Scialog Award. SL was funded by project FONDECYT 1191232.

This work is based on observations made with ESO Telescopes at the Paranal Observatory under programme ID 0104.A-0147(A), observations made with the 6.5m Magellan Telescopes located at Las Campanas Observatory, and spectroscopic data gathered under the HST-GO-15163.01A program using the NASA/ESA Hubble Space Telescope operated by the Space Telescope Science Institute and the Association of Universities for Research in Astronomy, Inc., under NASA contract NAS 5-26555. This research has made use of NASA’s Astrophysics Data System and the NASA/IPAC Extragalactic Database (NED) which is operated by the Jet Propulsion Laboratory, California Institute of Technology, under contract with the National Aeronautics and Space Administration.

\section*{Data Availability}

The data underlying this article will be shared on reasonable request to the corresponding author.

\bibliographystyle{mnras}
\bibliography{refs} % if your bibtex file is called example.bib
% Don't change these lines
\bsp	% typesetting comment
\label{lastpage}
\end{document}